\documentclass[11pt]{article}
\usepackage{amsmath}
\usepackage{changes}
\usepackage{natbib}
\usepackage{latexsym, epsfig, amssymb, amsmath, amsthm, graphicx, mathrsfs, booktabs}
\usepackage{rotating, tabularx, setspace,paralist}
\usepackage{color}
\usepackage{graphicx}
\usepackage[OT1]{fontenc}
\usepackage{hypernat}
\usepackage{anysize}
\renewcommand{\baselinestretch}{1.5}
\marginsize{1.2in}{1in}{0.55in}{1.55in} %

\makeatletter
\def\singlespace{\def\baselinestretch{1}\@normalsize}

\newtheorem{assumption}{Condition}
\newtheorem{lemma}{Lemma}
\newtheorem{proposition}{Proposition}
\newtheorem{theorem}{Theorem}
\renewcommand{\theequation}{
\arabic{equation}%
}

\makeatother

\newcommand{\bx}{\mbox{\bf x}}
\newcommand{\by}{\mbox{\bf y}}
\newcommand{\bz}{\mbox{\bf z}}
\newcommand{\bA}{\mbox{\bf A}}

\newcommand{\bD}{\mbox{\bf D}}

\newcommand{\bX}{\mbox{\bf X}}

\newcommand{\bZ}{\mbox{\bf Z}}
\newcommand{\bone}{\mbox{\bf 1}}
\newcommand{\bzero}{\mbox{\bf 0}}
\newcommand{\bveps}{\mbox{\boldmath $\varepsilon$}}

\newcommand{\bbeta}{\mbox{\boldmath $\beta$}}
\newcommand{\bdelta}{\mbox{\boldmath $\delta$}}
\newcommand{\btheta}{\mbox{\boldmath $\theta$}}
\newcommand{\bgamma}{\mbox{\boldmath $\gamma$}}

\newcommand{\bSig}{\mbox{\boldmath $\Sigma$}}
\newcommand{\tbx}{\widetilde \bx}

\newcommand{\var}{\mathrm{var}}
\newcommand{\cov}{\mathrm{cov}}
\newcommand{\corr}{\mathrm{corr}}

\newcommand{\veps}{\varepsilon}

\newcommand{\diag}{\mathrm{diag}}

\newcommand{\sgn}{\mathrm{sgn}}

\newcommand{\FS}{\text{FS}}
\newcommand{\argmin}{\mbox{argmin}}

\def\t{^T}

\begin{document}

\title{Interaction Pursuit with Feature Screening and Selection%
\date{\today}
\author{Yingying Fan, Yinfei Kong, Daoji Li and Jinchi Lv%
\thanks{Yingying Fan is Associate Professor, Data Sciences and Operations Department, Marshall School of Business, University of Southern California, Los Angeles, CA 90089 (E-mail: \textit{fanyingy@marshall.usc.edu}). Yinfei Kong is Assistant professor, Department of Information Systems and Decision Sciences, Mihaylo College of Business and Economics, California State University at Fullerton, CA 92831 (E-mail: \textit{yinfeiko@usc.edu}).
Daoji Li is Assistant Professor, Department of Statistics, University of Central Florida, Orlando, FL, 32816 (E-mail: \textit{daoji.li@ucf.edu}). Jinchi Lv is McAlister Associate Professor in Business Administration, Data Sciences and Operations Department, Marshall School of Business, University of Southern California, Los Angeles, CA 90089 (E-mail: \textit{jinchilv@marshall.usc.edu}). This work was supported by NSF CAREER Awards DMS-0955316 and DMS-1150318, USC Diploma in Innovation Grant, and a grant from the Simons Foundation. The authors sincerely thank Runze Li and Wei Zhong for sharing the R code for DC-SIS, and Jun S. Liu and Bo Jiang for providing the R code for SIRI. The authors also would like to thank the Joint Editor, Associate Editor, and referees for their valuable comments that have helped improve the paper significantly. Part of this work was completed while the first and last authors visited the Departments of Statistics at University of California, Berkeley and Stanford University. They sincerely thank both departments for their hospitality.}
\medskip\\
} %
}

\maketitle

\begin{abstract}
Understanding how features interact with each other is of paramount importance in many scientific discoveries and contemporary applications. Yet interaction identification becomes challenging even for a moderate number of covariates. In this paper, we suggest an efficient and flexible procedure, called the interaction pursuit (IP), for interaction identification in ultra-high dimensions. The suggested method first reduces the number of interactions and main effects to a moderate scale by a new feature screening approach, and then selects important interactions and main effects in the reduced feature space using regularization methods. 
Compared to existing approaches, our method screens interactions separately from main effects and thus can be more effective in interaction screening. Under a fairly general framework, we establish that for both interactions and main effects, the method enjoys the sure screening property in screening and oracle inequalities in selection. Our method and theoretical results are supported by several simulation and real data examples.
\end{abstract}

\textit{Running title}: Interaction Pursuit


\section{Introduction}\label{sec:Intro}
In many scientific discoveries, 
a fundamental question is how to identify important features within or across sources that may interact with each other in order to achieve better understanding of the risk factors. For instance, there is growing evidence in genome-wide association studies supporting the presence of interactions between different genes or single nucleotide polymorphisms (SNPs) towards the risks of complex diseases 
\citep{xu2004interaction, musani2007detection, schwender2008identification, cordell2009detecting}. It has also been increasingly recognized that the aetiology of most common diseases relates to not only genetic and environmental factors, but also interactions between the genes and environment \citep{hunter2005gene}. In these problems, ignoring interactions by considering main effects alone can lead to an inaccurate estimate of the population attributable risk associated with these factors. Identifying important interactions can also help improve model interpretability and prediction. 

Interaction identification with large-scale data sets poses great challenges since the number of pairwise interactions increases quadratically with the number of covariates $p$ and that of higher-order interactions grows even faster. In the low-dimensional setting, one may include all possible interactions in a model and find significant ones by multiple testing or variable selection methods. This simple strategy, however, becomes impractical or even infeasible when $p$ is moderate or large, owing to rapid increase in dimensionality incurred by interactions. 
There is a growing literature developing regularization methods to identify important interactions and main effects, with a focus on the low- or moderate-dimensional setting. Most of existing methods are rooted on a natural structural condition in certain applications, namely the strong or weak heredity assumption, and impose various constraints on coefficients to enforce the heredity assumption. Specifically, the strong heredity assumption requires that an interaction between two variables be included in the model only if both main effects are important, while the weak one relaxes such a constraint to the presence of at least one main effect being important. To name a few, \citet{yuan2009structured} employed the non-negative garrote \citep{breiman1995better} for structured variable selection and estimation by imposing multiple inequality constraints on coefficients. \citet{choi2010variable} reparameterized the coefficients of interactions to enforce the strong heredity constraint and showed that the resulting method enjoys the oracle property when $p=o(n^{1/10})$, where $n$ is the sample size. \citet{bien2013lasso} extended the Lasso \citep{tibshiranit1996regression} by adding a set of convex constraints to enforce the strong or weak heredity constraint.

The aforementioned methods with delicate design on the interaction structure are effective in identifying important interactions when the number of covariates $p$ is not large. In the regime of ultra-high dimensionality, that is, $p$ growing nonpolynomially with sample size $n$, those methods may, however, become inefficient or even fail, because they need to deal with complex penalty structures or multiple inequality constraints and thus the computational cost can be excessively expensive. In addition, it is unclear whether the theoretical results on variable selection for those methods can still hold when $p$ is ultra high. To reduce the computational cost, \citet{hall2014selecting} proposed a two-step recursive approach rooted on the strong heredity assumption 
to screen interactions based on the sure independence screening \citep{fan2008sure}. 
\citet{hao2014interaction} introduced a forward selection based procedure to identify interactions in a greedy fashion under the heredity assumption  and developed two algorithms iFORT and iFORM. 
\citet{hao2015model} studied regularization methods based on the Lasso for quadratic regression models 
under the heredity assumption and proposed a new algorithm RAMP for interaction identification.
Although the heredity assumption is desired and natural in many applications, it can also be easily violated in some situations as documented in the literature.  For example, \citet{Culverhouse2002} discussed the interaction models displaying no main effects and examined the extent to which pure epistatic interactions whose loci do not display any single-locus effects could account for the variation of the phenotype. In the Nature review paper \citet{cordell2009detecting}, concerns were raised that many existing methods may miss pure interactions in the absence of main effects. Efforts have already been made on detecting pure epistatic interactions in \cite{Ritchie2001},  where a real data example was presented to demonstrate the existence of such pure interactions.
In these applications, methods that are released from the heredity constraint can enjoy better flexibility and be more suitable for models with pure epistatic interactions.

To address the challenges of interaction identification in ultra-high dimensions and broader settings, we present our ideas by focusing on the linear interaction model
\begin{eqnarray}\label{eq:LM}
Y = \beta_0 + \sum_{j=1}^p\beta_j X_j + \sum_{k=1}^{p-1}\sum_{\ell = k+1}^p \gamma_{k\ell}X_k X_{\ell} + \varepsilon,
  \end{eqnarray}
where $Y$ is the response variable, $\bx=(X_1, \cdots, X_p)^T$ is a $p$-vector of covariates $X_j$'s, $\beta_0$ is the intercept, $\beta_j$'s and $\gamma_{k\ell}$'s are regression coefficients for main effects and interactions, respectively, and $\varepsilon$ is the mean zero random error independent of $X_j$'s. 
Denote by  $\bbeta_0 = (\beta_{0, j})_{1 \leq j \leq p}$ and $\bgamma_0 = (\gamma_{0, k\ell})_{1 \leq k < \ell \leq p}$  the true regression coefficient vectors for main effects and interactions, respectively. To ease the presentation, throughout the paper $X_k X_{\ell}$ is referred to as an \textit{important interaction} if its regression coefficient $\gamma_{0, k\ell}$ is nonzero, and $X_k$ is called an \textit{active interaction variable} if there exists some $1\leq \ell \neq k \leq p$ such that $X_kX_{\ell}$ is an important interaction.
Under the above model setting, we suggest a new approach, called the interaction pursuit (IP), for interaction identification using the ideas of feature screening and selection. The IP is a two-step procedure that first reduces the number of interactions and main effects to a moderate scale by a new feature screening approach, and then identifies important interactions and main effects in the reduced feature space, with interactions reconstructed based on the retained interaction variables, using regularization methods. A key innovation of IP is to screen interaction variables
instead of interactions directly and thus the computational cost can be reduced substantially from a factor of $O(p^2)$ to $O(p)$. 
Our interaction screening step shares a similar spirit to the SIRI proposed in \cite{jiang2014sliced} in the sense of detecting interactions by screening interaction variables. An important difference, however, lies in that SIRI was proposed under the sliced inverse index model and its theory relies heavily on the normality assumption.

The main contributions of this paper are threefold. First, the proposed procedure is computationally efficient thanks to the idea of interaction variable screening. Second, we provide theoretical justifications of the proposed procedure under mild interpretable conditions. Third, our procedure can deal with more general model settings without requiring the heredity or normality assumption, which provides more flexibility in applications. In particular, two key messages that we try to deliver in this paper are that a separate screening step for interactions can significantly improve the screening performance if one aims at finding important interactions, and screening interaction variables can be more effective and efficient than screening interactions directly due to the noise accumulation. We also would like to emphasize that although we advocate a separate screening step for interactions, we have no intension to downgrade the importance of main effect screening or even a joint screening of main effects and interactions. In fact, our interaction screening idea can be coupled with any main effect or joint screening procedure to boost the performance of feature screening in interaction models.

The rest of the paper is organized as follows. Section \ref{sec:Screening} introduces a new feature screening procedure for interaction models and investigates the theoretical properties of the proposed screening procedure. We exploit the regularization methods to further select important interactions and main effects and study the theoretical properties on variable selection in Section \ref{sec:Selection}. Section \ref{sec:Numerical} demonstrates the advantage of our proposed approach through simulation studies and a real data example. We discuss some implications and extensions of our method in Section \ref{sec:Discussion}. 
The proofs of all the results and technical details as well as some additional simulation studies are provided in the Supplementary Material.

\section{Interaction screening}\label{sec:Screening}
We begin with considering the problem of feature screening in interaction models with ultra-high dimensions. 
Define three sets of indices
\begin{align}
\mathcal{I} & =\left\{(k, \ell): 1\leq k < \ell\leq p \text{ with } \gamma_{0,k\ell}\neq 0\right\}, \nonumber\\
\mathcal{A} & =\left\{1\leq k\leq p: (k, \ell) \text{ or } (\ell, k) \in \mathcal{I} \text{ for some } \ell\right\}, \label{intvarsets} \\
\mathcal{B} & =\left\{1\leq j\leq p: \beta_{0,j}\neq 0\right\}.\nonumber
\end{align}
The set $\mathcal{I}$ contains all important interactions and the set $\mathcal{A}$ consists of all active interaction variables, while the set $\mathcal{B}$ is comprised of all important main effects. We combine sets $\mathcal{A}$ and $\mathcal{B}$, and define the set of important features as $\mathcal{M}=\mathcal{A}\cup \mathcal{B}$. As demonstrated in Section B of Supplementary Material, the sets $\mathcal{A}$, $\mathcal{I}$, and $\mathcal{M}$ are invariant under affine transformations 
 $X_j^{new} = b_j(X_j-a_j)$ with $a_j \in \mathbb{R}$ and $b_j \in \mathbb{R}\setminus \{0\}$ for $1\leq j\leq p$. 
{ We aim at recovering interactions in $\mathcal{I}$ and variables in $\mathcal{M}$} and thus there is no issue of identifiability. \footnote{We would like to thank a referee for helpful comments on the issue of invariance.} 

\subsection{A new interaction screening procedure}\label{sec: IP}
Without loss of generality, assume that $E X_j=0$ and $E X_j^2=1$ for each random covariate $X_j$. To ensure model identifiability and interpretability, we impose the sparsity assumption that only a small portion of the interaction and main effects are important with nonzero regression coefficients $\gamma_{k\ell}$ and $\beta_j$ in interaction model \eqref{eq:LM}. 
Our goal is to effectively identify all important interactions $\mathcal{I}$ and important features $\mathcal{M}$, and efficiently estimate the regression coefficients in  \eqref{eq:LM} and predict the future response.
Clearly, $\mathcal{I}$ is a subset of all pairwise interactions constructed from variables in $\mathcal{A}$. Thus, as mentioned before, to recover the set of important interactions $\mathcal{I}$ we first aim at screening the interaction variables while retaining active ones in set $\mathcal{A}$.

Let us develop some insights into the problem of interaction screening by considering the following specific case of interaction model \eqref{eq:LM}: 
\begin{eqnarray}\label{example}
Y = X_1X_2 + \varepsilon,
\end{eqnarray}
where $\bx$ is further assumed to be $N(\bzero, \bSig)$ with covariance matrix $\bSig$ having diagonal entries $1$ and off-diagonal entries $-1 < \rho < 1$. Simple calculations show that $\corr(X_j, Y)=0$ for each $j$. This entails that screening the main effects based on their marginal correlations with the response can easily miss the active interaction variable $X_1$. An interesting observation is, however, that taking the squares of all variables leads to $\cov(X_1^2, Y^2)=2+10\rho^2$ and $\cov(X_j^2, Y^2) = 4\rho^2(1+2\rho)$ for each $j \geq 3$, where the former is always larger than the latter in absolute value regardless of the value of $-1 < \rho < 1$. Thus, the active interaction variable $X_1$ can be safely retained by ranking the marginal correlations between the squared covariates and the squared response, that is, $X_j^2$ and $Y^2$. By symmetry, the same is true for the other active interaction variable $X_2$.

Model \eqref{example} is a specific model with only one interaction. The following proposition provides justification for more general interaction models.

\begin{proposition}\label{prop1}
In interaction model \eqref{eq:LM} with $\bx \sim N(\bzero, I_p)$, it holds that for each $j$,
\begin{eqnarray}\label{eq: cov}
\cov(X_j^2, Y^2)=2\Big(\beta_{0,j}^2+ \sum_{k=1}^{j-1} \gamma_{0,kj}^2 + \sum_{\ell=j+1}^{p} \gamma_{0,j\ell}^2 \Big).
\end{eqnarray}
\end{proposition}

Proposition \ref{prop1} shows that for the specific case of $\bSig = I_p$, the correlation between $X_j^2$ and $Y^2$ is always nonzero as long as $X_j$ is an active interaction variable, regardless of whether or not $X_j$ is an important main effect. In contrast, such a correlation becomes zero if $X_j$ is neither an important main effect nor an active interaction variable. In fact, it is seen from \eqref{eq: cov} that $\cov(X_j^2, Y^2)$ measures the cumulative effect of $X_j$ as an important main effect or an active interaction variable. 

Motivated by the simple interaction model \eqref{example} and Proposition \ref{prop1}, we propose to identify the set of active interaction variables $\mathcal{A}$ by first ranking the marginal correlations $\corr(X_k^2, Y^2)$ in magnitude, 
and then retaining the top ones with absolute correlations bounded from below by some positive threshold. This gives a new interaction screening procedure which is the first step of IP. 
More specifically, suppose we are given a sample $(\bx_i, y_i)_{i = 1}^n$ of $n$ independent and identically distributed (i.i.d.) observations from $(\bx, Y)$ in interaction model \eqref{eq:LM}. Observe that $\corr(X_k^2, Y^2) = \omega_k /\{\var(Y^2)\}^{1/2}$ with $\omega_k=\cov(X_k^2, Y^2)/\left\{\var(X_k^2)\right\}^{1/2}$.
Denote by $\widehat{\omega}_k$ the empirical version of the population quantity $\omega_k$ by plugging in the corresponding sample statistics, based on the sample $(\bx_i, y_i)_{i = 1}^n$. Then the screening step of IP is equivalent to thresholding the absolute values of $\widehat{\omega}_k$'s; that is, we estimate the set of active interaction variables $\mathcal{A}$ as
\begin{eqnarray}\label{eq: Ahat}
\widehat{\mathcal{A}}=\left\{1\leq k\leq p: \left|\widehat{\omega}_k\right|\geq \tau\right\}
\end{eqnarray}
for some threshold $\tau > 0$. 
The choice of threshold $\tau$ will be discussed later.
  Based on the retained interaction variables in $\widehat{\mathcal{A}}$, we can construct all pairwise interactions as
\begin{equation} \label{Ihatdef}
\widehat{\mathcal{I}}=\left\{(k,\ell): k, \ell\in\widehat{\mathcal{A}} \ \text{and } k < \ell\right\}.
\end{equation}
It is worth mentioning that $\widehat{\mathcal{I}}$ generally provides an overestimate of the set of important interactions $\mathcal{I}$, in the sense that some interactions in the constructed set $\widehat{\mathcal{I}}$ may be unimportant ones. This is, however, not an issue for the purpose of interaction screening and will be addressed later in the selection step of IP.

For completeness, we also briefly describe our procedure for main effect screening.
We adopt the SIS approach in \citet{fan2008sure} to screen unimportant main effects outside the set $\mathcal{B}$; that is, we first calculate the marginal correlations $\corr(X_j, Y)$ and then keep the ones with magnitude at or above some positive threshold $\widetilde{\tau}$.
Since we have assumed $E X_j=0$ and $E X_j^2=1$ for each covariate $X_j$, thresholding the marginal correlation between $X_j$ and $Y$ is equivalent to thresholding $\mathcal{\omega}_j^{\ast} = E(X_jY)$. Thus, we estimate the set
$\mathcal{B}$
by
\begin{eqnarray}\label{eq: Bhat}
\widehat{\mathcal{B}}=\left\{1\leq j\leq p: |\widehat{\omega}^{\ast}_j|\geq \widetilde{\tau}\right\},
\end{eqnarray}
where $\widehat{\omega}^{\ast}_j$ is the sample version of the population quantity $\mathcal{\omega}_j^{\ast}$ and $\widetilde{\tau} > 0$ is some threshold. Finally the set of important features $\mathcal{M}$ can then be estimated as $\widehat{\mathcal{M}}=\widehat{\mathcal{A}}\cup \widehat{\mathcal{B}}$. 
Although our approach for estimating the set 
$\mathcal{B}$ is the same as SIS,
the theoretical developments on the screening property for main effects are distinct from those in \citet{fan2008sure} due to the presence of interactions in our model.


\subsection{Sure screening property}\label{sec: Screening-Property}

We now turn our attention to the theoretical properties of the proposed screening procedure in IP.
It is desirable for a feature screening procedure to possess the sure screening property \citep{fan2008sure}, which means that all important variables are retained after screening with probability tending to one. We aim at establishing such a property for IP in terms of screening of both interactions and main effects. To this end, we need the following conditions. 

\begin{assumption}\label{con: sparsity}
There exist constants $0\leq \xi_1, \xi_2<1$ such that $s_1=|\mathcal{I}|=O(n^{\xi_1})$ and $s_2=|\mathcal{B}|=O(n^{\xi_2})$, and $|\beta_0|, \|\bbeta_0\|_{\infty}, \|\bgamma_0\|_{\infty} = O(1)$ with $\|\cdot\|_\infty$ denoting the vector $L_\infty$-norm.
\end{assumption}

\begin{assumption}\label{con: XYtail-new}
There exist constants $\alpha_1, \alpha_2, c_1>0$ such that for any $t>0$, $P(|X_j|> t) \leq c_1\exp(-c_1^{-1}t^{\alpha_1})$ for each $1\leq j\leq p$ and $P(|\varepsilon|> t) \leq c_1\exp(-c_1^{-1}t^{\alpha_2})$, and $\var(X_j^2)$ are uniformly bounded away from zero.
\end{assumption}

\begin{assumption}\label{con: signal}
There exist some constants $0\leq \kappa_1, \kappa_2 <1/2$ and $c_2>0$ such that $\min\nolimits_{k\in \mathcal{A}}|\omega_k|\\ \geq 2c_2n^{-\kappa_1}$ and \textbf {$\min\nolimits_{j\in \mathcal{B}}|\omega^{\ast}_j|\geq 2c_2n^{-\kappa_2}$. }
\end{assumption}

Condition \ref{con: sparsity} allows the numbers of important interactions and important main effects to grow with the sample size $n$, and imposes an upper bound on the magnitude of true regression coefficients. See, for example, \cite{cho2012high} and \citet{hao2014interaction} for similar assumptions. Clearly, Condition \ref{con: sparsity} entails that the number of active interaction variables is at most $2 s_1$, that is, $|\mathcal{A}| \leq 2s_1$.

The first part of Condition \ref{con: XYtail-new} is a usual assumption to control the tail behavior of the covariates and 
error, which is important for ensuring the sure screening property of our procedure. Similar assumptions have been made in such work as 
\cite{fan2010sure}, \cite{chang2013marginal}, and \cite{barut2016conditional}.
The scenario of $\alpha_1=\alpha_2=2$ corresponds to the case of sub-Gaussian covariates and error, including distributions with bounded support and light tails. 


Condition \ref{con: signal} puts constraints on the minimum marginal correlations, through different forms, for active interaction variables and important main effects, respectively. It is analogous to Condition 3 in \citet{fan2008sure}, and can be understood as an assumption on the minimum signal strength in the feature screening setting. Smaller constants $\kappa_1$ and $\kappa_2$ correspond to stronger marginal signals. This condition is crucial for ensuring that the marginal utilities carry enough information about the active interaction variables and important main effects. To gain more insights into Condition \ref{con: signal}, consider the specific case of $\bx\sim N(\bzero, I_p)$. Note that $\var(X_k^2)$ are  uniformly bounded by Condition \ref{con: XYtail-new}. Then it follows from Proposition \ref{prop1} that the constraint of $\min_{k\in\mathcal{A}}|\omega_k|\geq 2c_2n^{-\kappa_1}$ in Condition \ref{con: signal} is equivalent to that of 
    \begin{align*}
     \min_{k\in \mathcal{A}}\Big(\beta_{0, k}^2+\sum_{j=1}^{k-1}\gamma^2_{0, jk}+\sum_{\ell=k+1}^p\gamma^2_{0, k\ell}\Big)\geq c n^{-\kappa_1},
   \end{align*}
where $c$ is some positive constant which may be different from $c_2$. Thus Condition \ref{con: signal} can be understood as constraints imposed indirectly on the true nonzero regression coefficients.

Under these conditions, the following theorem shows that the sample estimates of the marginal utilities are sufficiently close to the population ones with significant probability, and establishes the sure screening property for both interaction and main effect screening.

\begin{theorem}\label{Th: Sure Screening-new}
(a) Under Conditions \ref{con: sparsity}--\ref{con: XYtail-new}, 
if $0\leq \max\{2\kappa_1+4\xi_1, 2\kappa_1+4\xi_2\}<1$  and $E(Y^4)=O(1)$, then for any $C>0$, there exists some constant $C_1>0$ depending on $C$ such that for $\log p=o(n^{\alpha_1\eta})$ with $\eta=\min\{(1-2\kappa_1-4\xi_2)/(8+\alpha_1),\,(1-2\kappa_1-4\xi_1)/(12+\alpha_1)\}$,
   \begin{eqnarray}\label{eq: bound-omega-new}
       P(\max_{1\leq k\leq p}|\widehat{\omega}_k-\omega_k|\geq Cn^{-\kappa_1})
     =o(n^{-C_1}).
   \end{eqnarray}

(b) Under Conditions \ref{con: sparsity}--\ref{con: XYtail-new}, 
if $0\leq \max\{2\kappa_2+2\xi_1, 2\kappa_2+2\xi_2\}<1$ and $E(Y^2)=O(1)$, then for any $C>0$, there exists some constant $C_2>0$ depending on $C$ such that
   \begin{eqnarray}\label{eq: bound-omega-star-new}
       P(\max_{1\leq j\leq p}|\widehat{\omega}^{\ast}_j-\omega^{\ast}_j|\geq Cn^{-\kappa_2})
     =o(n^{-C_2})
   \end{eqnarray}
 for $\log p=o(n^{\alpha_1\eta'})$ with $\eta'=\min\{(1-2\kappa_2-2\xi_2)/(4+\alpha_1), (1-2\kappa_2-2\xi_1)/(6+\alpha_1)\}$.

(c) Under Conditions \ref{con: sparsity}--\ref{con: signal} and the choices of $\tau = c_2n^{-\kappa_1}$ and $\widetilde{\tau}= c_2n^{-\kappa_2}$, if $0 \leq \xi_1, \xi_2 < \min\{1/4-\kappa_1/2, 1/2-\kappa_2\}$  and $E(Y^4)=O(1)$, then we have
\begin{eqnarray}\label{eq: sure1}
      P\Big(\mathcal{I}\subset\widehat{\mathcal{I}} \ \text{ and } \ \mathcal{M}\subset\widehat{\mathcal{M}}\Big)
     &=& 1-o\Big(n^{-\min\{C_1, C_2\}}\Big)
  \end{eqnarray}
 for $\log p = o(n^{\alpha_1 \min\{\eta, \eta'\}})$ with constants $C_1$ and $C_2$ given in \eqref{eq: bound-omega-new} and \eqref{eq: bound-omega-star-new}, respectively. 
 In addition, it holds that
\begin{align}
      P & \left(|\widehat{\mathcal{I}}| \leq O\{n^{4\kappa_1} \lambda_{\max}^2(\bSig^{\ast})\} \text{ and }
                 |\widehat{\mathcal{M}}|
 \leq O\{n^{2\kappa_1}\lambda_{\max}(\bSig^{\ast})+n^{2\kappa_2} \lambda_{\max}(\bSig)\} \right)  \nonumber\\
  & = 1-o\left(n^{-\min\{C_1, C_2\}}\right) \label{eq: model-size-1},
\end{align}
where $\lambda_{\max}(\cdot)$ denotes the largest eigenvalue, $\bSig=\mathrm{cov}(\bx)$, and $\bSig^{\ast}=\mathrm{cov}(\bx^{\ast})$ for $\bx^{\ast}=(X_1^{\ast}, \cdots, X_p^{\ast})^T$ with $X_k^{\ast}=(X_k^2-E X_k^2)/\{\var(X_k^2)\}^{1/2}$.
\end{theorem}

Comparing the results from the first two parts of Theorem \ref{Th: Sure Screening-new} on interactions and main effects, respectively, we see that interaction screening generally requires more restrictive assumption on dimensionality $p$. This reflects that the task of interaction screening is intrinsically more challenging than that of main effect screening. In particular, 
 when $\alpha_1=2$, IP can handle ultra-high dimensionality up to
\begin{equation} \label{neweq001}
\log p
   = o\left(n^{\min\{(1-2\kappa_1-4\xi_2)/5, \,(1-2\kappa_1-4\xi_1)/7, \, (1-2\kappa_2-2\xi_2)/3, \,(1-2\kappa_2-2\xi_1)/4\}}\right).
\end{equation}
It is worth mentioning that both constants $C_1$ and $C_2$ in the probability bounds \eqref{eq: bound-omega-new}--\eqref{eq: bound-omega-star-new} can be chosen arbitrarily large without affecting the order of $p$ and ranges of constants $\kappa_1$ and $\kappa_2$. We also observe that stronger marginal signal strength for interaction variables and main effects, in terms of smaller values of $\kappa_1$ and $\kappa_2$, can enable us to tackle higher dimensionality.

The third part of Theorem \ref{Th: Sure Screening-new} shows that IP enjoys the sure screening property for both interaction and main effect screening, and admits an explicit bound on the size of the reduced model after screening. More specifically, an upper bound of the reduced model size is controlled by the choices of both thresholds $\tau$ and $\widetilde{\tau}$, and the largest eigenvalues of the two population covariance matrices $\bSig^{\ast}$ and $\bSig$. If we assume $\lambda_{\max}(\bSig^{\ast})=O(n^{\xi_3})$ and $\lambda_{\max}(\bSig)=O(n^{\xi_4})$ for some constants $\xi_3, \xi_4 \geq 0$, then with overwhelming probability the total number of interactions and main effects in the reduced model is at most of a polynomial order of sample size $n$. 

The thresholds $\tau=c_2n^{-\kappa_1}$ and $\tilde{\tau}=c_2n^{-\kappa_2}$ given in Theorem 1 depend on unknown constants $c_2$, $\kappa_1$, and $\kappa_2$, and thus are unavailable in practice.
In real applications, to estimate the set of active interaction variables $\mathcal{A}$, we sort $|\hat{\omega}_k|, 1\leq k\leq p$, in decreasing order and then retain the top $d$ variables. This strategy is also widely used in the existing literature; see, for example, \cite{fan2008sure}, \cite{li2012feature}, \cite{he2013quantile}, \cite{shao2014martingale}, and \cite{cui2015model}.
The set of main effects $\mathcal{B}$ is estimated similarly 
except that the marginal utility $|\hat\omega_k^*|$ is used.
Following the suggestion in \cite{fan2008sure}, one may choose the number of retained variables  for each of sets $\mathcal{A}$ and $\mathcal{B}$ in a screening procedure as $n-1$ or $[c n/(\log n)]$ with $c$ some positive constant, depending on the available sample size $n$. The parameter $c$ can be tuned using some data-driven method such as the cross-validation.

It is worth pointing out that our result is weaker than that in \citet{fan2008sure} in terms of growth of dimensionality, where one can allow $\log p = o(n^{1-2\kappa_2})$. This is mainly because they considered linear models without interactions, indicating the intrinsic challenges of feature screening in the presence of interactions. Moreover, our assumptions on the distributions for the covariates and errors are more flexible.

The results in Theorem \ref{Th: Sure Screening-new} can be improved in the case when the covariates $X_j$'s and the response $Y$ are uniformly bounded. An application of the proofs for \eqref{eq: bound-omega-new}--\eqref{eq: bound-omega-star-new} in Section D of Supplementary Material yields
\begin{eqnarray*}
    && P\Big(\max_{1\leq k\leq p}|\widehat{\omega}_k-\omega_k|\geq c_2n^{-\kappa_1}\Big)
   \leq pC_3\exp(-C_3^{-1}n^{1-2\kappa_1}), \\
   &&  P\Big(\max_{1\leq j\leq p}|\widehat{\omega}_j^{\ast}-\omega_j^{\ast}|\geq c_2n^{-\kappa_2}\Big)
   \leq pC_3\exp(-C_3^{-1}n^{1-2\kappa_2}),
\end{eqnarray*}
where $C_3$ is some positive constant. In this case, IP can handle ultra-high dimensionality $\log p = o(n^{\xi})$ with $\xi=\min\{1-2\kappa_1, 1-2\kappa_2\}$.

\section{Interaction selection}\label{sec:Selection}

\subsection{Interaction models in reduced feature space} \label{sec: ReducedModel}

We now focus on the problem of interaction and main effect selection in the reduced feature space identified by the screening step of IP. To ease the presentation, we rewrite interaction model (\ref{eq:LM}) in the matrix form
\begin{eqnarray}\label{eq:LM:Mat}
\by = \beta_0 \bone + \widetilde{\bX} \btheta + \bveps,
\end{eqnarray}
where $\by = (y_1, \cdots, y_n)\t$ is the response vector, $\btheta = (\theta_1, \cdots, \theta_{\widetilde{p}})\t$ is a parameter vector consisting of $\widetilde{p} = p(p+1)/2$ regression coefficients $\beta_j$ and $\gamma_{k \ell}$, $\widetilde{\bX}$ is the corresponding $n \times \widetilde{p}$ augmented design matrix incorporating the covariate vectors for $X_j$'s and their interactions in columns, and $\bveps$ is the error vector. Hereafter, for the simplicity of presentation and theoretical derivations, we slightly abuse the notation and still use $\by$ and $\widetilde{\bX}$ to denote the de-meaned response and column de-meaned design 
matrix, respectively, which leads to $\beta_0 = 0$. Denote by $\widehat{\mathcal{A}}=\{k_1, \cdots, k_{p_1}\}$ and 
$\widehat{\mathcal{B}}=\{j_1, \cdots, j_{p_2}\}$ the sets of retained interaction variables and main effects, respectively, and $\mathcal{H}$ a subset of $\{1, \cdots, \widetilde{p}\}$ given by the features in 
$\widehat{\mathcal{M}}=\widehat{\mathcal{A}}\cup \widehat{\mathcal{B}}$ and constructed interactions in $\widehat{\mathcal{I}}$ based on $\widehat{\mathcal{A}}$ as defined in \eqref{Ihatdef}. To estimate the true value $\btheta_0 = (\theta_{0, 1}, \cdots, \theta_{0, \widetilde{p}})\t$ of the parameter vector $\btheta$, we can consider the reduced feature space spanned by the 
$q=2^{-1}p_1(p_1-1)+p_3$ columns of the augmented design matrix $\widetilde{\bX}$ in $\mathcal{H}$ with $p_3$ the cardinality of $\widehat{\mathcal{M}}$, thanks to the sure screening property of IP shown in Theorem \ref{Th: Sure Screening-new}.

When the model dimensionality is reduced to a moderate scale $q$, one can apply any favorite variable selection procedure for effective selection of important interactions and main effects and efficient estimation of their effects. There is a large literature on the developments of various variable selection methods. Among all approaches, two classes of regularization methods, the convex ones (e.g., \cite{tibshiranit1996regression,zou2006adaptive,candes2007dantzig}) and the concave ones (e.g., \cite{fan2001variable,lv2009unified,zhang10mcp}), have been extensively investigated.
To combine the strengths of both classes, \citet{fan2014asymptotic} introduced the combined $L_1$ and concave regularization method. Such an approach can be understood as a coordinated intrinsic two-scale learning, in the sense that the Lasso component plays the screening role, in terms of reducing the complexity of intrinsic parameter space, whereas the concave component plays the selection role, in terms of refined estimation.

Following \citet{fan2014asymptotic}, we consider the following combined $L_1$ and concave regularization problem
\begin{eqnarray}\label{eq:IMObj}
\min_{\btheta\in \mathbb{R}^{\widetilde{p}}, \btheta_{\mathcal{H}^c} = \bzero}\left\{(2n)^{-1}\|\by-\widetilde{\bX}\boldsymbol\theta\|^2_2
         +\lambda_0\|\boldsymbol\theta_*\|_1+\|p_{\lambda}(\boldsymbol\theta_*)\|_1\right\},
\end{eqnarray}
where $\btheta_{\mathcal{H}^c}$ denotes a subvector of $\btheta$ given by components in the complement $\mathcal{H}^c$ of the reduced set $\mathcal{H}$,  $\lambda_0 \geq 0$ is the regularization parameter for the $L_1$-penalty, $p_\lambda(\btheta_*) = p_\lambda(|\btheta_*|) = (p_\lambda(|\theta_1^*|), \ldots, p_\lambda(|\theta_{\widetilde{p}}^*|))\t$ with $\btheta_* = (\theta_1^*, \ldots, \theta_{\widetilde{p}}^*)\t$, and $p_\lambda(t)$ is an increasing concave penalty function on $[0, \infty)$ indexed by regularization parameter $\lambda \geq 0$. Here, $\boldsymbol\theta_* = \bD\btheta = n^{-1/2}(\|\widetilde{\bx}_1\|_2 \theta_1$, $\cdots, \|\widetilde{\bx}_{\widetilde{p}}\|_2 \theta_{\widetilde{p}})\t$ is the coefficient vector corresponding to the design matrix with each column rescaled to have $L_2$-norm $n^{1/2}$, where $\widetilde{\bX} = (\widetilde{\bx}_1, \cdots, \widetilde{\bx}_{\widetilde{p}})$ and $\bD =\diag\{\bD_{11}, \cdots, \bD_{\widetilde{p}\widetilde{p}}\}$ with $\bD_{mm} = n^{-1/2} \|\widetilde{\bx}_m\|_2$, $m=1,\cdots, \widetilde p$, is the scale matrix.
The computational cost of solving the regularization problem \eqref{eq:IMObj} in $q$ dimensions after screening from ultra-high scale to moderate scale is substantially reduced compared to that of solving the same problem in $\widetilde{p}$ dimensions without screening. Moreover, important theoretical challenges arise in investigating the asymptotic properties of the resulting regularized estimator for IP. \citet{fan2014asymptotic} considered linear models with deterministic design matrix and no interactions, whereas we now need to study the interaction model with random design matrix. The presence of both interactions and additional randomness requires more delicate analyses.


 We remark that although the combined $L_1$ and concave penalty is used in \eqref{eq:IMObj}, one can in fact use any favorite variable selection method in the selection step of IP. In particular, note that \eqref{eq:IMObj} does not automatically enforce the heredity constraint. If one believes in such constraint, other penalties, such as the ones in     {\cite{yuan2009structured}, \cite{choi2010variable}, and \cite{bien2013lasso}, can be used in the selection step of IP to achieve this goal. 
As specified in the Introduction, one major goal of our paper is to provide a methodological framework such that effective and efficient interaction screening can be conducted. So the penalty in \eqref{eq:IMObj} is just for demonstration purpose. 

\subsection{Asymptotic properties of interaction and main effect selection} \label{sec: Theory-selection}

Before presenting the theoretical results, 
we state some mild regularity conditions that are needed in our analysis. Without loss of generality, assume that the first $s = \|\btheta_0\|_0$ components of the true regression coefficient vector $\btheta_0$ in \eqref{eq:LM:Mat} are nonzero. Throughout the paper, the regularization parameter for the $L_1$ component is fixed to be 
$\lambda_0 = \widetilde{c}_0 \{(\log p)/n^{\alpha_1\alpha_2/(\alpha_1+2\alpha_2)}\}^{1/2}$ with $\widetilde{c}_0$ some positive constant. Some insights into this choice of $\lambda_0$ will be provided later. Denote by $p_{\text{H},\lambda}(t) = 2^{-1}\{\lambda^2-(\lambda-t)^2_{+}\}$, $t \geq 0$, the hard-thresholding penalty, where $(\cdot)_{+}$ denotes the positive part of a number.


\begin{assumption}\label{con:RE-new}
There exist some constants $\kappa_0, \kappa, L_1, L_2 > 0$ such that with probability $1-a_n$ satisfying $a_n=o(1)$, it holds that $\min_{\|\bdelta\|_2=1,\, \|\bdelta\|_0<2s} n^{-1/2}\|\widetilde{\bX}\bdelta\|_2 \geq \kappa_0$,
\begin{eqnarray*}
\min\limits_{\bdelta\neq0,\, \|\bdelta_2\|_1\leq 7\|\bdelta_1\|_1}
        \left\{n^{-1/2}\|\widetilde{\bX}\bdelta\|_2/(\|\bdelta_1\|_2\vee \|\widetilde{\bdelta}_2\|_2)\right\} \geq \kappa
\end{eqnarray*}
for $\bdelta=(\bdelta_1^T, \bdelta_2^T)^T\in \mathbb{R}^{\widetilde{p}}$ with $\bdelta_1\in \mathbb{R}^s$ and $\widetilde{\bdelta}_2$ a subvector of $\bdelta_2$ consisting of the $s$ largest components in magnitude, and $\bD_{mm}$'s are bounded between $L_1\leq L_2$.
\end{assumption}


\begin{assumption}\label{con:pen-new}
The concave penalty satisfies that $p_{\lambda}(t)\geq p_{\emph{H},\lambda}(t)$
on $[0,\lambda]$, $p'_{\lambda}\{(1-{c_3})\lambda\}\leq \min\{\lambda_0/4, {c_3}\lambda\}$ for some constant ${c_3}\in [0, 1)$, and $-p''_{\lambda}(t)$ is
decreasing on $[0, (1-{c_3})\lambda]$. Moreover, $\min_{1 \leq j \leq s}|\theta_{0,j}|>L_1^{-1}\max\{(1-{c_3})  \lambda, 2 L_2 \kappa_0^{-1}p_{\lambda}^{1/2}(\infty)\}$
with $p_{\lambda}(\infty)=\lim\limits_{t\rightarrow \infty}p_{\lambda}(t)$.
\end{assumption}

Condition \ref{con:RE-new} is similar to Condition 1 in \citet{fan2014asymptotic} for the case of deterministic design matrix, except that the design matrix is now random in our setting and also augmented with interactions.
We provide in Section \ref{sec: RE condition} some sufficient conditions ensuring that Condition \ref{con:RE-new} holds.
Condition \ref{con:pen-new} puts some basic constraints on the concave penalty $p_\lambda(t)$ as in \citet{fan2014asymptotic}. 
Under these regularity conditions, the following theorem presents the selection properties of the IP estimator $\widehat{\btheta} = (\widehat{\theta}_1, \cdots, \widehat{\theta}_{\widetilde{p}})\t$ including an explicit bound on the number of falsely discovered signs $\FS(\widehat{\btheta})=|\{1\leq m \leq \widetilde{p}: \sgn(\widehat{\theta}_m)\neq \sgn(\theta_{0,m})\}|$, which provides a stronger measure on variable selection than the total number of false positives and false negatives.


\begin{theorem}\label{Th:global}
Assume that the conditions of part c) of Theorem \ref{Th: Sure Screening-new} and {Conditions \ref{con:RE-new}--\ref{con:pen-new} hold, $\log p = o\{n^{\alpha_1\alpha_2/(\alpha_1+2\alpha_2)}\}$ with $\alpha_1\alpha_2/(\alpha_1+2\alpha_2)\leq 1$}, and $p_{\lambda}(t)$ is continuously differentiable. Then the global minimizer $\widehat{\boldsymbol\theta}$ of (\ref{eq:IMObj}) has the hard-thresholding property that each component is either zero or of magnitude larger than $(1-{c_3})\lambda$, and with probability at least $1-a_n-o(n^{-\min\{C_1, C_2\}}+ p^{-{c_4}})$, it satisfies simultaneously that
\begin{align*}
   & n^{-1/2}\left\|\widetilde{\bX}(\widehat{\boldsymbol\theta}-\boldsymbol\theta_0)\right\|_2
       = O(\kappa^{-1}\lambda_0s^{1/2}), \\
   &  \left\|\widehat{\boldsymbol\theta}-\boldsymbol\theta_0\right\|_d
      = O(\kappa^{-2} \lambda_0s^{1/d}), \quad d\in [1, 2],    \\
   & \emph{\FS}(\widehat{\boldsymbol\theta})
       = O\left\{\kappa^{-4} (\lambda_0/\lambda)^2s\right\},
\end{align*}
and furthermore $\sgn(\widehat{\boldsymbol\theta})=\sgn(\boldsymbol\theta_0)$ if $\lambda\geq 56(1-{c_3})^{-1}\kappa^{-2}\lambda_0s^{1/2}$, where ${c_4}$ is some positive constant. Moreover, the same results hold with probability at least $1-a_n-o(p^{-{c_4}})$ for the regularized estimator $\widehat{\boldsymbol\theta}$ without prescreening, that is, without the constraint $\btheta_{\mathcal{H}^c} = \bzero$ in (\ref{eq:IMObj}).
\end{theorem}

The results in Theorem \ref{Th:global} also apply to the regularized estimator with $p_1 = p_2 = p$ and $q = \widetilde{p} = p(p+1)/2$, that is, without any screening of variables. Theorem \ref{Th:global} shows that if the tuning parameter $\lambda$ satisfies
$\lambda_0/\lambda\rightarrow 0$, then the number of falsely discovered signs $\FS(\widehat{\btheta})$
is of order $o(s)$ and thus the false sign rate $\FS(\widehat{\btheta})/s$
is asymptotically vanishing with probability tending to one. We also observe that the bounds for prediction and estimation losses are independent of the tuning parameter $\lambda$ for the concave penalty.

As shown in Theorem \ref{Th:global}, the regularization parameter for the $L_1$ component $\lambda_0 = \widetilde{c}_0 \{(\log p)/n^{\alpha_1\alpha_2/(\alpha_1+2\alpha_2)}\}^{1/2}$ plays a crucial role in characterizing the rates of convergence for the regularized estimator $\widehat{\boldsymbol\theta}$. Such a parameter basically measures the maximum noise level in interaction models. In particular, the exponent $\alpha_1\alpha_2/(\alpha_1+2\alpha_2)$ is a key parameter that reflects the level of difficulty in the problem of interaction selection. This quantity is determined by three sources of heavy-tailedness: covariates themselves, their interactions, and the error. To simplify the technical presentation, in this paper we have focused on the more challenging case of $\alpha_1\alpha_2/(\alpha_1+2\alpha_2)\leq 1$. Such a scenario includes two specific cases: 1) sub-Gaussian covariates and sub-Gaussian error, that is, $\alpha_1=\alpha_2=2$ and 2) sub-Gaussian covariates and sub-exponential error, that is,  $\alpha_1=2, \alpha_2=1$. We remark that in the lighter-tailed case of $\alpha_1\alpha_2/(\alpha_1+2\alpha_2)>1$, one can simply set $\lambda_0 = \widetilde{c}_0 \{(\log p)/n\}^{1/2}$ and the results in Theorem \ref{Th:global} can still hold for this choice of $\lambda_0$ by resorting to Lemma \ref{lemma-Hao-Zhang-extend} and similar arguments in the proof of Theorem \ref{Th:global}.

\subsection{Verification of Condition \ref{con:RE-new}} \label{sec: RE condition}

Since Condition \ref{con:RE-new} is a key assumption for proving Theorem \ref{Th:global}, we provide some sufficient conditions that ensures this assumption on the augmented random design matrix $\widetilde{\bX} = (\widetilde{\bx}_1, \cdots, \widetilde{\bx}_{\widetilde{p}})$. Denote by $\widetilde{\bSig}$ the population covariance matrix of the augmented covariate vector consisting of $p$ main effects $X_j$'s and $p(p-1)/2$ interactions $X_k X_{\ell}$'s.

\begin{assumption}\label{con: eigen}
There exists some constant $K > 0$ such that for $\bdelta=(\bdelta_1^T, \bdelta_2^T)^T\in \mathbb{R}^{\widetilde{p}}$,
\begin{eqnarray*}
      \min_{\|\bdelta\|_2=1,\,  \|\bdelta\|_0<2s}\bdelta^T\widetilde{\bSig}\bdelta  \geq K \ \  \mbox{and} \ \
     \min_{\bdelta \neq 0,\, \|\bdelta_2\|_1\leq 7\|\bdelta_1\|_1}     \bdelta^T\widetilde{\bSig}\bdelta/\left(\|\bdelta_1\|_2\vee \|\widetilde{\bdelta}_2\|_2\right) \geq K,
\end{eqnarray*}
where $\bdelta_1\in \mathbb{R}^s$ and $\widetilde{\bdelta}_2$ is a subvector of $\bdelta_2$ consisting of the $s$ largest components in magnitude.
\end{assumption}

Condition \ref{con: eigen} is satisfied if the smallest eigenvalue of $\widetilde{\bSig}$ is assumed to be bounded away from zero. Such a condition is in fact much weaker than the minimum eigenvalue assumption, since it is the population version of a mild sparse eigenvalue assumption and the restricted eigenvalue assumption. The following theorem shows that under some mild assumptions, Condition \ref{con:RE-new} holds for the full augmented design matrix $\widetilde{\bX}$ and thus holds naturally for any $n \times q$ sub-design matrix with $q \leq \widetilde{p}$ and the sure screening property. 


\begin{theorem}\label{Th:RE condition}
Assume that Condition \ref{con: eigen} holds, there exist some constants $\alpha_1, c_1 > 0$ such that for any $t>0$, $P(|X_j|> t) \leq c_1\exp(-c_1^{-1}t^{\alpha_1})$ for each $j$, $s=O(n^{\xi_0})$, and $\log p=o(n^{\min\{\alpha_1/4,\, 1\}- 2\xi_0})$ with constant $0\leq \xi_0<\min\{\alpha_1/8, 1/2\}$. Then Condition \ref{con:RE-new} holds with $n^{\min\{\alpha_1/4,\, 1\}-2\xi_0} = O(-\log a_n)$.
\end{theorem}

\section{Numerical studies}\label{sec:Numerical}

In this section, we design two simulation examples 
 to verify the theoretical results and examine the finite-sample performance of the suggested approach IP. We also present an analysis of a prostate cancer data set.

\subsection{Feature screening performance}\label{screening}
We start with comparing IP with several recent feature screening procedures: the SIS, DC-SIS \citep{li2012feature}, and SIRI \citep{jiang2014sliced}. The SIRI is an iterative procedure that
alternates between a large-scale variable screening step and a moderate-scale variable
selection step when the dimensionality $p$ is large. Since all other screening methods are non-iterative, in this section, we compare the initial screening step of SIRI with other methods and name the screening only procedure as SIRI*. The full iterative SIRI will be included in Section \ref{sec: selection} later for comparison of variable selection. SIRI*, SIS,  and DC-SIS each return a set of variables without distinguishing between important main effects and active interaction variables. Thus, for each method, we construct interactions using all possible pairwise interactions of the recruited variables. By doing so, the strong heredity assumption is enforced. We name the resulting procedures as SIRI*2, SIS2, and DC-SIS2 to distinguish them from their original versions. 

For IP, as mentioned in Section \ref{sec: Screening-Property}, we retain the top $[n/(\log n)]$ variables in each of sets $\widehat{\mathcal{A}}$ and $\widehat{\mathcal{B}}$ defined in (\ref{eq: Ahat}) and (\ref{eq: Bhat}), respectively.
The features in the union set $\widehat{\mathcal{M}} = \widehat{\mathcal{A}} \cup \widehat{\mathcal{B}}$ are used as main effects while variables in set $\widehat{\mathcal{A}}$ are used to build interactions in the selection step of IP. To ensure a fair comparison, the numbers of variables kept in SIRI*2, SIS2, and DC-SIS2 are all equal to the cardinality of $\widehat{\mathcal{M}}$, which is up to $2[n/(\log n)]$.


\medskip

\textbf{Example 1} (Gaussian distribution). We consider the following four interaction models linking the covariates $X_j$'s to the response $Y$:
\begin{itemize}
\item M1 (strong heredity): $Y =2X_1+2X_5 + 3 X_1X_5+\varepsilon_1$,

\item M2 (weak heredity): $Y=2X_1 +2X_{10}+ 3X_1X_5+\varepsilon_2$,

\item M3 (anti-heredity): $Y=2X_{10} + 2X_{15} + 3X_1X_5+\varepsilon_3$,

\item M4 (interactions only): $Y=3X_1X_5 + 3X_{10}X_{15}+\varepsilon_4$,
\end{itemize}
where the covariate vector $\bx=(X_1, \cdots, X_p)^T \sim N(\bzero, \bSig)$ with
$\bSig=(\rho^{|j-k|})_{1 \leq j, k \leq p}$ and the errors $\varepsilon_1 \sim N(0, 2.5^2)$, $\varepsilon_2 \sim N(0, 2^2)$, $\varepsilon_3 \sim N(0, 2^2)$, and $\varepsilon_4 \sim N(0, 1.5^2)$ are independent of $\bx$. The first two models M1 and M2 satisfy the heredity assumption (either strong or weak), while the last two M3 and M4 do not obey such an assumption. Different levels of error variance are considered since the difficulty of feature screening varies across the four models. A sample of $n$ i.i.d. observations was generated from each of the four models. We further considered four different settings of $(n, p, \rho)=(200, 2000, 0)$,
$(200, 2000, 0.5)$, $(300, 5000, 0)$, and $(300, 5000, 0.5)$, and repeated each experiment 100 times.

\begin{center}
[Table \ref{tab:Ex1Screen} about here.]
\end{center}

Table \ref{tab:Ex1Screen} lists the comparison results for all screening methods in recovering each important interaction or main effect, and retaining all important ones. For model M1 satisfying the strong heredity assumption, all procedures performed rather similarly and all retaining percentages were either equal or close to 100\%. 
Both DC-SIS2 and IP performed similarly and improved over SIS2 and SIRI*2 in model M2 in which the weak heredity assumption holds. In models M3 and M4, IP significantly outperformed all other methods in detecting interactions across all four settings, showing its advantage when the heredity assumption is not satisfied. We also observe that SIS2 failed to detect interactions, whereas SIRI*2 improved over DC-SIS2 in these two models. These results suggest that a separate screening step should be designed specifically for interactions to improve the screening accuracy, which is indeed one of the main innovations of IP.

\medskip

\textbf{Example 2} (Non-Gaussian distribution). The second example adopts the same four models as in Example 1, but with different distributions for the covariates $X_j$'s and error $\varepsilon$. We added an independently generated random variable $U_j$ to each covariate $X_j$ as given in Example 1 to obtain new covariates, where $U_j$'s are i.i.d. and follow the uniform distribution on $[-0.5, 0.5]$. The errors $\varepsilon_1 \sim t_{(3)}$, $\varepsilon_2 \sim t_{(4)}$, $\varepsilon_3 \sim t_{(4)}$, and $\varepsilon_4 \sim t_{(8)}$ are independent of $\bx$.

\begin{center}
[Table \ref{tab:Ex2Screen} about here.]
\end{center}

The screening results of all the methods are summarized in Table \ref{tab:Ex2Screen}. Similarly as in Example 1, IP outperformed SIS2 in interaction screening. 
{When the heredity assumption is satisfied, IP performed comparably to DC-SIS2. In particular,  both approaches were better than SIS2 and SIRI*2 when the weak heredity assumption is satisfied.} The improvement of IP over all other methods in detecting interactions became substantial when the heredity assumption is violated.

We also calculated the overall signal-to-noise ratio (SNR) and the individual SNR for each model, where the former is defined as $\var(\widetilde{\bx}\t\btheta)/\var(\varepsilon)$ with $\widetilde{\bx}$ the augmented covariate vector defined in Section \ref{sec: ReducedModel}, $\varepsilon$ the error term and $\btheta$ given in model (\ref{eq:LM:Mat}), and the latter 
is defined similarly by replacing $\var(\widetilde{\bx}\t\btheta)$ with the variance of each individual term. The overall and individual SNRs for the models considered in both Examples 1 and 2 are listed in Table \ref{tab:snr}. In particular, we see that although the overall SNRs are at decent levels, the individual ones are weaker, reflecting the general difficulty of retaining all important features for screening.

\begin{center}
[Table \ref{tab:snr} about here.]
\end{center}


\subsection{Variable selection performance}\label{sec: selection}
We further assess the variable selection performance of IP.  For all screening methods but  
SIRI*2, with each data set generated in Examples 1 and 2, we can employ regularization methods such as the Lasso and the combined $L_1$ and concave method to select important interactions and main effects after the screening step. As shown in \citet{fan2014asymptotic}, different choices of the concave penalty gave rise to similar performance. We thus implemented the combined $L_1$ and SICA ($L_1$+SICA) for simplicity. The approach of SIS2 followed by Lasso is referred to as SIS2-Lasso for short. All other combinations of screening and selection methods are defined similarly. We also paired up the hierNet \citep{bien2013lasso} with the IP for interaction identification. 
For SIRI, we used the full iterative procedure as described in \cite{jiang2014sliced}. Since SIRI only returns a set of important variables, we added an additional refitting step using the selected variables to calculate model performance measures. We also included additional competitor   
methods iFORT and iFORM in \cite{hao2014interaction} and RAMP in \cite{hao2015model} in our simulation studies.
The oracle procedure based on the true underlying interaction model was used as a reference point for comparisons. The cross-validation (CV) was used to select tuning parameters for all the methods, except that the BIC was applied to $L_1$+SICA related procedures for computational efficiency since two regularization parameters are involved.

To evaluate the variable selection performance of each method, we employed three performance measures. The first one is the prediction error (PE), which was calculated using an independent test sample of size 10,000. The second and third measures are the numbers of false positives (FP) and false negatives (FN), which are defined as the numbers of included noise variables and missed important variables in the final model, respectively.

\begin{center}
[Tables \ref{tab: Ex1Select} and \ref{tab: Ex2Select} about here.]
\end{center}

Table \ref{tab: Ex1Select} presents the medians and robust standard deviations (RSD) of these measures based on 100 simulations
for different models in Example 1. The RSD is defined as
the interquartile range (IQR) divided by 1.34. We used the median and RSD instead of the mean and standard deviation since these robust measures are better suited to summarize the results due to the existence of outliers.  When the strong heredity assumption holds (model M1), both DC-SIS2-$L_1$+SICA and IP-$L_1$+SICA followed closely the oracle procedure,
and outperformed the other methods in terms of PE, FP, and FN across all four settings. In model M2 with the weak heredity assumption, variable selection methods based on both DC-SIS2 and IP performed fairly well. In the cases when the heredity assumption does not hold (models M3 and M4), the IP-$L_1$+SICA still mimicked the oracle procedure and uniformly outperformed the other methods over all settings. The inflated RSDs, relative to medians, in model M4 were due to the relatively low sure screening probabilities (see Tables \ref{tab:Ex1Screen} and \ref{tab:Ex2Screen}). When the sure screening probability is low, a nonnegligible number of replications can have nonzero false negatives, which inflated the corresponding prediction errors. The comparison results of variable selection for Example 2 are summarized in Table \ref{tab: Ex2Select}. The conclusions are similar to those for Example 1.

\subsection{Real data analysis}
In addition, we illustrate our procedure IP through an analysis of the prostate cancer data studied originally in \citet{singh2002gene} and analyzed also in \citet{fan2008high} and \citet{hall2014selecting}. This data set, which is available at http://www.broad.mit.edu/cgi-bin/cancer/datasets.cgi, contains $136$ samples with $77$ from the tumor group and $59$ from the normal group, each of which records the expression levels measured for $12,600$ genes. \citet{hall2014selecting} applied a four-step procedure to preprocess the data. Their procedure includes the truncation of intensities to make them positive, the removal of genes having little variation in intensity, the transformation of intensities to base $10$ logarithms, and the standardization of each data vector to have zero mean and unit variance. An application of the four-step procedure results in a total of $p=3,239$ genes.

We treated the disease status as the response and the resulting $3,239$ genes as covariates. The data set was randomly split into a training set
and a test set. Each training set consists of $69$ samples from the tumor group and $53$ samples from the normal group, and the test set is formed by the remaining samples. For each split, we applied the screening method IP to the training data and retained the top 
$d=[cn/(\log n)] = [25.4 c]$ genes in each of sets $\widehat{\mathcal{A}}$ and $\widehat{\mathcal{B}}$ with $c$ chosen from the grid $\{0.5, 1, 2\}$. For SIS2 and DC-SIS2, we retained the top $|\widehat{\mathcal{M}}|= |\widehat{\mathcal{A}}\cup \widehat{\mathcal{B}}|$ variables in the screening step. Because of the limited sample size, to increase the stability we constructed interactions in a more conservative way by using variables in set $\widehat{\mathcal{M}}$ instead of only $\widehat{\mathcal{A}}$ to build interactions in the selection step of IP. In addition, to overcome the difficulty caused by potential high
collinearity, in our real data analysis we used the elastic net penalty introduced in \cite{zou2005regularization}. We then tuned $c$ in terms of minimizing the classification error calculated using the test data. We also repeated the random split 100 times.

\begin{center}
[Table \ref{tab:real} about here.]
\end{center}

Three competing 
methods SIS2-Enet, DC-SIS2-Enet, and IP-Enet were considered, where SIS2-Enet denotes the approach of
SIS2 followed by the elastic net, and the latter two methods are defined similarly. Since the same penalty is used for the step of variable selection, the difference in performance should come mainly from the screening step.
Table \ref{tab:real} summarizes the classification results and median model sizes for each method. We observe that the approach of IP-Enet yielded lower classification errors.
Paired $t$-tests of classification errors on the 100 splits of IP-Enet against SIS2-Enet and DC-SIS2-Enet gave $p$-values $4.95\times 10^{-10}$ and $1.62\times 10^{-8}$, respectively.
These results show that our proposed method outperformed significantly SIS2-Enet and DC-SIS2-Enet in classification error.

\begin{center}
[Table \ref{genes} about here.]
\end{center}

We also present in Table \ref{genes}  
the top 10 interactions and top 10 main effects that were most frequently selected over 100 splits. 
We see from Table \ref{genes} that a set of genes, 
such as SERINC5, HPN, HSPD1, LMO3, and TARP,
were selected by all methods as main effects, revealing that those genes may play a significant role in the etiology of prostate cancer. 
 For example, \citet{Holt2010} claimed Hepsin (HPN) as one of the most consistently overexpressed genes in prostate cancer. In addition, evidence of the association between TARP gene variants and prostate cancer risk has been shown in \citet{Wolfgang2000}, \citet{Oh2004}, and \citet{Hillerdal2012}. Note that the gene ERG was missed by both SIS2-Enet and DC-SIS2-Enet in the top 10 main effects, but it was selected by IP-Enet as a main effect and part of an interaction (SLC7A1$\times$ERG). 
There are a wide range of studies investigating the effect of ERG on prostate cancer \citep{Klezovitch2008, Furusato2010}.


The most frequently selected interaction DPT$\times$S100A4 by SIS2-Enet and DC-SIS2-Enet is also among the top 10 list by IP-Enet. Two more interactions, RARRES2$\times$KLK3 and MAF$\times$NELL2, are also among the top 10 lists by both IP-Enet and SIS2-Enet. However, some interactions involving PRKDC (PRKDC$\times$CFD and PRKDC$\times$KLK3) were very often selected by IP-Enet but missed by the other two methods.
There are studies showing that PRKDC is associated with prostate cancer (McCarthy et al., 2013). Such a finding favors the results of IP that the  interactions PRKDC$\times$CFD and PRKDC$\times$KLK3 were identified to be associated with the phenotype.

}





\section{Discussion}\label{sec:Discussion}
We have considered in this paper the problem of interaction identification in ultra-high dimensions. The proposed method IP based on a new interaction screening procedure and post-screening variable selection is computationally efficient, and capable of reducing dimensionality from a large scale to a moderate one and recovering important interactions and main effects. To simplify the technical presentation, our analysis has been focused on the linear pairwise interaction models. Screening for main effects in more general model settings has been explored by many researchers; see, for example, \cite{fan2010sure}, \cite{fan2011nonparametric}, \cite{chang2013marginal}, and \cite{cheng2014nonparametric}. It would be interesting to extend the interaction screening idea of IP to 
these and other more general model frameworks such as the generalized linear models, nonparametric models, and survival models with interactions.

The key idea of IP is to use different marginal utilities to screen interactions and main effects separately. As such, it can suffer from the same potential issues as the SIS. First, some noise interactions or main effects that are highly correlated with the important ones can have higher marginal utilities and thus priority to be selected than other important ones that are relatively weakly related to the response. Second, some important interactions or main effects that are jointly correlated but marginally uncorrelated with the response can be missed after screening. To address these issues, we next briefly discuss two extensions of IP that enable us to exploit more fully the joint information among the covariates.

Our first extension of IP, the iterative IP (IIP), is motivated by the idea of two-scale learning with the iterative SIS (ISIS) in \citet{fan2008sure} and \citet{fan2009ultrahigh}. The IIP works as follows by applying large-scale screening and moderate-scale selection in an iterative fashion. First, apply IP to the original sample $(\bx_i, y_i)_{i = 1}^n$ to obtain two sets $\widehat{\mathcal{I}}_1$ of interactions and 
$\widehat{\mathcal{B}}_1$ of main effects,
and construct a set $\widehat{\mathcal{A}}_1$ of interaction variables based on $\widehat{\mathcal{I}}_1$ as in (\ref{intvarsets}).
Second, update the sets of candidate interaction variables as $\{1, \cdots, p\}\setminus \widehat{\mathcal{A}}_1$ and candidate main effects as 
$\{1, \cdots, p\}\setminus \widehat{\mathcal{B}}_1$,
treat the residual vector from the previous iteration as the new response, and apply IP to the updated sample to obtain new sets $\widehat{\mathcal{I}}_2$, 
$\widehat{\mathcal{B}}_2$, and $\widehat{\mathcal{A}}_2$ defined similarly as before. Third, iteratively update the feature space for candidate interaction variables and main effects and the response, and apply IP to the updated sample to similarly obtain sequences of sets $(\widehat{\mathcal{I}}_k)$, 
$(\widehat{\mathcal{B}}_k)$, and $(\widehat{\mathcal{A}}_k)$, until the total number of selected interactions and main effects in sets $\widehat{\mathcal{I}}_k$'s and 
$\widehat{\mathcal{B}}_k$'s
reaches a prespecified threshold. Fourth, finally select important interactions and main effects using a regularization method in the reduced feature space given by the union of $\widehat{\mathcal{I}}_k$'s and 
$\widehat{\mathcal{B}}_k$'s.

The second extension of IP, the conditional IP (CIP), exploits the idea of the conditional SIS (CSIS) in \citet{barut2016conditional}, which replaces the simple marginal correlation with the conditional marginal correlation to assess the importance of covariates when some variables are known in advance to be important. Suppose we have some prior knowledge that two given sets $\mathcal{A}_0$, 
$\mathcal{B}_0 \subset\{1, \cdots, p\}$ contain some active interaction variables and important main effects, respectively. For interaction screening, the CIP regresses the squared response $Y^2$ on each squared covariate $X_k^2$ with $k$ outside $\mathcal{A}_0$ by conditioning on $(X_\ell^2)_{\ell \in \mathcal{A}_0}$, and retains top ones in the conditional marginal utilities as interaction variables. Similarly, in main effect screening it employs marginal regression of the response $Y$ on each covariate $X_k$ with $k$ outside 
$\mathcal{B}_0$ conditional on $(X_\ell)_{\ell \in \mathcal{B}_0}$.
After screening, CIP further selects important interactions and main effects using a variable selection procedure in the reduced feature space. The approach of CIP can also be incorporated into IIP by conditioning on selected variables in previous steps when calculating the marginal utilities along the course of iteration.

The investigation of these extensions is beyond the scope of the current paper and will be interesting topics for future research.

\newpage

\begin{table}
\caption{The percentages of retaining each important interaction or main effect, and all important ones (All) by all the screening methods over different models and settings in Example 1.} \vspace{0.1in}
\centering
\scalebox{0.75}{
\begin{tabular}{ l c c c c c c c c ccccccc }
  \toprule
   Method & \multicolumn{4}{c}{M1} & \multicolumn{4}{c}{M2} & \multicolumn{4}{c}{M3}  & \multicolumn{3}{c}{M4} \\
    \cmidrule(lr{.75em}){2-5}  \cmidrule(lr{.75em}){6-9} \cmidrule(lr|{.75em}){10-13}  \cmidrule(lr|{.75em}){14-16}
    & $X_1$ & $X_5$ &  $X_1X_5$  & All & $X_1$ & $X_{10}$ & $X_1X_5$  &All & $X_{10}$ & $X_{15}$ & $X_1X_5$  & All  & $X_1X_5$ & $X_{10}X_{15}$ & All  \\
  \midrule
    \multicolumn{15}{c}{{Setting 1}: $(n, p, \rho)=(200, 2000, 0)$} \\
  SIS2      & 1.00  & 1.00  & 1.00  & 1.00   & 1.00 & 1.00 & 0.09 & 0.09  & 1.00 & 1.00 & 0.02 & 0.02   & 0.02 & 0.02 & 0.00\\
  DC-SIS2 & 1.00  & 1.00  & 1.00  & 1.00   & 1.00 & 1.00 & 0.88 & 0.88  & 1.00 & 1.00 & 0.04 & 0.04   & 0.15 & 0.16 & 0.03\\
SIRI$^*2$   & 1.00  & 1.00  & 1.00  & 1.00   & 1.00 & 1.00 & 0.67 & 0.67  & 1.00 & 1.00 & 0.13 & 0.13   & 0.34 & 0.29 & 0.13\\
  IP        & 1.00  & 1.00  & 0.97  & 0.97   & 1.00 & 1.00 & 0.88 & 0.88  & 1.00 & 1.00 & 0.93 & 0.93   & 0.80 & 0.79 & 0.59\\
\addlinespace
    \multicolumn{15}{c}{{Setting 2}: $(n, p, \rho)=(200, 2000, 0.5)$} \\
  SIS2      & 1.00  & 1.00  & 1.00  & 1.00   & 1.00 & 1.00 & 0.15 & 0.15  & 1.00 & 1.00 & 0.01 & 0.01   & 0.01 & 0.04 & 0.00\\
  DC-SIS2 & 1.00  & 1.00  & 1.00  & 1.00   & 1.00 & 1.00 & 0.85 & 0.85  & 1.00 & 1.00 & 0.03 & 0.03   & 0.14 & 0.11 & 0.03\\
SIRI$^*2$   & 1.00  & 1.00  & 1.00  & 1.00   & 1.00 & 1.00 & 0.62 & 0.62  & 1.00 & 1.00 & 0.09 & 0.09   & 0.36 & 0.31 & 0.11\\
  IP        & 1.00  & 1.00  & 0.96  & 0.96   & 1.00 & 1.00 & 0.85 & 0.85  & 1.00 & 1.00 & 0.84 & 0.84   & 0.75 & 0.84 & 0.59\\
\addlinespace
      \multicolumn{15}{c}{{Setting 3}: $(n, p, \rho)=(300, 5000, 0)$} \\
  SIS2      & 1.00  & 1.00  & 1.00  & 1.00  & 1.00 & 1.00 & 0.07 & 0.07   & 1.00 & 1.00 & 0.01 & 0.01   & 0.00 & 0.00 & 0.00\\
  DC-SIS2 & 1.00  & 1.00  & 1.00  & 1.00   & 1.00 & 1.00 & 0.93 & 0.93  & 1.00 & 1.00 & 0.03 & 0.03   & 0.14 & 0.16 & 0.01\\
SIRI$^*2$   & 1.00  & 1.00  & 1.00  & 1.00   & 1.00 & 1.00 & 0.72 & 0.72 & 1.00 & 1.00 & 0.15 & 0.15    & 0.40 & 0.43 & 0.16\\
  IP        & 1.00  & 1.00  & 0.97  & 0.97   & 1.00 & 1.00 & 0.90 & 0.90  & 1.00 & 1.00 & 0.96 & 0.96   & 0.83 & 0.82 & 0.65\\
\addlinespace
    \multicolumn{15}{c}{{Setting 4}: $(n, p, \rho)=(300, 5000, 0.5)$} \\
  SIS2      & 1.00  & 1.00  & 1.00  & 1.00   & 1.00 & 1.00 & 0.17 & 0.17 & 1.00 & 1.00 & 0.04 & 0.04   & 0.02 & 0.00 & 0.00\\
  DC-SIS2 & 1.00  & 1.00  & 1.00  & 1.00    & 1.00 & 1.00 & 0.95 & 0.95 & 1.00 & 1.00 & 0.07 & 0.07   & 0.13 & 0.18 & 0.02\\
SIRI$^*2$  & 1.00  & 1.00  & 1.00  & 1.00   & 1.00 & 1.00 & 0.83 & 0.83  & 1.00 & 1.00 & 0.18 & 0.18   & 0.46 & 0.47 & 0.18\\
  IP        & 1.00  & 1.00  & 0.99  & 0.99    & 1.00 & 1.00 & 0.90 & 0.90  & 1.00 & 1.00 & 0.94 & 0.94  & 0.79 & 0.85 & 0.64\\
  \midrule
\end{tabular}
\label{tab:Ex1Screen}
}
\end{table}

\begin{table}
\caption{The percentages of retaining each important interaction or main effect, and all important ones (All) by all the screening methods over different models and settings in Example 2.} \vspace{0.1in}
\centering
\scalebox{0.75}{
\begin{tabular}{ l c c c c c c c c ccccccc }
  \toprule
   Method & \multicolumn{4}{c}{M1} & \multicolumn{4}{c}{M2} & \multicolumn{4}{c}{M3}  & \multicolumn{3}{c}{M4} \\
    \cmidrule(lr{.75em}){2-5}  \cmidrule(lr{.75em}){6-9} \cmidrule(lr|{.75em}){10-13}  \cmidrule(lr|{.75em}){14-16}
    & $X_1$ & $X_5$ &  $X_1X_5$  & All & $X_1$ & $X_{10}$ & $X_1X_5$  &All & $X_{10}$ & $X_{15}$ & $X_1X_5$  & All  & $X_1X_5$ & $X_{10}X_{15}$ & All  \\
  \midrule
    \multicolumn{15}{c}{{Setting 1}: $(n, p, \rho)=(200, 2000, 0)$} \\
  SIS2      & 1.00  & 1.00  & 1.00  & 1.00   & 1.00 & 1.00 & 0.13 & 0.13  & 1.00 & 1.00 & 0.02 & 0.02   & 0.00 & 0.01 & 0.00\\
  DC-SIS2 & 1.00  & 1.00  & 1.00  & 1.00    & 1.00 & 1.00 & 0.91 & 0.91 & 1.00 & 1.00 & 0.06 & 0.06   & 0.17 & 0.20 & 0.01\\
SIRI$^*2$  & 1.00  & 1.00  & 1.00  & 1.00   & 1.00 & 1.00 & 0.75 & 0.75 & 1.00 & 1.00 & 0.18 & 0.18    & 0.36 & 0.40 & 0.11\\
  IP        & 1.00  & 1.00  & 0.96  & 0.96    & 1.00 & 1.00 & 0.95 & 0.95 & 1.00 & 1.00 & 0.97 & 0.97   & 0.80 & 0.83 & 0.63\\
\addlinespace
    \multicolumn{15}{c}{{Setting 2}: $(n, p, \rho)=(200, 2000, 0.5)$} \\
  SIS2      & 1.00  & 1.00  & 1.00  & 1.00   & 1.00 & 1.00 & 0.18 & 0.18 & 1.00 & 1.00 & 0.01 & 0.01    & 0.00 & 0.00 & 0.00\\
  DC-SIS2 & 1.00  & 1.00  & 1.00  & 1.00   & 1.00 & 1.00 & 0.95 & 0.95 & 1.00 & 1.00 & 0.10 & 0.10     & 0.14 & 0.14 & 0.02\\
SIRI$^*2$  & 1.00  & 1.00  & 1.00  & 1.00   & 1.00 & 1.00 & 0.85 & 0.85 & 1.00 & 1.00 & 0.16 & 0.16    & 0.41 & 0.43 & 0.18\\
  IP        & 1.00  & 1.00  & 0.95  & 0.95    & 1.00 & 1.00 & 0.97 & 0.97 & 1.00 & 1.00 & 0.96 & 0.96   & 0.80 & 0.81 & 0.61\\
\addlinespace
      \multicolumn{15}{c}{{Setting 3}: $(n, p, \rho)=(300, 5000, 0)$} \\
  SIS2      & 1.00  & 1.00  & 1.00  & 1.00   & 1.00 & 1.00 & 0.14 & 0.14   & 1.00 & 1.00 & 0.02 & 0.02   & 0.00 & 0.01 & 0.00\\
  DC-SIS2 & 1.00  & 1.00  & 1.00  & 1.00    & 1.00 & 1.00 & 1.00 & 1.00 & 1.00 & 1.00 & 0.10 & 0.10   & 0.18 & 0.20 & 0.01\\
SIRI$^*2$   & 1.00  & 1.00  & 1.00  & 1.00     & 1.00 & 1.00 & 0.93 & 0.93 & 1.00 & 1.00 & 0.32 & 0.32  & 0.63 & 0.65 & 0.43\\
  IP        & 1.00  & 1.00  & 0.98  & 0.98     & 1.00 & 1.00 & 0.98 & 0.98 & 1.00 & 1.00 & 0.97 & 0.97  & 0.86 & 0.85 & 0.71\\
\addlinespace
    \multicolumn{15}{c}{{Setting 4}: $(n, p, \rho)=(300, 5000, 0.5)$} \\
  SIS2      & 1.00  & 1.00  & 1.00  & 1.00    & 1.00 & 1.00 & 0.09 & 0.09 & 1.00 & 1.00 & 0.02 & 0.02   & 0.00 & 0.01 & 0.00\\
  DC-SIS2 & 1.00  & 1.00  & 1.00  & 1.00    & 1.00 & 1.00 & 1.00 & 1.00  & 1.00 & 1.00 & 0.16 & 0.16  & 0.32 & 0.25 & 0.05\\
SIRI$^*2$   & 1.00  & 1.00  & 1.00  & 1.00     & 1.00 & 1.00 & 0.92 & 0.92 & 1.00 & 1.00 & 0.34 & 0.34  & 0.70 & 0.57 & 0.36\\
  IP        & 1.00  & 1.00  & 0.99  & 0.99     & 1.00 & 1.00 & 0.94 & 0.94 & 1.00 & 1.00 & 0.97 & 0.97  & 0.81 & 0.89 & 0.70\\
  \midrule
\end{tabular}
\label{tab:Ex2Screen}
}
\end{table}

\begin{table}
\caption{The overall and individual signal-to-noise ratios (SNRs) of each model in Examples 1 and 2.} \vspace{0.1in}
\centering
\scalebox{1}{
\begin{tabular}{ l c c c c cc }
  \toprule
  && \multicolumn{2}{c}{Example 1} & \multicolumn{2}{c}{Example 2}\\
  \cmidrule(lr{.5em}){3-4}  \cmidrule(lr{.5em}){5-6}
  & & Settings 1,\,3 & Settings 2,\,4 & Settings 1,\,3 & Settings 2,\,4 \\
  \midrule
 M1 & $X_1$  & 0.64 & 0.64 & 1.44 & 1.44 \\
             & $X_5$  & 0.64 & 0.64 & 1.44 & 1.44 \\
             & $X_1X_5$  & 1.44 & 1.45 & 3.52 & 3.53 \\
             & Overall  & 2.72 & 2.81 & 6.41 & 6.59 \\
  \addlinespace
  M2 &  $X_1$  & 1.00 & 1.00 & 2.17 & 2.17\\
              & $X_{10}$  & 1.00 & 1.00 & 2.17 & 2.17 \\
              & $X_1X_5$  & 2.25 & 2.26 & 5.28 & 5.30 \\
              & Overall  & 4.25 & 4.26 & 9.61 & 9.64 \\
  \addlinespace
  M3 &  $X_{10}$  & 1.00 & 1.00 & 2.17 & 2.17 \\
  & $X_{15}$  & 1.00 & 1.00 & 2.17 & 2.17 \\
  & $X_1X_5$  & 2.25 & 2.26 & 5.28 & 5.30 \\
  & Overall  & 4.25 & 4.32 & 9.61 & 9.76 \\
  \addlinespace
   M4 & $X_1X_5$  & 4.00 & 4.02 & 7.92 & 7.95 \\
  & $X_{10}X_{15}$  & 4.00 & 4.00 & 7.92 & 7.93 \\
  & Overall  & 8.00 & 8.02 & 15.84 & 15.88 \\
  \midrule
\end{tabular}
\label{tab:snr}
}
\end{table}

\begin{sidewaystable}
\caption{Variable selection results for all the selection methods in terms of medians and robust standard deviations (in parentheses) of various performance measures in Example 1.} \vspace{0.1in}
  \centering
\scalebox{0.65}{
    \begin{tabular}{l rcc rcc rcc rcc}
    \toprule
    Method       & \multicolumn{3}{c}{M1}          &      \multicolumn{3}{c}{M2 }   &      \multicolumn{3}{c}{M3 }   & \multicolumn{3}{c}{M4 } \\
  \cmidrule(lr{.75em}){2-4}  \cmidrule(lr{.75em}){5-7} \cmidrule(lr|{.75em}){8-10}  \cmidrule(lr|{.75em}){11-13}
                       & \multicolumn{1}{c}{PE}   &  FP  &  FN  &   \multicolumn{1}{c}{PE}  & FP  &  FN  &  \multicolumn{1}{c}{PE}  &  FP  &  FN  & \multicolumn{1}{c}{PE}    & FP    & FN \\
       \midrule
          \multicolumn{13}{c}{Setting 1: $(n, p, \rho)=(200, 2000, 0)$} \\
SIS2-Lasso         &8.002 (0.560) & 81 (10.8)   & 0 (0)    & 15.928 (1.003)  & 88.5 (8.6) & 1 (0)   & 15.877 (0.981) & 90 (8.6)    & 1 (0)   & 22.673 (0.672) & 96 (7.1)   & 2 (0) \\
SIS2-$L_1$+SICA    &9.705 (2.388) & 12.5 (10.4) & 0 (0)    & 21.013 (3.354)  & 15 (10.4)  & 1 (0)   & 21.320 (3.220) & 20.5 (11.2) & 1 (0)   & 32.605 (3.190) & 16 (9.7)   & 2 (0)\\
DC-SIS2-Lasso      &7.957 (0.475) & 82 (9.7)    & 0 (0)    & 5.151 (0.328)   & 82 (10.1)  & 0 (0)   & 15.859 (0.969) & 90 (7.1)    & 1 (0)   & 22.440 (6.808) & 94 (8.2)   & 2 (0.7)\\
DC-SIS2-$L_1$+SICA &9.043 (2.174) & 10.5 (9.7)  & 0 (0)    & 6.332 (1.721)   & 11.5 (8.2) & 0 (0)   & 20.693 (3.704) & 15.5 (10.4) & 1 (0)   & 31.637 (9.652) & 17 (11.2)  & 2 (0.7)\\
SIRI	           &6.472 (0.193) &	3 (9.0)	    & 0 (0)    & 4.326 (0.319)	 & 7 (6.7)	  & 0 (0)   & 14.443 (0.874) & 8 (6.7)	   & 1 (0)   & 21.026 (6.344) & 6 (2.1)	   & 2 (0.7) \\
RAMP	           &6.370 (0.089) & 0 (0)	    & 0 (0)	   &13.692 (0.245)	 & 0 (0)	  & 1 (0)   & 13.653 (0.219) & 0 (0)	   & 1 (0)   & 22.856 (0.906) & 2 (1.5)	   & 2 (0) \\
iFORT	           &6.387 (2.020) &	0 (0.2)	    & 0 (0.3)  &13.211 (0.408)	 & 1 (0.2)	  & 2 (0.1)	& 13.379 (0.561) & 1 (0.5)	   & 2 (0.1) & 20.304 (0.422) & 0 (0.1)	   & 2 (0) \\
iFORM	           &6.374 (0.147) &	0 (0.2)	    & 0 (0.3)  &13.199 (0.398)	 & 1 (0.2)	  & 2 (0.1)	& 13.266 (1.092) & 1 (0.5)	   & 2 (0.1) & 20.304 (0.377) & 0 (0.1)	   & 2 (0) \\
IP-hierNet         &8.525 (0.836) & 55.5 (34.7) & 0 (0)    & 6.557 (0.853)   & 82 (28.5)  & 0 (0)   & 7.181 (0.912)  & 95 (27.2)   & 0 (0)   &  6.149 (8.394) & 115 (21.3) & 0 (0.7)\\
IP-Lasso           &8.429 (0.807) & 79.5 (16.8) & 0 (0)    & 5.302 (0.455)   & 75 (14.6)  & 0 (0)   & 5.386 (0.422)  & 74.5 (13.1) & 0 (0)   &  3.135 (7.909) & 79 (11.2)  & 0 (0.7)\\
IP-$L_1$+SICA      &7.391 (1.509) & 3 (5.2)     & 0 (0)    & 4.358 (0.964)   & 1 (4.1)    & 0 (0)   & 4.640 (0.890)  & 2 (5.2)     & 0 (0)   &  3.108 (8.847) & 3 (6.0)    & 0 (0.7)\\
Oracle             &6.340 (0.111) & 0 (0)       & 0 (0)    & 4.051 (0.080)   & 0 (0)      & 0 (0)   & 4.058 (0.081)  & 0 (0)       & 0 (0)   &  2.269 (0.037) & 0 (0)      & 0 (0)\\
       \addlinespace
         \multicolumn{13}{c}{Setting 2: $(n, p, \rho)=(200, 2000, 0.5)$} \\
SIS2-Lasso         &7.993 (0.472) & 83 (10.4)   & 0 (0)    & 15.625 (1.358) & 89 (10.4)   & 1 (0)   & 15.997 (0.965) & 89 (8.6)    & 1 (0)   & 22.973 (0.724) & 94 (5.6)   & 2 (0)\\
SIS2-$L_1$+SICA    &9.355 (2.791) & 12 (11.6)   & 0 (0)    & 20.120 (4.690) & 17.5 (11.6) & 1 (0)   & 21.274 (2.612) & 18 (9.0)    & 1 (0)   & 33.225 (3.600) & 17 (10.4)  & 2 (0)\\
DC-SIS2-Lasso      &7.907 (0.518) & 82 (9.3)    & 0 (0)    & 5.154 (0.417)  & 83 (12.7)   & 0 (0)   & 15.943 (1.115) & 89 (9.0)    & 1 (0)   & 22.804 (1.492) & 92 (6.7)   & 2 (0)\\
DC-SIS2-$L_1$+SICA &8.661 (2.526) & 9 (9.7)     & 0 (0)    & 6.046 (1.995)  & 10 (8.2)    & 0 (0)   & 20.477 (3.891) & 15.5 (11.9) & 1 (0)   & 31.691 (8.200) & 16 (10.4)  & 2 (0)\\
SIRI	           & 6.550 (0.353) & 3 (5.2)	& 0 (0)	   & 4.293 (0.258)	& 7 (3.7)	  & 0 (0)	& 14.097 (0.690) & 8 (6.2)	   & 1 (0)	 & 21.740 (1.209) & 6 (3.7)	   & 2 (0) \\
RAMP	           & 6.360 (0.076) & 0 (0)	    & 0 (0)	   &13.397 (0.265)	& 0 (0)	      & 1 (0)	& 13.423 (0.223) & 0 (0)	   & 1 (0)	 & 22.862 (0.836) & 1 (0.7)	   & 2 (0) \\
iFORT	           & 6.397 (1.699) & 0 (0.2)	& 0 (0.3)  &13.288 (0.471)	& 1 (0.2)	  & 2 (0.1)	& 13.400 (0.594) & 1 (0.5)	   & 2 (0.1) & 20.309 (0.306) & 0 (0)	   & 2 (0) \\
iFORM	           & 6.375 (0.103) & 0 (0.2)	& 0 (0.3)  &13.288 (0.457)	& 1 (0.2)	  & 2 (0.1)	& 13.266 (1.291) & 1 (0.5)	   & 2 (0.1) & 20.309 (0.306) & 0 (0)	   & 2 (0) \\
IP-hierNet         & 8.310 (0.705) & 37 (32.5)  & 0 (0)    & 6.441 (1.004)  & 71 (22.4)   & 0 (0)   & 6.891 (1.179)  & 85.5 (22.6) & 0 (0)   & 5.467 (8.654)  & 109 (21.1) & 0 (0.7)\\
IP-Lasso           & 8.487 (0.698) & 73.5 (16.4)& 0 (0)    & 5.423 (0.482)  & 76 (13.4)   & 0 (0)   & 5.375 (0.577)  & 71.5 (16.8) & 0 (0)   & 3.053 (8.036)  & 77 (16.0)  & 0 (0.7)\\
IP-$L_1$+SICA      & 7.343 (1.603) & 3 (6.0)    & 0 (0)    & 4.373 (0.970)  & 1 (3.7)     & 0 (0)   & 4.561 (1.380)  & 2 (5.6)     & 0 (0)   & 2.826 (9.292)  & 3.5 (6.7)  & 0 (0.7)\\
Oracle             & 6.335 (0.115) & 0 (0)      & 0 (0)    & 4.057 (0.073)  & 0 (0)       & 0 (0)   & 4.06 (0.082)   & 0 (0)       & 0 (0)   & 2.270 (0.041)  & 0 (0)      & 0 (0)\\
       \addlinespace
          \multicolumn{13}{c}{Setting 3: $(n, p, \rho)=(300, 5000, 0)$} \\
SIS2-Lasso         &7.686 (0.307)  & 123 (16.4) & 0 (0)    & 15.365 (0.798) & 138 (8.6)   & 1 (0)   & 15.491 (0.613) & 139 (10.4)  & 1 (0)   & 22.582 (0.622) & 145 (6.3)    & 2 (0)\\
SIS2-$L_1$+SICA    &10.047 (1.277) & 19 (5.6)   & 0 (0)    & 21.666 (2.105) & 29 (7.8)    & 1 (0)   & 21.748 (2.205) & 29.5 (8.6)  & 1 (0)   & 32.559 (3.175) & 32.5 (14.9)  & 2 (0)\\
DC-SIS2-Lasso      &7.660 (0.293)  & 129 (16.8) & 0 (0)    & 4.919 (0.221)  & 127 (14.9)  & 0 (0)   & 15.419 (0.608) & 136 (9.7)   & 1 (0)   & 22.383 (7.327) & 139.5 (10.4) & 2 (0.7)\\
DC-SIS2-$L_1$+SICA &10.248 (1.705) & 20.5 (8.2) & 0 (0)    & 6.098 (0.986)  & 14 (5.6)    & 0 (0)   & 21.694 (2.437) & 30 (8.2)    & 1 (0)   & 31.714 (10.652)& 30 (12.3)    & 2 (0.7)\\
SIRI	           & 6.360 (0.192) & 3 (5.2)	& 0 (0)	   & 4.156 (0.158)	& 7 (6.7)	  & 0 (0)	& 14.011 (0.607) & 4 (3.0)	   & 1 (0)	 & 13.357 (7.306) & 5 (4.5)	     & 1 (0.7) \\
RAMP	           & 6.306 (0.053) & 0 (0)	    & 0 (0)	   & 13.603 (0.137)	& 0 (0)	      & 1 (0)	& 13.576 (0.081) & 0 (0)	   & 1 (0)	 & 22.461 (0.754) & 1 (0.7)	     & 2 (0) \\
iFORT	           & 6.316 (0.074) & 0 (0.7)	& 0 (0)	   & 13.067 (0.224)	& 1 (0.1)	  & 2 (0)	& 13.285 (0.306) & 1 (0.4)	   & 2 (0)	 & 20.284 (0.374) & 0 (0.1)	     & 2 (0) \\
iFORM	           & 6.316 (0.091) & 0 (0.7)	& 0 (0)	   & 13.067 (0.217)	& 1 (0.1)	  & 2 (0)	& 13.123 (1.608) & 1 (0.4)	   & 2 (0)	 & 20.284 (0.355) & 0 (0.1)	     & 2 (0) \\
IP-hierNet         &8.435 (0.903)  & 103 (44.8) & 0 (0)    & 5.879 (0.608)  & 105 (37.7)  & 0 (0)   & 6.241 (0.660)  & 122.5 (33.8)& 0 (0)   & 4.735 (8.426)  & 156.5 (29.9) & 0 (0.7)\\
IP-Lasso           &8.105 (0.474)  & 115.5 (17.5) & 0 (0)  & 5.140 (0.371)  & 109.5 (27.6)& 0 (0)   & 5.118 (0.371)  & 105 (25.7)  & 0 (0)   & 2.903 (8.092)  & 118 (17.2)   & 0 (0.7)\\
IP-$L_1$+SICA      &6.986 (1.404)  & 3 (7.8)      & 0 (0)  & 4.624 (1.325)  & 4 (9.7)     & 0 (0)   & 4.653 (1.243)  & 4 (9.7)     & 0 (0)   & 2.859 (9.151)  & 7 (9.3)      & 0 (0.7)\\
Oracle             &6.307 (0.093)  & 0 (0)        & 0 (0)  & 4.036 (0.054)  & 0 (0)       & 0 (0)   & 4.034 (0.056)  & 0 (0)       & 0 (0)   & 2.261 (0.033)  & 0 (0)        & 0 (0)\\
       \addlinespace
          \multicolumn{13}{c}{Setting 4: $(n, p, \rho)=(300, 5000, 0.5)$} \\
SIS2-Lasso         &7.733 (0.383)  & 123 (14.6)   & 0 (0)  & 15.250 (1.363)  & 133 (13.8)   & 1 (0)  & 15.519 (0.562) & 137 (11.2)  & 1 (0)   & 22.717 (0.638) & 143 (6.0)    & 2 (0)\\
SIS2-$L_1$+SICA    &9.999 (1.706)  & 19 (8.2)     & 0 (0)  & 20.051 (3.688) & 23.5 (11.2)  & 1 (0)  & 21.494 (3.002) & 29 (10.4)   & 1 (0)   & 33.411 (2.945) & 34 (11.2)    & 2 (0)\\
DC-SIS2-Lasso      &7.633 (0.376)  & 123 (18.7)   & 0 (0)  & 4.852 (0.238)  & 127.5 (13.8) & 0 (0)  & 15.486 (0.601) & 134 (12.7)  & 1 (0)   & 22.349 (7.051) & 138.5 (10.8) & 2 (0.7)\\
DC-SIS2-$L_1$+SICA &10.013 (1.510) & 19 (6.7)     & 0 (0)  & 6.082 (0.834)  & 15 (6.3)     & 0 (0)  & 21.585 (3.663) & 27 (10.4)   & 1 (0)   & 31.448 (10.427)& 32 (9.3)     & 2 (0.7)\\
SIRI	           &6.374 (0.206)  & 3 (5.2)	  & 0 (0)   & 4.143 (0.133)	 & 7 (6.7)	   & 0 (0)	& 13.865 (0.730) & 8 (6.7)	   & 1 (0)	 & 12.710 (7.244) & 5 (4.5)	     & 1 (0.7) \\
RAMP	           &6.305 (0.063)  & 0 (0)	      & 0 (0)   & 13.374 (0.156) & 0 (0)	   & 1 (0)	& 13.355 (0.083) & 0 (0)	   & 1 (0)	 & 22.616 (0.705) & 1 (0.7)	     & 2 (0) \\
iFORT	           &6.325 (1.375)  & 0 (0.1)	  & 0 (0.2) & 13.090 (0.225) & 1 (0.1)	   & 2 (0)	& 13.237 (0.281) & 1 (0.4)	   & 2 (0)	 & 20.257 (0.376) & 0 (0.1)	     & 2 (0) \\
iFORM	           &6.325 (0.095)  & 0 (0.1)	  & 0 (0.2)	& 13.071 (0.220) & 1 (0.1)	   & 2 (0)	& 13.110 (1.664) & 1 (0.4)	   & 2 (0)	 & 20.257 (0.349) & 0 (0.1)	     & 2 (0) \\
IP-hierNet         &8.097 (0.470)   & 76 (41.0)    & 0 (0)   & 5.776 (0.536)  & 96 (25.6)   & 0 (0)  & 5.978 (0.538)  & 105 (29.9)  & 0 (0)   & 4.647 (8.700)  & 151 (27.6)   & 0 (0.7)\\
IP-Lasso           &7.975 (0.465)  & 112 (19.0)   & 0 (0)   & 5.115 (0.417)  & 109.5 (30.6)& 0 (0)  & 5.095 (0.285)  & 106 (28.0)  & 0 (0)   & 2.860 (7.827)  & 113.5 (19.4) & 0 (0.7)\\
IP-$L_1$+SICA      &6.753 (1.271)  & 1 (6.7)      & 0 (0)   & 4.470 (1.182)   & 2.5 (9.0)   & 0 (0)  & 4.450 (0.801)  & 3 (7.5)     & 0 (0)   & 3.121 (9.126)  & 7.5 (9.0)    & 0 (0.7)\\
Oracle             &6.305 (0.091)  & 0 (0)        & 0 (0)   & 4.039 (0.060)  & 0 (0)       & 0 (0)  & 4.033 (0.067)  & 0 (0)       & 0 (0)   & 2.261 (0.033)  & 0 (0)        & 0 (0)\\
    \bottomrule
    \end{tabular}
  \label{tab: Ex1Select}
}
\end{sidewaystable}

\begin{sidewaystable}
\caption{Variable selection results for all the selection methods in terms of medians and robust standard deviations (in parentheses) of various performance measures in Example 2.} \vspace{0.1in}
  \centering
\scalebox{0.65}{
    \begin{tabular}{l rcc rcc rcc rcc}
    \toprule
    Method       & \multicolumn{3}{c}{M1 }          &      \multicolumn{3}{c}{M2 }   &      \multicolumn{3}{c}{M3 }   & \multicolumn{3}{c}{M4 } \\
  \cmidrule(lr{.75em}){2-4}  \cmidrule(lr{.75em}){5-7} \cmidrule(lr|{.75em}){8-10}  \cmidrule(lr|{.75em}){11-13}
                       & \multicolumn{1}{c}{PE}   &  FP  &  FN  &   \multicolumn{1}{c}{PE}  & FP  &  FN  &  \multicolumn{1}{c}{PE}  &  FP  &  FN  & \multicolumn{1}{c}{PE}    & FP    & FN \\
       \midrule
          \multicolumn{13}{c}{Setting 1: $(n, p, \rho)=(200, 2000, 0)$} \\
SIS2-Lasso         &3.652 (0.422) & 73.5 (15.3) & 0 (0)  & 15.093 (1.077) & 88 (12.7)   & 1 (0)   & 15.317 (0.779) & 88.5 (9.0)  & 1 (0) & 25.252 (0.812)  & 93 (6.3)   & 2 (0)\\
SIS2-$L_1$+SICA    &3.081 (0.584) & 0 (3.0)     & 0 (0)  & 19.573 (3.202) & 13.5 (11.9) & 1 (0)   & 20.181 (3.209) & 14 (10.4)   & 1 (0) & 35.674 (3.862)  & 14 (10.1)  & 2 (0)\\
DC-SIS2-Lasso      &3.678 (0.393) & 75.5 (13.1) & 0 (0)  & 2.470 (0.248)  & 73 (17.5)   & 0 (0)   & 15.395 (1.029) & 87 (9.7)    & 1 (0) & 24.959 (8.620)  & 93 (8.6)   & 2 (0.7)\\
DC-SIS2-$L_1$+SICA &3.092 (0.800) & 0 (4.5)     & 0 (0)  & 2.089 (0.532)  & 0 (4.9)     & 0 (0)   & 20.506 (3.690) & 15.5 (11.9) & 1 (0) & 32.404 (12.479) & 15.5 (9.0) & 2 (0.7)\\
SIRI	           &2.973 (0.235) & 0 (0)	    & 0 (0)	 & 2.1689 (0.084) & 3 (3.0)	    & 0 (0)	  & 13.169 (0.763) & 4 (3.0)	 & 1 (0) & 21.832 (8.210)  & 4.5 (2.8)	& 2 (0.7) \\
RAMP	           &2.980 (0.277) & 0 (0)	    & 0 (0)	 & 12.709 (0.353) & 0 (0)	    & 1 (0)	  & 12.604 (0.172) & 0 (0)	     & 1 (0) & 23.535 (0.837)  & 1 (0.7)	& 2 (0) \\
iFORT	           &2.974 (1.725) & 0 (0.2)	    & 0 (0.2)& 13.732 (0.415) & 1 (0.2)	    & 2 (0.1) & 13.961 (0.543) & 1 (0.5)	 & 2 (0) & 24.107 (0.551)  & 0 (0.1)	& 2 (0) \\
iFORM	           &2.968 (1.715) & 0 (0.2)	    & 0 (0.2)& 13.728 (0.578) & 1 (0.2)	    & 2 (0.1) & 13.775 (2.646) & 1 (0.5)	 & 2 (0) & 24.107 (0.551)  & 0 (0.1)	& 2 (0) \\
IP-hierNet         &4.487 (0.624) & 46.5 (38.4) & 0 (0)  & 3.108 (0.545)  & 57 (24.3)   & 0 (0)   & 3.399 (0.583)  & 72 (21.3)   & 0 (0) & 3.557 (10.816)  & 110 (21.1) & 0 (0.7)\\
IP-Lasso           &3.777 (0.438) & 76.5 (14.9) & 0 (0)  & 2.595 (0.275)  & 71 (19.4)   & 0 (0)   & 2.609 (0.292)  & 72 (13.8)   & 0 (0) & 1.719 (9.417)   & 64 (19.0)  & 0 (0.7)\\
IP-$L_1$+SICA      &3.061 (0.579) & 0 (1.9)     & 0 (0)  & 2.076 (0.342)  & 0 (2.2)     & 0 (0)   & 2.058 (0.399)  & 0 (3.4)     & 0 (0) & 1.543 (10.135)  & 1.5 (4.5)  & 0 (0.7)\\
Oracle             &2.929 (0.237) & 0 (0)       & 0 (0)  & 2.002 (0.069)  & 0 (0)       & 0 (0)   & 2.017 (0.065)  & 0 (0)       & 0 (0) & 1.339 (0.035)   & 0 (0)      & 0 (0)\\
       \addlinespace
          \multicolumn{13}{c}{Setting 2: $(n, p, \rho)=(200, 2000, 0.5)$} \\
SIS2-Lasso         &3.643 (0.430) & 76.5 (14.2) & 0 (0) & 15.314 (1.863) & 87 (10.8)    & 1 (0)   & 15.285 (0.958) & 88 (7.5)    & 1 (0) & 25.151 (0.881)  & 95 (7.1)    & 2 (0)\\
SIS2-$L_1$+SICA    &3.152 (0.887) & 0 (5.2)     & 0 (0) & 19.485 (3.944) & 13 (11.9)    & 1 (0)   & 20.731 (3.278) & 19 (9.7)    & 1 (0) & 36.600 (2.672)  & 16.5 (10.4) & 2 (0)\\
DC-SIS2-Lasso      &3.668 (0.422) & 78.5 (15.7) & 0 (0) & 2.481 (0.192)  & 74.5 (21.6)  & 0 (0)   & 15.163 (1.150) & 87 (12.7)   & 1 (0) & 24.997 (7.947)  & 90.5 (10.8) & 2 (0.7)\\
DC-SIS2-$L_1$+SICA &3.183 (0.717) & 0 (3.7)     & 0 (0) & 2.226 (0.560)  & 1 (4.9)      & 0 (0)   & 18.800 (5.117) & 12 (12.3)   & 1 (0) & 34.958 (10.563) & 19 (10.1)   & 2 (0.7)\\
SIRI	           & 3.001 (0.236) & 0 (0)	    & 0 (0)	  &  2.198 (0.064) & 3 (3.0)	& 0 (0)	  & 13.312 (0.577) & 4 (3.0)	 & 1 (0)   & 22.190 (8.575)	& 4 (3.9)	& 2 (0.7) \\
RAMP	           & 2.992 (0.250) & 0 (0)	    & 0 (0)	  & 12.938 (0.293) & 0 (0)	    & 1 (0)	  & 12.853 (0.177) & 0 (0)	     & 1 (0)   & 24.072 (1.078)	& 2 (0.9)	& 2 (0) \\
iFORT	           & 2.984 (2.020) & 0 (0.2)	& 0 (0.3) &	12.834 (0.379) & 1 (0.3)	& 2 (0.1) & 13.104 (1.493) & 1 (0.5)	 & 2 (0.1) & 23.197 (0.401)	& 0 (0.1)	& 2 (0) \\
iFORM	           & 2.984 (2.508) & 0 (0.2)	& 0 (0.3) & 12.815 (0.330) & 1 (0.3)	& 2 (0.1) & 12.876 (2.248) & 1 (0.5)	 & 2 (0.1) & 23.197 (0.401)	& 0 (0.1)	& 2 (0) \\
IP-hierNet         &4.306 (0.560) & 28 (29.5)   & 0 (0) & 2.881 (0.497)  & 52 (21.5)    & 0 (0)   & 3.286 (0.390)  & 64 (24.3)   & 0 (0) & 3.442 (10.716)  & 107 (22.9)  & 0 (0.7)\\
IP-Lasso           &3.829 (0.406) & 71 (19.4)   & 0 (0)   & 2.516 (0.266)  & 68.5 (15.7)& 0 (0)   & 2.543 (0.247)  & 71 (14.5)   & 0 (0) & 1.727 (9.245)    & 60 (17.2) & 0 (0.7)\\
IP-$L_1$+SICA      &3.028 (0.549) & 0 (3.7)     & 0 (0)   & 2.079 (0.220)  & 0 (2.3)    & 0 (0)   & 2.056 (0.303)  & 0 (3.0)     & 0 (0) & 1.492 (9.988)    & 1 (4.5)   & 0 (0.7)\\
Oracle             &2.941 (0.238) & 0 (0)       & 0 (0)   & 2.021 (0.072)  & 0 (0)      & 0 (0)   & 2.007 (0.061)  & 0 (0)       & 0 (0) & 1.345 (0.033)    & 0 (0)     & 0 (0)\\
       \addlinespace
          \multicolumn{13}{c}{Setting 3: $(n, p, \rho)=(300, 5000, 0)$} \\
SIS2-Lasso         &3.481 (0.361) & 115 (24.6)   & 0 (0)  & 14.708 (0.678) & 133 (14.2) & 1 (0)   & 14.861 (0.639) & 132 (11.2)  & 1 (0) & 24.988 (0.751)  & 143.5 (6.7) & 2 (0)\\
SIS2-$L_1$+SICA    &3.146 (0.792) & 0 (5.2)      & 0 (0)  & 19.613 (3.332) & 24 (14.6)  & 1 (0)   & 20.765 (1.990) & 28.5 (5.2)  & 1 (0) & 36.296 (3.241)  & 33 (14.9)   & 2 (0)\\
DC-SIS2-Lasso      &3.475 (0.345) & 109.5 (26.5) & 0 (0)  & 2.396 (0.135)  & 126 (26.5) & 0 (0)   & 14.724 (0.703) & 131 (15.3)  & 1 (0) & 24.635 (8.786)  & 140 (12.7)  & 2 (0.7)\\
DC-SIS2-$L_1$+SICA &3.092 (0.671) & 0 (4.5)      & 0 (0)  & 2.136 (0.327)  & 1 (4.1)    & 0 (0)   & 19.987 (3.224) & 26 (13.1)   & 1 (0) & 34.206 (13.298) & 27 (12.7)   & 2 (0.7)\\
SIRI	           &2.943 (0.251) &	0 (0)	     & 0 (0)  &  2.170 (0.040)	& 3 (3.0)	& 0 (0)	  & 13.212 (0.507) & 7 (6.0)	 & 1 (0)   & 12.263 (2.749)	& 5 (5.4)	 & 1 (0.2) \\
RAMP	           &2.934 (0.246) & 0 (0)	     & 0 (0)  & 12.848 (0.192)	& 0 (0)	    & 1 (0)	  & 12.790 (0.105) & 0 (0)	     & 1 (0)   & 23.749 (0.631)	& 1 (0.7)	 & 2 (0) \\
iFORT	           &2.925 (0.644) &	0 (0.2)	     & 0 (0)  & 12.671 (0.287)	& 1 (0.3)	& 2 (0)	  & 12.836 (0.889) & 1 (0.4)	 & 2 (0.7) & 23.159 (0.411)	& 0 (0.1)	 & 2 (0) \\
iFORM	           &2.934 (0.946) & 0 (0.2)	     & 0 (0)  & 12.651 (0.262)	& 1 (0.3)	& 2 (0)	  & 12.648 (2.042) & 1 (0.4)	 & 2 (0.7) & 23.152 (0.394)	& 0 (0.1)	 & 2 (0) \\
IP-hierNet         &3.753 (0.534) & 42 (27.6)    & 0 (0)  & 2.725 (0.253)  & 61.5 (26.9)& 0 (0)   & 2.955 (0.366)  & 79 (34.5)   & 0 (0) & 2.587 (10.023)  & 138 (36.0)  & 0 (0.7)\\
IP-Lasso           &3.620 (0.400) & 112.5 (24.6) & 0 (0)  & 2.441 (0.192)   & 96 (21.3) & 0 (0)   & 2.445 (0.185)  & 98.5 (19.4) & 0 (0)   & 1.574 (8.975)  & 78 (35.4)  & 0 (0.7)\\
IP-$L_1$+SICA      &3.117 (0.850) & 0 (6.7)      & 0 (0)  & 2.071 (0.171)   & 0 (2.2)   & 0 (0)   & 2.074 (0.228)  & 0 (3.0)     & 0 (0)   & 1.377 (9.704)  & 0 (3.4)    & 0 (0.7)\\
Oracle             &2.924 (0.251) & 0 (0)        & 0 (0)  & 2.006 (0.055)   & 0 (0)     & 0 (0)   & 2.007 (0.064)  & 0 (0)       & 0 (0)   & 1.347 (0.031)  & 0 (0)      & 0 (0)\\
       \addlinespace
          \multicolumn{13}{c}{Setting 4: $(n, p, \rho)=(300, 5000, 0.5)$} \\
SIS2-Lasso         &3.457 (0.329) & 109.5 (25.4) & 0 (0)  & 14.505 (1.194) & 133 (14.9)   & 1 (0) & 14.947 (0.596) & 136 (9.7)   & 1 (0)   & 25.174 (0.702) & 144 (7.1)   & 2 (0)\\
SIS2-$L_1$+SICA    &3.095 (0.823) & 0 (4.5)      & 0 (0)  & 19.140 (3.153)  & 26 (9.0)     & 1 (0) & 20.824 (2.960) & 28.5 (10.4) & 1 (0)   & 36.423 (3.332) & 31.5 (14.6) & 2 (0)\\
DC-SIS2-Lasso      &3.492 (0.358) & 112 (26.1)   & 0 (0)  & 2.384 (0.134)  & 118.5 (35.1) & 0 (0) & 14.929 (0.784) & 135 (10.4)  & 1 (0)   & 14.477 (8.994) & 140 (14.9)  & 1 (0.7)\\
DC-SIS2-$L_1$+SICA &3.204 (0.963) & 0.5 (7.8)    & 0 (0)  & 2.074 (0.274)  & 0 (3.4)      & 0 (0) & 19.962 (3.679) & 26 (13.4)   & 1 (0)   & 21.558 (13.215)& 27 (13.4)   & 1 (0.7)\\
SIRI	           &3.017 (0.034) & 0 (2.2)	     & 0 (0)   & 2.166 (0.0384)  & 3 (3.0)	  & 0 (0) & 13.238 (0.598) & 7.5 (6.0)	 & 1 (0)   & 12.350 (7.679) & 5 (4.478)   & 1 (0.7) \\
RAMP	           &2.961 (0.254) &	0 (0)	     & 0 (0)   & 12.902 (0.182) & 0 (0)	  & 1 (0) & 12.861 (0.133) & 0 (0)	     & 1 (0)   & 24.173 (0.694) & 1 (0.7)	  & 2 (0) \\
iFORT	           &2.937 (1.669) & 0 (0.2)	     & 0 (0.2) & 12.066 (0.296) & 1 (0.2)	  & 2 (0) & 12.179 (0.434) & 1 (0.4)	 & 2 (0)   & 22.540 (0.368) & 0 (0)	      & 2 (0) \\
iFORM	           &2.926 (1.113) & 0 (0.2)	     & 0 (0.2) & 12.066 (0.275) & 1 (0.2)	  & 2 (0) & 12.097 (1.764) & 1 (0.4)	 & 2 (0)   & 22.540 (0.368) & 0 (0)	      & 2 (0) \\
IP-hierNet         &3.727 (0.484) & 38 (30.6)    & 0 (0)   & 2.762 (0.284)   & 58 (23.1)  & 0 (0) & 2.863 (0.357)  & 69.5 (31.0) & 0 (0)   & 2.483 (10.133) & 130.5 (28.0)& 0 (0.7)\\
IP-Lasso           &3.590 (0.319) & 109 (20.5)   & 0 (0)   & 2.491 (0.220)   & 100 (20.5) & 0 (0) & 2.433 (0.203)  & 97 (18.3)   & 0 (0)   & 1.555 (9.057)  & 71 (39.2)   & 0 (0.7)\\
IP-$L_1$+SICA      &3.108 (0.769) & 0 (4.5)      & 0 (0)   & 2.061 (0.255)   & 0 (3.0)    & 0 (0) & 2.072 (0.173)  & 0 (2.2)     & 0 (0)   & 1.381 (9.257)  & 0 (4.5)     & 0 (0.7)\\
Oracle             &2.921 (0.245) & 0 (0)        & 0 (0)   & 2.009 (0.064)   & 0 (0)      & 0 (0) & 2.009 (0.070)  & 0 (0)       & 0 (0)   & 1.343 (0.028)  & 0 (0)       & 0 (0)\\
    \bottomrule
    \end{tabular}
  \label{tab: Ex2Select}
}
\end{sidewaystable}

\clearpage

\begin{table}
\caption{\label{tab:real} The means and standard errors (in parentheses) of classification errors and median model sizes in prostate cancer data analysis.}
\centering
\begin{tabular}{lcc}
\midrule
 Method        &   Classification error  &  Median model size   \\
\midrule
SIS2-Enet      &   0.0754 (0.0030)  &  75 \\
DC-SIS2-Enet   &   0.0745 (0.0031)  &  70 \\
IP-Enet        &   0.0681 (0.0033)  & 106 \\
\midrule
\end{tabular}
\end{table}

\begin{table}
\caption{\label{genes} List of top 10 genes in main effects and top 10 gene-gene interactions selected by SIS2-Enet, DC-SIS2-Enet, and IP-Enet in prostate cancer data analysis.}
\centering
\scalebox{0.85}{
\begin{tabular}{cccccc}
  \toprule
 \multicolumn{2}{c}{SIS2-Enet} &  \multicolumn{2}{c}{DC-SIS2-Enet} & \multicolumn{2}{c}{IP-Enet}  \\
 \cmidrule(lr{.75em}){1-2}     \cmidrule(lr{.75em}){3-4}    \cmidrule(lr{.75em}){5-6}
   \multicolumn{6}{c}{Main effects}  \\
  Gene name  & Frequency  & Gene name     & Frequency   & Gene name     & Frequency  \\
    \cmidrule(lr{.75em}){1-2}     \cmidrule(lr{.75em}){3-4}    \cmidrule(lr{.75em}){5-6}
 SERINC5 & 100	 & SERINC5	& 100	& HPN	    & 100   \\
 HPN	 & 100	 & HPN	    & 100	& HSPD1	    & 100  \\
 HSPD1	 & 100	 & HSPD1	& 100	& LMO3	    & 100  \\
 LMO3	 & 100	 & LMO3	    & 100	& ERG	    & 100  \\
 TARP	 & 100	 & ANGPT1	& 100	& TARP	    & 98    \\
 ANGPT1	 & 99	 & TARP	    & 100	& SERINC5	& 86   \\
 S100A4	 & 95	 & PDLIM5	& 98	& ANGPT1	& 86   \\
 CALM1	 & 93	 & CALM1	& 97	& RBP1	    & 85   \\
 PDLIM5	 & 89	 & RBP1	    & 92	& CALM1	    & 82   \\
 RBP1	 & 85	 & S100A4	& 89	& S100A4	& 70   \\
    \cmidrule(lr{.75em}){1-2}     \cmidrule(lr{.75em}){3-4}    \cmidrule(lr{.75em}){5-6}
 \multicolumn{6}{c}{Interactions}  \\
 Interaction  & Frequency   & Interaction  & Frequency  & Interaction  & Frequency  \\
  \cmidrule(lr{.75em}){1-2}     \cmidrule(lr{.75em}){3-4}    \cmidrule(lr{.75em}){5-6}
 DPT$\times$S100A4   & 70  & DPT$\times$S100A4	 & 71  &  SLC7A1$\times$ERG       &  75\\
 GUCY1A3$\times$MAF  & 64  & DPT$\times$CFD	 & 57  &  PRKDC$\times$CFD	      &  67\\
 RARRES2$\times$KLK3 & 60	& HSPD1$\times$LMO3	 & 56  &  PRKDC$\times$KLK3	      &  64\\
 AGR2$\times$EPCAM	  & 60	& PDLIM5$\times$CFD	 & 53  &  AFFX-CreX-3$\times$CHPF &  64\\
 FOXA1$\times$SIM2	  & 51	& LMOD1$\times$RGS10 & 53  &  RASSF7$\times$PRKDC	  &  59\\
 RBP1$\times$TGFB3	  & 49	& PENK$\times$GSTP1	 & 50  &  DPT$\times$S100A4	      &  54\\
 MAF$\times$NELL2	  & 49	& RBP1$\times$EPCAM	 & 50  &  KANK1$\times$ERG	      &  51\\
 HSPD1$\times$LMO3	  & 46	& DPYSL2$\times$EPCAM   & 49   & RBP1$\times$MAF	  &  49\\
 FOXA1$\times$EPCAM  & 44	& ALCAM$\times$EPCAM    & 45   & MAF$\times$NELL2	  &  49\\
 DPT$\times$CFD	  & 44	& SLC25A6$\times$KLK3   & 44   & RARRES2$\times$KLK3  &  48\\
  \toprule
\end{tabular}
}
\end{table}

\newpage
\quad

\newpage
\setcounter{page}{1}
\setcounter{section}{0}
\setcounter{equation}{0}

\renewcommand{\theequation}{A.\arabic{equation}}
\setcounter{equation}{0}
\setcounter{table}{7}
\appendix

\begin{center}{\bf \Large Supplementary Material to ``Interaction Pursuit with\\ Feature Screening and Selection''}

\bigskip

Yingying Fan, Yinfei Kong, Daoji Li and Jinchi Lv
\end{center}



\noindent This Supplementary Material consists of five parts. Section A presents some additional simulation studies. We establish the invariance of the three sets $\mathcal{A}$, $\mathcal{I}$, and $\mathcal{M}$ under affine transformations in Section B. Section C illustrates that in the presence of correlation among covariates, using $\corr(X_j^2, Y^2)$ as the marginal utility still has differentiation power between interaction variables (that is, variables contributing to interactions) and noise variables (variables contributing to neither interactions nor main effects). We provide the proofs of Proposition \ref{prop1} and Theorems \ref{Th: Sure Screening-new}--\ref{Th:RE condition} in Section D. Section E contains some technical lemmas and their proofs. Hereafter we use $\widetilde{C}_i$ with $i = 1, 2, \cdots$ to denote some generic positive or nonnegative constants whose values may vary from line to line. For any set $\mathcal{D}$, denote by $|\mathcal{D}|$ its cardinality.

\section*{Appendix A:  Additional simulation studies}\label{AppA}

\renewcommand{\theequation}{A.\arabic{equation}}
\setcounter{equation}{0}

\section*{A.1. Lower signal-to-noise ratios in Example 1}

In Section \ref{screening}, we investigated the screening performance of each procedure at certain noise levels. It is also interesting to test the robustness of those methods when the signal-to-noise ratio (SNR) becomes  smaller. Therefore, keeping all the settings in Example 1 the same as before, we now consider three more sets of noise level:
\begin{enumerate}
\item[] \hspace{-0.25in} Case 1: $\varepsilon_1 \sim N(0, 3^2)$, $\varepsilon_2 \sim N(0, 2.5^2)$, $\varepsilon_3 \sim N(0, 2.5^2)$, $\varepsilon_4 \sim N(0, 2^2)$;
\item[] \hspace{-0.25in} Case 2: $\varepsilon_1 \sim N(0, 3.5^2)$, $\varepsilon_2 \sim N(0, 3^2)$, $\varepsilon_3 \sim N(0, 3^2)$, $\varepsilon_4 \sim N(0, 2.5^2)$;
\item[] \hspace{-0.25in} Case 3: $\varepsilon_1 \sim N(0, 4^2)$, $\varepsilon_2 \sim N(0, 3.5^2)$, $\varepsilon_3 \sim N(0, 3.5^2)$, $\varepsilon_4 \sim N(0, 3^2)$.
\end{enumerate}
%
%
%

Following the same definition of SNR as in Section 4.1, the SNRs in the settings above are listed in Table \ref{tab:snr-supp} and far lower than before. For example, the SNRs in the third set of noise levels for models M1--M4 are 
$0.39$, $0.33$, $0.33$, and $0.25$ times as large as before, respectively.

\begin{center}
[Table \ref{tab:snr-supp} about here.]
\end{center}

The corresponding screening results for those three sets of noise levels are summarized in Table \ref{supp:screen}.  It is seen that our approach IP performed better than all others across three settings in models M2--M4, where the strong heredity assumption is not satisfied. In model M1, the IP did not perform as well as other methods, since it kept only $[n / (\log n)]$ variables for constructing interactions while the other methods kept up to $2[n / (\log n)]$ interaction variables in the screening step.

\begin{center}
[Table  \ref{supp:screen} about here.]
\end{center}

%
%
%

\section*{A.2. Computation time}

To demonstrate the effect of interaction screening on the computational cost, we consider model M2 in Example 1 with $n=200$, $\rho=0.5$ and 
$p = 200, 300,$ and $500$, and calculate the average computation time of hierNet and IP-hierNet. The only difference between these two methods is that IP-hierNet has the screening step whereas hierNet does not. Table \ref{tab:comptime} reports the average computation time of hierNet and IP-hierNet based on 100 replications.  We see from Table \ref{tab:comptime} that when the dimensionality gets higher, the ratio of average computation time of hierNet over IP-hierNet becomes larger. In particular, the average computation time for hierNet reaches $292.77$ minutes for a single repetition when $p=500$, while that for IP-hierNet is only $6.05$ minutes. As expected, our proposed procedure IP is computationally much more efficient thanks to the additional screening step.

\begin{center}
[Table \ref{tab:comptime} about here.]
\end{center}

\section*{A.3. Feature screening with main effect only model}

As suggested by the AE and one referee, we now consider the following additional simulation example to compare the feature screening performance when the model contains no interactions 
  \begin{align*}
     \mbox{M5}: Y= X_1 + X_5 + X_{10} + X_{15} + \varepsilon,
  \end{align*}
where the covariate vector $\bx=(X_1, \cdots, X_p)^T \sim N(\bzero, \bSig)$ with
$\bSig=(\rho^{|j-k|})_{1 \leq j, k \leq p}$, and the random error $\varepsilon$ is independent of $\bx$ and generated from $N(0, 2^2)$ or $t_{(3)}$. Four different settings of $(n, p, \rho)=(200, 2000, 0)$,
$(200, 2000, 0.5)$, $(300, 5000, 0)$, and $(300, 5000, 0.5)$ are considered and we repeated each experiment 100 times.

Table \ref{tab:no-interaction-screen} below presents the feature  screening results. As expected, SIS2, DC-SIS2, and IP performed very similarly and were able to retain almost all important main effects across all the settings. Interestingly, these three methods also outperformed SIRI*2 when the error follows Gaussian distribution $N(0, 2^2)$ in settings 1 and 2.

\begin{center}
[Table \ref{tab:no-interaction-screen} about here.]
\end{center}

\section*{A.4. Feature screening with equal correlation model}

Following the suggestion of the AE and one referee, we also consider the following additional simulation example with equal correlation among covariates 
\begin{itemize}
\item  M3$'$: $Y=2X_{10} + 2X_{15} + 3X_1X_5+\varepsilon_3$,

\item  M4$'$: $Y=3X_1X_5 + 3X_{10}X_{15}+\varepsilon_4$,
\end{itemize}
where $\bx=(X_1, \cdots, X_p)^T\sim N(\bzero, \bSig)$ with $\bSig$ having diagonal entries 1 and off-diagonal entries $0.2$, and $(n, p)=(200, 2000)$. Here, the equal correlation 0.2 was suggested by a referee. Models M3$'$ and M4$'$ are the same as settings 1 and 2 of models M3 and M4 in the main text, respectively, except for the covariance matrix $\bSig$.  Table \ref{tab:screen-CS} below summarizes the feature screening performance of all methods. Comparing Table \ref{tab:screen-CS}  with Table 1 (settings 1 and 2) in the main text, we see that the problem of interaction screening becomes more difficult in this new setting. This is reasonable and expected because of higher collinearity in models M3$'$ and M4$'$. Nevertheless, IP still improved over other methods in retaining active interaction variables. 

\begin{center}
[Table \ref{tab:screen-CS} about here.]
\end{center}


\section*{Appendix B:  Invariance of sets $\mathcal{A}$, $\mathcal{I}$, and $\mathcal{M}$}

\label{AppB}

\renewcommand{\theequation}{B.\arabic{equation}}
\setcounter{equation}{0}

Consider the linear interaction model
\begin{align*}
Y=\beta_0+\sum_{j=1}^p\beta_jX_j+\sum_{k=1}^{p-1}\sum_{\ell=k+1}^p\gamma_{k\ell}X_{k}X_{\ell}+\veps
\end{align*}
given in (\ref{eq:LM}).
For any $ k, \ell\in \{1, \cdots, p\}$, define $\gamma_{k\ell}^{*}=\gamma_{k\ell}/2$ for $k<\ell$, $\gamma_{k\ell}^{*}=0$ for $k=\ell$, and $\gamma_{k\ell}^{*}=\gamma_{\ell k}/2$ for $k>\ell$. Then $\gamma_{k\ell}^{*}=\gamma_{\ell k}^{*}$ and our model can be rewritten as
   \begin{align*}
         Y=\beta_0+\sum_{j=1}^p\beta_jX_j+\sum_{k, \ell=1}^p\gamma_{k\ell}^{*}X_{k}X_{\ell}+\veps.
    \end{align*}

Under affine transformations 
$X_j^{new}=b_j(X_j-a_j)$ with $a_j \in \mathbb{R}$ and $b_j\in \mathbb{R}\setminus \{0\}$ for $j=1, \cdots, p$, our model becomes
   \begin{align*}
         Y&=\beta_0+\sum_{j=1}^p\beta_j(b_j^{-1}X_j^{new}+a_j)
               +\sum_{k, \ell=1}^p\gamma_{k\ell}^{*}(b_k^{-1}X_k^{new}+a_k)(b_{\ell}^{-1}X^{new}_{\ell}
               +a_{\ell})+\veps\\
          &=(\beta_0+\sum_{j=1}^p\beta_ja_j+\sum_{k, \ell=1}^p\gamma_{k\ell}^{*}a_ka_{\ell})
              +\sum_{j=1}^p(\beta_j+\sum_{\ell=1}^p\gamma_{j\ell}^{*}a_{\ell}
              +\sum_{k=1}^p\gamma_{kj}^{*}a_{k})b_j^{-1}X_j^{new}  \\
          &\quad    +\sum_{k, \ell=1}^p\gamma_{k\ell}^{*}b_k^{-1}b_{\ell}^{-1}X_k^{new}X^{new}_{\ell}
              +\veps\\
          &=\widetilde{\beta}_0+\sum_{j=1}^p\widetilde{\beta}_jX^{new}_j
              +\sum_{k=1}^{p-1}\sum_{\ell=k+1}^p \widetilde{\gamma}_{k\ell}X^{new}_{k}X^{new}_{\ell}+\veps,
    \end{align*}
where
   \begin{align}
    \widetilde{\beta}_0
  &=\beta_0+\sum_{j=1}^p\beta_ja_j+\sum_{k,\ell=1}^p\gamma_{k\ell}^{*}a_ka_{\ell}
    =\beta_0+\sum_{j=1}^p\beta_ja_j+\sum_{1\leq k<\ell\leq p}\gamma_{k\ell}a_ka_{\ell}, \label{eq-resp-A1}\\
    \widetilde{\beta}_j
&=(\beta_j+\sum_{\ell=1}^p\gamma_{j\ell}^{*}a_{\ell}
              +\sum_{k=1}^p\gamma_{kj}^{*}a_{k})b_j^{-1}
 =(\beta_j+\sum_{1\leq k<j}\gamma_{kj}a_{k}
              +\sum_{j< k\leq p}\gamma_{jk}a_{k})b_j^{-1}, \label{eq-resp-A2}\\
    \widetilde{\gamma}_{k\ell} &=\gamma_{k\ell}b_k^{-1}b_{\ell}^{-1}.\label{eq-resp-A3}
    \end{align}
Similar to the definitions of sets $\mathcal{I}$, $\mathcal{A}$, $\mathcal{B}$, and $\mathcal{M}$ in
\eqref{intvarsets}, we define index sets
\begin{align*}
\widetilde{\mathcal{I}} & =\left\{(k, \ell): 1\leq k < \ell\leq p \text{ with } \widetilde{\gamma}_{k\ell}\neq 0\right\}, \nonumber\\
\widetilde{\mathcal{A}}& =\left\{1\leq k\leq p: (k, \ell) \text{ or } (\ell, k) \in \mathcal{I} \text{ for some } \ell\right\},  \\
\widetilde{\mathcal{B}} & =\left\{1\leq j\leq p: \widetilde{\beta}_{j}\neq 0\right\}.\nonumber
\end{align*}
Then from \eqref{eq-resp-A3}, we have $\widetilde{\mathcal{I}}=\mathcal{I}$ and thus $\widetilde{\mathcal{A}}=\mathcal{A}$.

\quad Next we show $\widetilde{\mathcal{M}}=\mathcal{M}$. 
It is equivalent to show that $\widetilde{\mathcal{A}}^c\cap\widetilde{\mathcal{B}}^c=\mathcal{A}^c\cap\mathcal{B}^c$. To this end, we first prove $\mathcal{A}^c\cap\mathcal{B}^c \subset \widetilde{\mathcal{A}}^c\cap\widetilde{\mathcal{B}}^c $. For any $j\in \mathcal{A}^c\cap\mathcal{B}^c$, we have $\beta_j=0$ and $\gamma_{jk}=0$ for all $1\leq k\neq j\leq p$. In view of \eqref{eq-resp-A2} and \eqref{eq-resp-A3}, we have $\widetilde{\beta}_j=0$ and $\widetilde{\gamma}_{jk}=0$, which means $j\in\widetilde{\mathcal{A}}^c\cap\widetilde{\mathcal{B}}^c$. Thus  $\mathcal{A}^c\cap\mathcal{B}^c \subset \widetilde{\mathcal{A}}^c\cap\widetilde{\mathcal{B}}^c $ holds. Similarly, we can also show that $\widetilde{\mathcal{A}}^c\cap\widetilde{\mathcal{B}}^c\subset \mathcal{A}^c\cap\mathcal{B}^c$.  Combining these results yields $\widetilde{\mathcal{A}}^c\cap\widetilde{\mathcal{B}}^c=\mathcal{A}^c\cap\mathcal{B}^c$ and thus 
$\widetilde{\mathcal{M}}=\mathcal{M}$.

Therefore, the three sets $\mathcal{A}$, $\mathcal{I}$, and $\mathcal{M}$ are invariant under affine transformations $X_j^{new}=b_j(X_j-a_j)$ with $a_j \in \mathbb{R}$ and $b_j\in \mathbb{R}\setminus \{0\}$ for $j=1, \cdots, p$.

\section*{Appendix C:  $\cov(X_j^2, Y^2)$ under specific models} \label{AppC}

\renewcommand{\theequation}{C.\arabic{equation}}
\setcounter{equation}{0}

Without loss of generality, we assume that $\beta_0 = 0$ and the $s$ true main effects concentrate at the first $s$ coordinates, that is, $\mathcal{B} = \{1, \cdots, s\}$. Here we slightly abuse the notation $s$ for simplicity. Due to the existence of $O(p^2)$ interaction terms, it is generally too complicated to calculate $\cov(X_j^2, Y^2)$ explicitly. Since our purpose is to illustrate that in the presence of correlation among covariates, using 
$\corr(X_j^2, Y^2)$ as the marginal utility still has differentiation power between interaction variables (i.e., variables contributing to interactions) and noise variables (variables contributing to neither interactions nor main effects), we consider the specific case when there is only one interaction and $\bx=(X_1, \cdots, X_p)^T\sim N(\bzero, \bSig)$ with $\bSig=(\sigma_{k\ell})$ being tridiagonal, that is, $\sigma_{k\ell}=1$ for $k=\ell$, $\sigma_{k\ell}=\rho \in [-1, 1]$ for $|k-\ell|=1$, and $\sigma_{k\ell}=0$ for $|k-\ell|>1$. In addition, assume that all nonzero main effect coefficients take the same value $\beta$, that is, $\beta_{0,1} = \cdots = \beta_{0,s}  = \beta \neq 0 $.

We consider the following three different settings according to whether or not the heredity assumption holds:
\begin{enumerate}
	\item[] Case 1: $\mathcal{A} = \{1,2\}$ -- strong heredity  if $s\geq 2$,
	\item[] Case 2: $\mathcal{A} = \{1,s+1\}$ -- weak heredity,
	\item[] Case 3: $\mathcal{A} = \{s+1,s+2\}$ -- anti-heredity.
\end{enumerate}
Here, in each case, the set of active interaction variables $\mathcal{A}$ is  chosen without loss of generality. For the ease of presentation, denote by $J_1 = \sum_{j=1}^s\beta_{0,j}X_j$ and $J_2=\gamma X_kX_\ell$ with $k,\ell \in \mathcal{A}$ and $k\neq \ell$. Then, $Y = J_1 + J_2 + \veps$ and
\begin{align*}
       \cov(X_j^2, Y^2)
     =\cov(X_j^2, J_1^2)+\cov(X_j^2, J_2^2).
   \end{align*}

Direct calculations yield
\begin{align*}
       \cov(X_j^2, J_1^2)
  =\left\{\begin{array}{ll}
         2\beta^2,  & j=1 \\
         2\beta^2\rho^2,  & j=2\\
         0,        & j\geq 3
       \end{array}
      \right.\quad\mbox{when}\,\, s=1,
   \end{align*}
\begin{align*}
       \cov(X_j^2, J_1^2)
  =\left\{\begin{array}{ll}
         2\beta^2(1+\rho)^2,  & j=1, 2 \\
         2\beta^2\rho^2,  & j=3 \\
         0,        & j\geq 4
       \end{array}
      \right.\quad\mbox{when}\,\, s=2,
   \end{align*}
\begin{align*}
       \cov(X_j^2, J_1^2)
  =\left\{\begin{array}{ll}
         2\beta^2(1+\rho)^2,  & j=1\,\mbox{or}\,s \\
         2\beta^2(1+2\rho)^2,  & 2\leq j \leq s-1\\
         2\beta^2\rho^2,  & j = s+1\\
         0,        & j\geq s+2
       \end{array}
      \right.\quad\mbox{when}\,\, s\geq 3.
   \end{align*}
Next, we deal with $\cov(X_j^2, J_2^2)$. By Isserlis' Theorem, we have
   \begin{align*}
      E(X_j^2X_{k}X_{\ell}X_{k'}X_{\ell'})
    =& \sigma_{jj}\sigma_{k\ell}\sigma_{k'\ell'} + \sigma_{jj}\sigma_{kk'}\sigma_{\ell\ell'}
      +\sigma_{jj}\sigma_{k\ell'}\sigma_{\ell k'}  \\
     &+ \sigma_{jk}\sigma_{j\ell}\sigma_{k'\ell'} + \sigma_{jk}\sigma_{jk'}\sigma_{\ell\ell'}
      +\sigma_{jk}\sigma_{j\ell'}\sigma_{\ell k'}  \\
     & +\sigma_{j\ell}\sigma_{jk}\sigma_{k'\ell'} + \sigma_{j\ell}\sigma_{jk'}\sigma_{k\ell'}
      +\sigma_{j\ell}\sigma_{j\ell'}\sigma_{k k'}  \\
     &+\sigma_{jk'}\sigma_{jk}\sigma_{\ell\ell'} + \sigma_{jk'}\sigma_{j\ell}\sigma_{k\ell'}
      +\sigma_{jk'}\sigma_{j\ell'}\sigma_{k\ell}  \\
     &+\sigma_{j\ell'}\sigma_{jk}\sigma_{\ell k'} + \sigma_{j\ell'}\sigma_{j\ell}\sigma_{kk'}
      +\sigma_{j\ell'}\sigma_{jk'}\sigma_{k\ell}
   \end{align*}
and $ E(X_{k}X_{\ell}X_{k'}X_{\ell'})
    =\sigma_{k\ell}\sigma_{k'\ell'}+\sigma_{kk'}\sigma_{\ell\ell'}+\sigma_{k\ell'}\sigma_{\ell k'}$.
Combining these two results above gives
   \begin{align*}
      & \cov(X_j^2, X_{k}X_{\ell}X_{k'}X_{\ell'})\\
    = &  E(X_j^2X_{k}X_{\ell}X_{k'}X_{\ell'})-E(X_j^2)E(X_{k}X_{\ell}X_{k'}X_{\ell'}) \\
    = & 2(\sigma_{jk}\sigma_{j\ell}\sigma_{k'\ell'} + \sigma_{jk}\sigma_{jk'}\sigma_{\ell\ell'}
      +\sigma_{jk}\sigma_{j\ell'}\sigma_{\ell k'} + \sigma_{j\ell}\sigma_{jk'}\sigma_{k\ell'}
      +\sigma_{j\ell}\sigma_{j\ell'}\sigma_{k k'}  + \sigma_{jk'}\sigma_{j\ell'}\sigma_{k\ell}).
   \end{align*}
Next, we calculate the value of $\cov(X_j^2, J_2^2)$ according to the three different model settings discussed above.

\textbf{\underline{Case 1: $\mathcal{A} = \{1,2\}$}}.
Then  $J_2=\gamma X_1X_2$ and $\cov(X_j^2, J_2^2)
=2\gamma^2(\sigma_{j1}^2\sigma_{22}+4\sigma_{j1}\sigma_{j2}\sigma_{12}+\sigma_{j2}^2\sigma_{11})$. Thus
\begin{align*}
       \cov(X_j^2, J_2^2)
   =  \left\{\begin{array}{ll}
         2\gamma^2(1+5\rho^2),  & j=1\,\,\mbox{or}\,\,2, \\
         2\gamma^2\rho^2,  & j=3,\\
         0,        & j\geq 4.
       \end{array}
      \right.
   \end{align*}
In summary, $\cov(X_1^2, Y^2)>0$ and $\cov(X_2^2, Y^2)>0$ for all $-1\leq \rho\leq 1$, while $\cov(X_j^2, Y^2)=0$ 
 for $j \geq \max\{s+2, 4\}$.

\textbf{\underline{Case 2: $\mathcal{A} = \{1,s+1\}$}}. Then $J_2=\gamma X_1X_{s+1}$ and $\cov(X_j^2, J_2^2)
 = 2\gamma^2(\sigma_{j1}^2\sigma_{s+1, s+1}+4\sigma_{j1}\sigma_{j, s+1}\sigma_{1, s+1}+\sigma_{j, s+1}^2\sigma_{11})$. Thus
\begin{align*}
       \cov(X_j^2, J_2^2)
  = & 2\gamma^2(\sigma_{j1}^2+4\sigma_{j1}\sigma_{j2}\rho+\sigma_{j2}^2)
 =  \left\{\begin{array}{ll}
         2\gamma^2(1+5\rho^2),  & j=1\,\,\mbox{or}\,\, 2 \\
         2\gamma^2\rho^2,  & j=3\\
         0,        & j\geq 4
       \end{array}
      \right.\quad\mbox{when}\,\, s=1,
   \end{align*}
\begin{align*}
       \cov(X_j^2, J_2^2)
 = & 2\gamma^2(\sigma_{j1}^2+\sigma_{j3}^2)
  = \left\{\begin{array}{ll}
         2\gamma^2,  & j=1\,\,\mbox{or}\,\,3 \\
         4\gamma^2\rho^2,  & j=2\\
         2\gamma^2\rho^2,  & j=4\\
         0,        & j\geq 5
       \end{array}
      \right.\quad\mbox{when}\,\, s=2,
   \end{align*}
\begin{align*}
       \cov(X_j^2, J_2^2)
   =2\gamma^2(\sigma_{j1}^2+\sigma_{j, s+1}^2)
  = & \left\{\begin{array}{ll}
         2\gamma^2,  & j=1\,\,\mbox{or}\,\,s+1 \\
         2\gamma^2\rho^2,  & j=2 \,\,\mbox{or}\,\,s \,\,\mbox{or}\,\,s+2\\
         0,        & 3\leq j\leq s-1\,\,\mbox{or}\,\, j\geq s+3
       \end{array}
      \right.\quad\mbox{when}\,\, s\geq 3.
   \end{align*}
So it holds that $\cov(X_j^2, Y^2)=0$ for all $j\geq s+3$, and
$\cov(X_j^2, Y^2)>0$ for 
 $j \in \mathcal{A} = \{1, s+1\}$.

\textbf{\underline{Case 3: $\mathcal{A} = \{s+1,s+2\}$}}.   Then $J_2=\gamma X_sX_{s+1}$ and $\cov(X_j^2, J_2^2)=2\gamma^2(\sigma_{js}^2\sigma_{s+1, s+1}+4\sigma_{js}\sigma_{j, s+1}\sigma_{s, s+1}+\sigma_{j, s+1}^2\sigma_{ss})$.
Thus
\begin{align*}
       \cov(X_j^2, J_2^2)
   =  \left\{\begin{array}{ll}
         2\gamma^2(1+5\rho^2),  & j=s\,\,\mbox{or}\,\,s+1,\\
         2\gamma^2\rho^2,  & j=s-1\,\,\mbox{or}\,\,s+2,\\
         0,        & \mbox{otherwise}.
       \end{array}
      \right.
   \end{align*}
So we have that $\cov(X_j^2, Y^2)=0$ for all $j\geq s+3$, and $\cov(X_j^2, Y^2)>0$ for 
 $j \in \mathcal{A} = \{s+1, s+2\}$.

Therefore, $\cov(X_j^2, Y^2)>0$ for all $j\in \mathcal{A}$, whereas $\cov(X_j^2, Y^2)=0$ for all $j\geq \max\{s+2, 4\}$ for Case 1, and $\cov(X_j^2, Y^2)=0$ for all $j\geq s+3$ for Cases 2 and 3.  Note that  $\corr(X_j^2, Y^2)=\cov(X_j^2, Y^2)/\sqrt{\var(X_j^2)\var(Y^2)}$. This ensures that the correlations between $X_j^2$ and $Y^2$ are nonzero for those active interaction variables. In other words, using $\corr(X_j^2, Y^2)$ as the marginal utility can still single out active interaction variables.

\section*{Appendix D:  Proofs of Proposition \ref{prop1} and Theorems \ref{Th: Sure Screening-new}--\ref{Th:RE condition}} \label{AppD}

\renewcommand{\theequation}{D.\arabic{equation}}
\setcounter{equation}{0}

\subsection*{D.1. Proof of Proposition \ref{prop1}}

Let $J_1=\sum_{j=1}^p\beta_j X_j$ and $J_2=\sum_{k=1}^{p-1}\sum_{\ell=k+1}^p\gamma_{k\ell}X_kX_{\ell}$.
Then our interaction model \eqref{eq:LM} can be written as $Y=\beta_0+J_1+J_2+\varepsilon$.
For each $j\in\{1, \cdots, p\}$,  the covariance between $X_j^2$ and $Y^2$ can be expressed as
  \begin{align}\label{eq: cov-Xsq-Ysq}
     \cov(X_j^2, Y^2) &= \cov(X_j^2,  J_1^2) + \cov(X_j^2,  J_2^2) + \cov(X_j^2, \varepsilon^2) + 2\beta_0\cov(X_j^2, J_1) \nonumber   \\
                              \nonumber  &\quad +2\beta_0\cov(X_j^2, J_2)+2\beta_0\cov(X_j^2, \varepsilon) +2\cov(X_j^2, J_1J_2)\\
                           & \quad  +  2\cov(X_j^2, J_1\varepsilon)  + 2\cov(X_j^2, J_2\varepsilon).
  \end{align}

Recall that $\varepsilon$ is independent of $X_j$. Thus $\cov(X_j^2, \varepsilon^2) =0$ and $\cov(X_j^2, \varepsilon)=0$.  With the assumption of $E(\varepsilon)=0$, we have
    \begin{equation*}
      \cov(X_j^2, J_1\varepsilon)
   =E(X_j^2J_1\varepsilon)-E(X_j^2)E(J_1\varepsilon)
   =E(X_j^2J_1)E(\varepsilon)-E(X_j^2)E(J_1)E(\varepsilon)
   =0.
   \end{equation*}
Similarly, $\cov(X_j^2, J_2\varepsilon)=0$.  Note that $\cov(X_j^2, J_1J_2)=E(X_j^2J_1J_2)-E(X_j^2)E(J_1J_2)$.  Since $X_1, \cdots, X_p$ are i.i.d. $N(0, 1)$, direct calculation yields
$E(X_j^2J_1J_2)=E(J_1J_2)=0$, which leads to $\cov(X_j^2, J_1J_2)=0$.  Similarly, $\cov(X_j^2, J_1)=0$.
Thus, \eqref{eq: cov-Xsq-Ysq} reduces to
    \begin{equation}\label{eq: cov-Xsq-Ysq-short}
      \cov(X_j^2, Y^2) = \cov(X_j^2,  J_1^2) + \cov(X_j^2,  J_2^2)+ 2\beta_0\cov(X_j^2, J_2).
   \end{equation}
It remains to calculate the three terms on the right hand side of \eqref{eq: cov-Xsq-Ysq-short}.

We first consider $\cov(X_j^2, J_1^2)$.  For each fixed $j=1, \cdots, p$, denote by $J_3=\sum_{k\neq j}\beta_kX_k$. Then  $J_1=\beta_jX_j+J_3$ and
$\cov(X_j^2, J_1^2) =\cov(X_j^2, \beta_j^2X_j^2)  + \cov(X_j^2, 2\beta_jX_jJ_3)+\cov(X_j^2, J_3^2)$. Since $X_j$ is independent of $J_3$, it follows that $\cov(X_j^2, J_3^2) =0$.
Note that $\cov(X_j^2, \beta_j^2X_j^2)=\beta_j^2\var(X_j^2)=2\beta_j^2$ and
  \begin{align*}
     & \cov(X_j^2, 2\beta_jX_jJ_3)
     = 2\beta_j[E(X_j^3J_3)-E(X_j^2)E(X_jJ_3)]  \nonumber \\
     =& 2\beta_j[E(X_j^3)E(J_3)-E(X_j^2)E(X_j)E(J_3)] = 0.
  \end{align*}
Therefore, we obtain
    \begin{equation}\label{eq: cov-Xsq-J1sq-final}
         \cov(X_j^2, J_1^2)=2\beta_j^2.
  \end{equation}

Next, we deal with $\cov(X_j^2, J_2^2)$.  For a fixed $j=1, \cdots, p$, let $J_4=\sum_{k=1}^{j-1}\gamma_{kj}X_k+\sum_{\ell=j+1}^{p}\gamma_{j\ell}X_{\ell}$ and $J_5=\sum_{k=1, k\neq j}^{p-1}\sum_{\ell=k+1, \ell\neq j}^{p}\gamma_{k\ell}X_kX_{\ell}$. Then $J_2=J_4X_j+J_5$.  Since $X_j$ is independent of $J_4$ and $J_5$,  we have $\cov(X_j^2, J_5^2)=0$ and
    \begin{equation}\label{eq: cov-Xsq-J2sq}
          \cov(X_j^2, J_2^2)=\cov(X_j^2, J_4^2X_j^2)+\cov(X_j^2, 2J_4X_jJ_5).
  \end{equation}
The first term on the right hand side of \eqref{eq: cov-Xsq-J2sq} can be further calculated as
  \begin{align*}
          \cov(X_j^2, J_4^2X_j^2)
    &= E(X_j^4J_4^2)-E(X_j^2)E(J_4^2X_j^2)
     =E(X_j^4)E(J_4^2)-E(X_j^2)E(J_4^2)E(X_j^2)   \nonumber \\
     &= 2E(J_4^2) =  2\var(J_4) = 2(\sum_{k=1}^{j-1}\gamma_{kj}^2+\sum_{\ell=k+1}^{p}\gamma_{j\ell}^2).
  \end{align*}
The second term on
the right hand side of \eqref{eq: cov-Xsq-J2sq} is
  \begin{align*}
          \cov(X_j^2, 2J_4X_jJ_5)
    &= 2E(X_j^3J_4J_5)-2E(X_j^2)E(J_4X_jJ_5) \nonumber \\
    &= 2E(X_j^3)E(J_4J_5)-2E(X_j^2)E(J_4J_5)E(X_j) = 0,
  \end{align*}
since $E(X_j^3)=E(X_j)=0$.   Therefore, it holds that
    \begin{equation}\label{eq: cov-Xsq-J2sq-final}
          \cov(X_j^2, J_2^2)=2(\sum_{k=1}^{j-1}\gamma_{kj}^2+\sum_{\ell=k+1}^{p}\gamma_{j\ell}^2).
  \end{equation}

Finally, we handle $\cov(X_j^2, J_2)$.  
Recall that $J_2=J_4X_j+J_5$ and $X_j$ is independent of $J_4$ and $J_5$, we have
  \begin{align*}
          \cov(X_j^2, J_2)
     &=\cov(X_j^2, J_4X_j) + \cov(X_j^2, J_5)=E(X_j^3J_4)-E(X_j^2)E(J_4X_j) \nonumber \\
     &=E(X_j^3)E(J_4)-E(X_j^2)E(J_4)E(X_j)=0,
  \end{align*}
which together with \eqref{eq: cov-Xsq-Ysq-short},  \eqref{eq: cov-Xsq-J1sq-final}, and \eqref{eq: cov-Xsq-J2sq-final} completes the proof of Proposition \ref{prop1}.

\subsection*{D.2. Proof of part a) of Theorem \ref{Th: Sure Screening-new}}

Let $S_{k1}=n^{-1}\sum\limits_{i=1}^n X_{ik}^2Y_i^2$, $S_{k2}=n^{-1}\sum\limits_{i=1}^n X_{ik}^2$,  $S_{k3}=n^{-1}\sum\limits_{i=1}^n X_{ik}^4$, and $S_{4}=n^{-1}\sum\limits_{i=1}^nY_i^2$.  Then $\omega_k$ and $\widehat{\omega}_k$ can be written as
  \begin{eqnarray*}
       \omega_k=\frac{E(S_{k1})-E(S_{k2})E(S_{4})}{\sqrt{E(S_{k3})-E^2(S_{k2})}}
       \quad\mbox{and}\quad
       \widehat{\omega}_k=\frac{S_{k1}-S_{k2}S_{4}}{\sqrt{S_{k3}-S_{k2}^2}}.
   \end{eqnarray*}
To prove \eqref{eq: bound-omega-new}, the key step is to show that for any positive constant $C$, there exist some  constants $\widetilde{C}_1, \cdots, \widetilde{C}_4>0$ 
such that the following probability bounds
  \begin{align}
       P(\max_{1\leq k\leq p}|S_{k1}-E(S_{k1})| \geq Cn^{-\kappa_1})
   &\leq    p\widetilde{C}_1\exp\left(-\widetilde{C}_2n^{\alpha_1\eta_1}\right)
       +\widetilde{C}_3\exp\left(-\widetilde{C}_4n^{\alpha_2\eta_1}\right),   \label{eq: Sk1-Prob}\\
      P(\max_{1\leq k\leq p}|S_{k2}-E(S_{k2})| \geq Cn^{-\kappa_1})
      &\leq  p\widetilde{C}_1\exp[-\widetilde{C}_2n^{\alpha_1(1-2\kappa_1)/(4+\alpha_1)}], \label{eq: Sk2-Prob}\\
      P(\max_{1\leq k\leq p}|S_{k3}-E(S_{k3})| \geq Cn^{-\kappa_1})
      &\leq  p\widetilde{C}_1\exp[-\widetilde{C}_2n^{\alpha_1(1-2\kappa_1)/(8+\alpha_1)}], \label{eq: Sk3-Prob}\\
     P(|S_4-E(S_4)|\geq Cn^{-\kappa_1})
    &\leq \widetilde{C}_1\exp\left(-\widetilde{C}_2n^{\alpha_1\zeta_1}\right)
       +\widetilde{C}_3\exp\left(-\widetilde{C}_4n^{\alpha_2\zeta_2'}\right)      \label{eq: S4-Prob}
  \end{align}
 hold for all $n$ sufficiently large when $0\leq 2\kappa_1+4\xi_1<1$ and $0\leq 2\kappa_1+4\xi_2<1$,
where $\eta_1=\min\{(1-2\kappa_1-4\xi_2)/(8+\alpha_1),\,(1-2\kappa_1-4\xi_1)/(12+\alpha_1)\}$,
$\zeta_1=\min\{(1-2\kappa_1-4\xi_2)/(4+\alpha_1),\,(1-2\kappa_1-4\xi_1)/(8+\alpha_1)\}$,  $\zeta_2=\min\{(1-2\kappa_1-2\xi_2)/(4+\alpha_1),\,(1-2\kappa_1-2\xi_1)/(6+\alpha_1)\}$, and 
$\zeta_2'=\min\{\zeta_2, (1-2\kappa_1)/(4+\alpha_2)\}$.
Define  $\eta=\min\{\eta_1, (1-2\kappa_1)/(4+\alpha_1),  (1-2\kappa_1)/(8+\alpha_1), \zeta_1\}$ and
$\zeta=\min\{\eta_1, \zeta_2'\}$. Then $\eta=\eta_1$ and $\zeta=\min\{\eta_1, (1-2\kappa_1)/(4+\alpha_2)\}$.
Thus, by Lemmas \ref{lem: Ajsq}--\ref{lem: AjBj-ratio}, we have
  \begin{align}\label{eq:omega-max}
        P(\max_{1\leq k\leq p}|\widehat{\omega}_k-\omega_k|\geq Cn^{-\kappa_1})
   \leq  p\widetilde{C}_1\exp(-\widetilde{C}_2n^{\alpha_1\eta})
        + \widetilde{C}_3\exp(-\widetilde{C}_4n^{\alpha_2\zeta}).
  \end{align}
Thus, if $\log p=o\{n^{\alpha_1 \eta}\}$, the result of the part (a) in Theorem \ref{Th: Sure Screening-new} follows immediately.

It thus remains to prove the probability bounds \eqref{eq: Sk1-Prob}--\eqref{eq: S4-Prob}.  Since the proofs
of \eqref{eq: Sk1-Prob}--\eqref{eq: S4-Prob} are similar, here we focus on \eqref{eq: Sk1-Prob} to save space.  Throughout the proof, the same notation $\widetilde{C}$ is used to denote a generic positive constant without loss of generality, which may take different values at each appearance.

Recall that $Y_i=\beta_0+\bx_i^T\bbeta_0+\bz_i^T\bgamma_0+\varepsilon_i=\beta_0+\bx_{i,\,\mathcal{B}} ^T\bbeta_{0,\mathcal{B}}+\bz_{i,\,\mathcal{I}}^T\bgamma_{0,\mathcal{I}}+\varepsilon_i$, where
 $\bx_{i}=(X_{i1}, \cdots, X_{ip})^T$, $\bz_{i}=(X_{i1}X_{i2}, \cdots, X_{i, p-1}X_{i, p})^T$,
$\bx_{i,\,\mathcal{B}}=(X_{ij}, j\in \mathcal{B})^T$, $\bz_{i,\,\mathcal{I}}=(X_{ik}X_{i\ell}, (k, \ell)\in \mathcal{I})^T$, 
 $\bbeta_{0,\mathcal{B}}=(\beta_{0, j}\in \mathcal{B})^T$, and $\bgamma_{0,\mathcal{I}}=(\gamma_{0,\, k\ell}, (k,\ell)\in \mathcal{I})^T$. To simplify the presentation,  we assume that the intercept $\beta_0$ is zero without loss of generality. Thus
  \begin{align*}
      S_{k1}
   =& n^{-1}\sum\limits_{i=1}^n X_{ik}^2Y_i^2
   =n^{-1}\sum\limits_{i=1}^n X_{ik}^2(\bx_{i,\,\mathcal{B}} ^T\bbeta_{0,\mathcal{B}}+\bz_{i,\,\mathcal{I}}^T\bgamma_{0,\mathcal{I}}+\varepsilon_i)^2 \\
   =& n^{-1}\sum\limits_{i=1}^n X_{ik}^2(\bx_{i,\,\mathcal{B}} ^T\bbeta_{0,\mathcal{B}}+\bz_{i,\,\mathcal{I}}^T\bgamma_{0,\mathcal{I}})^2 + 2n^{-1}\sum\limits_{i=1}^n X_{ik}^2(\bx_{i,\,\mathcal{B}} ^T\bbeta_{0,\mathcal{B}}+\bz_{i,\,\mathcal{I}}^T\bgamma_{0,\mathcal{I}})\varepsilon_i + n^{-1}\sum\limits_{i=1}^n X_{ik}^2\varepsilon_i^2 \\
  \triangleq &S_{k1, 1} +2S_{k1, 2} +S_{k1, 3}.
   \end{align*}
Similarly, $E(S_{k1})$ can be written as $E(S_{k1})=E(S_{k1, 1}) +2E(S_{k1, 2}) +E(S_{k1, 3})$. So $S_{k1}-E(S_{k1})$ can be expressed as $S_{k1}-E(S_{k1})
     =[S_{k1, 1}-E(S_{k1, 1})] + 2 [S_{k1, 2}-E(S_{k1, 2})] +[S_{k1, 3}-E(S_{k1, 3})]$.
By the triangle inequality and the union bound we have
  \begin{align}\label{eq: Sk1-bound}
      P(\max_{1\leq k\leq p}|S_{k1}-E(S_{k1})|\geq Cn^{-\kappa_1})
   \leq & P(\bigcup_{j=1}^3\{\max_{1\leq k\leq p}|S_{k1, j}-E(S_{k1, j})|\geq Cn^{-\kappa_1}/4\}) \nonumber\\
   \leq & \sum_{j=1}^3P(\max_{1\leq k\leq p}|S_{k1, j}-E(S_{k1, j})|\geq Cn^{-\kappa_1}/4).
   \end{align}
In what follows, we will provide details on deriving an exponential tail probability bound for each term on the right hand side above.  To enhance readability, we split the proof into three steps.

{\bf Step 1.} We start with the first term $\max_{1\leq k\leq p}|S_{k1, 1}-E(S_{k1, 1})|$.
Define the event $\Omega_i=\{|X_{ij}|\leq M_1\,\mbox{for all}\, j\in\mathcal{M}\cup\{k\}\}$ with $\mathcal{M}=\mathcal{A}\cup \mathcal{B}$ and $M_1$ a large positive number that will be specified later.
Let $T_{k1}=n^{-1}\sum\limits_{i=1}^n X_{ik}^2(\bx_{i,\,\mathcal{B}} ^T\bbeta_{0,\mathcal{B}}+\bz_{i,\,\mathcal{I}}^T\bgamma_{0,\mathcal{I}})^2\mathbb{I}_{\Omega_i}$ and $T_{k2}=n^{-1}\sum\limits_{i=1}^n X_{ik}^2(\bx_{i,\,\mathcal{B}} ^T\bbeta_{0,\mathcal{B}}+\bz_{i,\,\mathcal{I}}^T\bgamma_{0,\mathcal{I}})^2\mathbb{I}_{\Omega_i^c}$,
where $\mathbb{I}(\cdot)$ is the indicator function and $\Omega_i^c$ is the complement of
the set $\Omega_i$.  Then
  \begin{align}\label{eq: Sk1-part1-decomp}
     S_{k1, 1}-E(S_{k1, 1})
    = [T_{k1} -E(T_{k1})] + T_{k2} -E(T_{k2}).
   \end{align}
Note that $E(T_{k2})=E[X_{1k}^2(\bx_{1,\,\mathcal{B}} ^T\bbeta_{0,\mathcal{B}}+\bz_{1,\,\mathcal{I}}^T\bgamma_{0,\mathcal{I}})^2\mathbb{I}_{\Omega_1^c}]$.  By the fact $(a+b)^2\leq 2(a^2+b^2)$ for two real numbers $a$ and $b$, the Cauchy-Schwarz inequality, and 
 Condition \ref{con: sparsity}, we have
  \begin{align}\label{eq: Ymean-bound1}
     (\bx_{1,\,\mathcal{B}} ^T\bbeta_{0,\mathcal{B}}+\bz_{1,\,\mathcal{I}}^T\bgamma_{0,\mathcal{I}})^2
  & \leq 2[(\bx_{1,\,\mathcal{B}} ^T\bbeta_{0,\mathcal{B}})^2+(\bz_{1,\,\mathcal{I}}^T\bgamma_{0,\mathcal{I}})^2] \nonumber \\
  &
   \leq 2C_0^2(s_2\|\bx_{1,\,\mathcal{B}}\|^2+s_1\|\bz_{1,\,\mathcal{I}}\|^2),
   \end{align}
 where $C_0$ is some positive constant and $\|\cdot\|$ denotes the Euclidean norm. This ensures that $E(T_{k2})$ is bounded by $2C_0^2[s_2E(X_{1k}^2\|\bx_{1,\,\mathcal{B}}\|^2\mathbb{I}_{\Omega_1^c})+s_1E(X_{1k}^2\|\bz_{1,\,\mathcal{I}}\|^2\mathbb{I}_{\Omega_1^c})]$.
By the Cauchy-Schwarz inequality, the union bound, and the inequality $(a+b)^2\leq 2(a^2+b^2)$, we obtain that
  \begin{align*}
     E(X_{1k}^2\|\bx_{1,\,\mathcal{B}}\|^2\mathbb{I}_{\Omega_1^c})
    \leq & \left[E(X_{1k}^4\|\bx_{1,\,\mathcal{B}}\|^4)P(\Omega_1^c)\right]^{1/2}
    \leq \left\{\left[s_2\sum_{j\in\mathcal{B}}E(X_{1k}^4X_{1j}^4)\right]P(\Omega_1^c)\right\}^{1/2} \\
    \leq & \left\{2^{-1}s_2\sum_{j\in\mathcal{B}}[E(X_{1k}^8)+E(X_{1j}^8)]\right\}^{1/2}\left[\sum_{j\in \mathcal{M}\cup\{k\}}P(|X_{ij}|>M_1)\right]^{1/2} \\
     \leq &\widetilde{C}s_2(1+s_2+2s_1)^{1/2}\exp[-M_1^{\alpha_1}/(2c_1)]
   \end{align*}
for some positive constant $\widetilde{C}$, where the last inequality follows from Condition \ref{con: XYtail-new} and Lemma  \ref{lem: Sub-exp-bound}. Similarly, we have
$E(X_{1k}^2\|\bz_{1,\,\mathcal{I}}\|^2\mathbb{I}_{\Omega_1^c})\leq\widetilde{C}s_1(1+s_2+2s_1)^{1/2}\exp[-M_1^{\alpha_1}/(2c_1)]$.  
This together with the above inequalities entails that
  \begin{align*}
     0\leq E(T_{k2})
  \leq 2C_0^2\widetilde{C}(s_1^2+s_2^2)(1+s_2+2s_1)^{1/2}\exp[-M_1^{\alpha_1}/(2c_1)].
   \end{align*}
If we choose $M_1=n^{\eta_1}$ with $\eta_1>0$, then by Condition \ref{con: sparsity}, for any positive constant $C$, when $n$ is sufficiently large,
   \begin{eqnarray}\label{eq: TK2-expectation}
      |E(T_{k2})|
  \leq 2C_0^2\widetilde{C}(n^{2\xi_1}+n^{2\xi_2})(1+n^{\xi_2}+2n^{\xi_1})^{1/2}\exp[-n^{\alpha_1\eta_1}/(2c_1)]<Cn^{-\kappa_1}/12
   \end{eqnarray}
holds uniformly for all $1\leq k\leq p$.  The above inequality together with \eqref{eq: Sk1-part1-decomp} ensures that
  \begin{align}\label{eq: Sk1-part1-prob}
          & P(\max_{1\leq k\leq p}|S_{k1,1}-E(S_{k1, 1})|\geq Cn^{-\kappa_1}/4) \nonumber\\
   \leq & P(\max_{1\leq k\leq p}|T_{k1}-E(T_{k1})|\geq Cn^{-\kappa_1}/12)
          + P(\max_{1\leq k\leq p}|T_{k2}|\geq Cn^{-\kappa_1}/12)
   \end{align}
for all $n$ sufficiently large. Thus we only need to establish the probability bound for each term on the right hand side of \eqref{eq: Sk1-part1-prob}.

First consider $\max_{1\leq k\leq p}|T_{k1}-E(T_{k1})|$.  
 Using similar arguments for proving \eqref{eq: Ymean-bound1}, we have $(\bx_{i,\,\mathcal{B}} ^T\bbeta_{0,\mathcal{B}}+\bz_{i,\,\mathcal{I}}^T\bgamma_{0,\mathcal{I}})^2
   \leq 2C_0^2(s_2\|\bx_{i,\,\mathcal{B}}\|^2+s_1\|\bz_{i,\,\mathcal{I}}\|^2)$ and thus
  \begin{align*}
      0\leq X_{ik}^2(\bx_{i,\,\mathcal{B}} ^T\bbeta_{0,\mathcal{B}}+\bz_{i,\,\mathcal{I}}^T\bgamma_{0,\mathcal{I}})^2\mathbb{I}_{\Omega_i}
   \leq 2C_0^2X_{ik}^2(s_2\|\bx_{i,\,\mathcal{B}}\|^2+s_1\|\bz_{i,\,\mathcal{I}}\|^2)\mathbb{I}_{\Omega_i}
\leq 2C_0^2M_1^4(s_2^2+s_1^2M_1^2).
   \end{align*}
For any $\delta>0$, by Hoeffding's inequality \citep{hoeffding1963probability}, we obtain
  \begin{align*}
       P(|T_{k1}-E(T_{k1})|\geq \delta)
     \leq & 2\exp\left[-\frac{n\delta^2}{2C_0^4M_1^8(s_2^2+s_1^2M_1^2)^2}\right]
     \leq 2\exp\left[-\frac{n\delta^2}{4C_0^4M_1^8(s_2^4+s_1^4M_1^4)}\right] \\
     \leq & 2\exp\left(-\frac{n\delta^2}{8C_0^4M_1^8s_2^4}\right) + 2\exp\left(-\frac{n\delta^2}{8C_0^4M_1^{12}s_1^4}\right),
   \end{align*}
where we have used the fact that $(a+b)^2\leq 2(a^2+b^2)$ for any real numbers $a$ and $b$, and $\exp[-c/(a+b)]\leq \exp[-c/(2a)]+\exp[-c/(2b)]$ for any $a, b, c>0$.  Recall that $M_1=n^{\eta_1}$. Under Condition \ref{con: sparsity}, taking $\delta=Cn^{-\kappa_1}/12$ gives that
  \begin{align}\label{eq: Tk1-prob}
       & P(\max_{1\leq k\leq p}|T_{k1}-E(T_{k1})|\geq Cn^{-\kappa_1}/12)
     \leq  \sum_{k=1}^pP(|T_{k1}-E(T_{k1})|\geq Cn^{-\kappa_1}/12) \nonumber\\
     \leq & 2p\exp\left(-\widetilde{C}n^{1-2\kappa_1-8\eta_1-4\xi_2}\right) + 2p\exp\left(-\widetilde{C}n^{1-2\kappa_1-12\eta_1-4\xi_1}\right).
   \end{align}

Next, consider $\max_{1\leq k\leq p}|T_{k2}|$. Recall that $T_{k2}=n^{-1}\sum\limits_{i=1}^n X_{ik}^2(\bx_{i,\,\mathcal{B}} ^T\bbeta_{0,\mathcal{B}}+\bz_{i,\,\mathcal{I}}^T\bgamma_{0,\mathcal{I}})^2\mathbb{I}_{\Omega_i^c}\geq 0$. By Markov's inequality, for any $\delta>0$, we have $ P(|T_{k2}|\geq \delta)\leq \delta^{-1}E(|T_{k2}|)=\delta^{-1}E(T_{k2})$. In view of the first inequality in \eqref{eq: TK2-expectation}, taking $\delta=Cn^{-\kappa_1}/12$ leads to
   \begin{align*}
      P(|T_{k2}|\geq Cn^{-\kappa_1}/12)
 \leq & 24C^{-1}C_0^2\widetilde{C}n^{\kappa_1}(n^{2\xi_1}+n^{2\xi_2})(1+n^{\xi_2}+2n^{\xi_1})^{1/2}\exp[-n^{\alpha_1\eta_1}/(2c_1)]
   \end{align*}
for all $1\leq k\leq p$. Therefore,
   \begin{align}\label{eq: Tk2-prob}
      & P(\max_{1\leq k\leq p}|T_{k2}|\geq Cn^{-\kappa_1}/12)
  \leq \sum_{k=1}^pP(|T_{k2}|\geq Cn^{-\kappa_1}/12) \nonumber\\
 \leq & 24pC^{-1}C_0^2\widetilde{C}n^{\kappa_1}(n^{2\xi_1}+n^{2\xi_2})(1+n^{\xi_2}+2n^{\xi_1})^{1/2}\exp[-n^{\alpha_1\eta_1}/(2c_1)].
   \end{align}

Combining \eqref{eq: Sk1-part1-prob}, \eqref{eq: Tk1-prob}, and \eqref{eq: Tk2-prob} yields that for sufficiently large $n$,
   \begin{align}\label{eq: Sk1-part1-bound}
      & P(\max_{1\leq k\leq p}|S_{k1,1}-E(S_{k1, 1})|\geq Cn^{-\kappa_1}/4) \nonumber\\
     \leq  & 2p\exp\left(-\widetilde{C}n^{1-2\kappa_1-8\eta_1-4\xi_2}\right)  + 2p\exp\left(-\widetilde{C}n^{1-2\kappa_1-12\eta_1-4\xi_1}\right) \nonumber\\
     &\quad \quad+ 24pC^{-1}C_0^2\widetilde{C}n^{\kappa_1}(n^{2\xi_1}+n^{2\xi_2})(1+n^{\xi_2}+2n^{\xi_1})^{1/2}\exp[-n^{\alpha_1\eta_1}/(2c_1)].
   \end{align}
To balance the three terms on the right hand side of \eqref{eq: Sk1-part1-bound}, we choose $\eta_1=\min\{(1-2\kappa_1-4\xi_2)/(8+\alpha_1),\,(1-2\kappa_1-4\xi_1)/(12+\alpha_1)\}>0$ and the probability bound
\eqref{eq: Sk1-part1-bound} becomes
   \begin{align}\label{eq: Sk1-part1-final}
      P(\max_{1\leq k\leq p}|S_{k1,1}-E(S_{k1, 1})|\geq Cn^{-\kappa_1}/4)
     \leq p\widetilde{C}_5\exp\left(-\widetilde{C}_6n^{\alpha_1\eta_1}\right)
   \end{align}
 for all $n$ sufficiently large, where $\widetilde{C}_5$ and $\widetilde{C}_6$ are two positive constants.

{\bf Step 2.} We establish the probability bound for $\max_{1\leq k\leq p}|S_{k1, 2}-E(S_{k1, 2})|$.  Define the event $\Psi_i=\{|X_{ij}|\leq M_2\,\mbox{for all}\, j\in\mathcal{M}\cup\{k\}\}$ with $\mathcal{M}=\mathcal{A}\cup \mathcal{B}$ and let
   \begin{align*}
     T_{k3} &=n^{-1}\sum\limits_{i=1}^n X_{ik}^2(\bx_{i,\,\mathcal{B}} ^T\bbeta_{0,\mathcal{B}}+\bz_{i,\,\mathcal{I}}^T\bgamma_{0,\mathcal{I}})\varepsilon_i\mathbb{I}_{\Psi_i}\mathbb{I}(|\varepsilon_i|\leq M_3), \\
     T_{k4} &=n^{-1}\sum\limits_{i=1}^n X_{ik}^2(\bx_{i,\,\mathcal{B}} ^T\bbeta_{0,\mathcal{B}}+\bz_{i,\,\mathcal{I}}^T\bgamma_{0,\mathcal{I}})\varepsilon_i\mathbb{I}_{\Psi_i}\mathbb{I}(|\varepsilon_i|> M_3),\\
     T_{k5}&=n^{-1}\sum\limits_{i=1}^n X_{ik}^2(\bx_{i,\,\mathcal{B}} ^T\bbeta_{0,\mathcal{B}}+\bz_{i,\,\mathcal{I}}^T\bgamma_{0,\mathcal{I}})\varepsilon_i\mathbb{I}_{\Psi_i^c},
   \end{align*}
where $M_2$ and $M_3$ are two large positive numbers which will be specified later.  Then $S_{k1, 2}=T_{k3}+T_{k4}+T_{k5}$. Similarly,  $E(S_{k1, 2})$ can be written as $E(S_{k1, 2})=E(T_{k3})+E(T_{k4})+E(T_{k5})$. 
Since $\varepsilon_1$ has mean zero and is independent of 
 $X_{1, 1}, \cdots, X_{1, p}$, we have
$E(T_{k5})=E[X_{1k}^2(\bx_{1,\,\mathcal{B}} ^T\bbeta_{0,\mathcal{B}}+\bz_{1,\,\mathcal{I}}^T\bgamma_{0,\mathcal{I}})\varepsilon_1\mathbb{I}_{\Psi_1^c}]=E[X_{1k}^2(\bx_{1,\,\mathcal{B}} ^T\bbeta_{0,\mathcal{B}}+\bz_{1,\,\mathcal{I}}^T\bgamma_{0,\mathcal{I}})\mathbb{I}_{\Psi_1^c}]E(\varepsilon_1)=0$.
Thus $S_{k1, 2}-E(S_{k1, 2})$ can be expressed as
  \begin{align}\label{eq: Sk1-part2-decomp}
     S_{k1, 2}-E(S_{k1, 2})
    = [T_{k3} -E(T_{k3})] + T_{k4} + T_{k5}-E(T_{k4}).
   \end{align}

Note that $E(T_{k4})=E[X_{1k}^2(\bx_{1,\,\mathcal{B}} ^T\bbeta_{0,\mathcal{B}}+\bz_{1,\,\mathcal{I}}^T\bgamma_{0,\mathcal{I}})\varepsilon_1\mathbb{I}_{\Psi_1}\mathbb{I}(|\varepsilon_1|> M_3)]$. Thus
  \begin{align*}
     |E(T_{k4})|
 \leq  E[X_{1k}^2|\bx_{1,\,\mathcal{B}} ^T\bbeta_{0,\mathcal{B}}+\bz_{1,\,\mathcal{I}}^T\bgamma_{0,\mathcal{I}}|\mathbb{I}_{\Psi_1}|\varepsilon_1|\mathbb{I}(|\varepsilon_1|> M_3)].
   \end{align*}
It follows from the triangle inequality and 
 Condition \ref{con: sparsity} that
  \begin{align}\label{eq:main-inter-bound}
     X_{1k}^2|\bx_{1,\,\mathcal{B}} ^T\bbeta_{0,\mathcal{B}}+\bz_{1,\,\mathcal{I}}^T\bgamma_{0,\mathcal{I}}|\mathbb{I}_{\Psi_1}
    & \leq X_{1k}^2(|\bx_{1,\,\mathcal{B}} ^T\bbeta_{0,\mathcal{B}}|+|\bz_{1,\,\mathcal{I}}^T\bgamma_{0,\mathcal{I}}|)\mathbb{I}_{\Psi_1} \nonumber \\
   & \leq C_0M_2^3(s_2+s_1M_2)
   \end{align}
 for all $1\leq k\leq p$ and some positive constant $C_0$. By the Cauchy-Schwarz inequality, Condition \ref{con: XYtail-new}, and Lemma \ref{lem: Sub-exp-bound}, we have
  \begin{align}\label{eq: error-tail-M3}
      E[|\varepsilon_1|\mathbb{I}(|\varepsilon_1|> M_3)]
   \leq  [E(\varepsilon_1^2)P(|\varepsilon_1|> M_3)]^{1/2}
   \leq \widetilde{C}\exp[-M_3^{\alpha_2}/(2c_1)].
  \end{align}
This together with the above inequalities entails that
  \begin{align*}
  |E(T_{k4})|
 \leq C_0M_2^3(s_2+s_1M_2)E[|\varepsilon_1|\mathbb{I}(|\varepsilon_1|> M_3)]
\leq  C_0\widetilde{C}M_2^3(s_2+s_1M_2)\exp[-M_3^{\alpha_2}/(2c_1)].
   \end{align*}
If we choose $M_2=n^{\eta_2}$ and $M_3=n^{\eta_3}$ with $\eta_2>0$ and $\eta_3>0$, then under Condition \ref{con: sparsity}, for any positive constant $C$, when $n$ is sufficiently large,
   \begin{eqnarray*}
      |E(T_{k4})|
 \leq C_0\widetilde{C}n^{3\eta_2}(n^{\xi_2}+ n^{\xi_1+\eta_2})\exp[-n^{\alpha_2\eta_3}/(2c_1)]
  \leq Cn^{-\kappa_1}/16
   \end{eqnarray*}
holds uniformly for all $1\leq k\leq p$.  This together with \eqref{eq: Sk1-part2-decomp} ensures that
  \begin{align}\label{eq: Sk1-part2-prob}
          & P(\max_{1\leq k\leq p}|S_{k1,2}-E(S_{k1, 2})|\geq Cn^{-\kappa_1}/4)
   \leq  P(\max_{1\leq k\leq p}|T_{k3}-E(T_{k3})|\geq Cn^{-\kappa_1}/16)  \nonumber\\
         & \quad\quad + P(\max_{1\leq k\leq p}|T_{k4}|\geq Cn^{-\kappa_1}/16)
          + P(\max_{1\leq k\leq p}|T_{k5}|\geq Cn^{-\kappa_1}/16)
   \end{align}
for all $n$ sufficiently large. In what follows, we will provide details on establishing the probability bound for each term on the right hand side of
\eqref{eq: Sk1-part2-prob}.

First consider $\max_{1\leq k\leq p}|T_{k3}-E(T_{k3})|$. In view of \eqref{eq:main-inter-bound}, we have 
$|X_{ik}^2(\bx_{i,\,\mathcal{B}} ^T\bbeta_{0,\mathcal{B}}+\bz_{i,\,\mathcal{I}}^T\bgamma_{0,\mathcal{I}})\varepsilon_i\mathbb{I}_{\Psi_i}\mathbb{I}(|\varepsilon_i|\leq M_3)|\leq C_0M_2^3M_3(s_2+s_1M_2)$.
For any $\delta>0$, by Hoeffding's inequality \citep{hoeffding1963probability}, it holds that
  \begin{align*}
       P(|T_{k3}-E(T_{k3})|\geq \delta)
     \leq & 2\exp\left[-\frac{n\delta^2}{2C_0^2M_2^6M_3^2(s_2+s_1M_2)^2}\right]
     \leq  2\exp\left[-\frac{n\delta^2}{4C_0^2M_2^6M_3^2(s_2^2+s_1^2M_2^2)}\right]\\
     \leq & 2\exp\left(-\frac{n\delta^2}{8C_0^2M_2^6M_3^2s_2^2}\right) + 2\exp\left(-\frac{n\delta^2}{8C_0^2M_2^{8}M_3^2s_1^2}\right),
   \end{align*}
where we have used the fact that $\exp[-c/(a+b)]\leq \exp[-c/(2a)]+\exp[-c/(2b)]$ for any $a, b, c>0$.  Recall that $M_2=n^{\eta_2}$ and $M_3=n^{\eta_3}$. Thus, taking $\delta=Cn^{-\kappa_1}/16$ gives
  \begin{align}\label{eq: Tk3-prob}
      & P(\max_{1\leq k\leq p}|T_{k3}-E(T_{k3})|\geq Cn^{-\kappa_1}/16)
      \leq  \sum_{k=1}^pP(|T_{k3}-E(T_{k3})|\geq Cn^{-\kappa_1}/16) \nonumber\\
     \leq & 2p\exp\left(-\widetilde{C}n^{1-2\kappa_1-6\eta_2-2\eta_3-2\xi_2}\right) + 2p\exp\left(-\widetilde{C}n^{1-2\kappa_1-8\eta_2-2\eta_3-2\xi_1}\right).
   \end{align}

Next we handle $\max_{1\leq k\leq p}|T_{k4}|$.  Using similar arguments as for proving \eqref{eq:main-inter-bound}, we have  
 $X_{ik}^2|\bx_{i,\,\mathcal{B}} ^T\bbeta_{0,\mathcal{B}}+\bz_{i,\,\mathcal{I}}^T\bgamma_{0,\mathcal{I}}|\mathbb{I}_{\Psi_i}\leq C_0M_2^3(s_2+s_1M_2)$ for all $1\leq i\leq n$ and $1\leq k\leq p$
and thus 
   \begin{align*}
      \max_{1\leq k\leq p}|T_{k4}|\leq C_0M_2^3(s_2+ s_1M_2)n^{-1}\sum\limits_{i=1}^n |\varepsilon_i|\mathbb{I}(|\varepsilon_i|> M_3).
   \end{align*}
It follows from Markov's inequality and \eqref{eq: error-tail-M3} that
   \begin{align*}
      P(\max_{1\leq k\leq p}|T_{k4}|\geq \delta)
   \leq & P\left\{C_0M_2^3(s_2+ s_1M_2)n^{-1}\sum\limits_{i=1}^n |\varepsilon_i|\mathbb{I}(|\varepsilon_i|> M_3)\geq \delta\right\} \\
  \leq & \delta^{-1}E\left[C_0M_2^3(s_2+ s_1M_2)n^{-1}\sum\limits_{i=1}^n |\varepsilon_i|\mathbb{I}(|\varepsilon_i|> M_3)\right] \\
  =&  \delta^{-1}C_0M_2^3(s_2+ s_1M_2)E[|\varepsilon_1|\mathbb{I}(|\varepsilon_1|> M_3)] \\
  \leq &  \delta^{-1}C_0\widetilde{C}M_2^3(s_2+ s_1M_2)\exp[-M_3^{\alpha_2}/(2c_1)].
   \end{align*}
Recall that $M_2=n^{\eta_2}$ and $M_3=n^{\eta_3}$. Thus, taking $\delta=Cn^{-\kappa_1}/16$ results in
   \begin{align}\label{eq: Tk4-prob}
      & P(\max_{1\leq k\leq p}|T_{k4}|\geq Cn^{-\kappa_1}/16) \nonumber\\
 \leq & 16C^{-1}C_0\widetilde{C}n^{3\eta_2+\kappa_1}(n^{\xi_2}+n^{\xi_1+\eta_2})\exp[-n^{\alpha_2\eta_3}/(2c_1)].
   \end{align}

We next consider $\max_{1\leq k\leq p}|T_{k5}|$. Since $|T_{k5}|\leq n^{-1}\sum\limits_{i=1}^n X_{ik}^2|(\bx_{i,\,\mathcal{B}} ^T\bbeta_{0,\mathcal{B}}+\bz_{i,\,\mathcal{I}}^T\bgamma_{0,\mathcal{I}})\varepsilon_i|\mathbb{I}_{\Psi_i^c}$, by Markov's inequality we have
   \begin{align*}
      P(|T_{k5}|\geq \delta)
   \leq & P\left\{n^{-1}\sum\limits_{i=1}^n X_{ik}^2|(\bx_{i,\,\mathcal{B}} ^T\bbeta_{0,\mathcal{B}}+\bz_{i,\,\mathcal{I}}^T\bgamma_{0,\mathcal{I}})\varepsilon_i|\mathbb{I}_{\Psi_i^c}\geq \delta\right\} \\
  \leq & \delta^{-1}E\left[n^{-1}\sum\limits_{i=1}^n X_{ik}^2|(\bx_{i,\,\mathcal{B}} ^T\bbeta_{0,\mathcal{B}}+\bz_{i,\,\mathcal{I}}^T\bgamma_{0,\mathcal{I}})\varepsilon_i|\mathbb{I}_{\Psi_i^c}\right] \\
  =&  \delta^{-1}E[X_{1k}^2|(\bx_{1,\,\mathcal{B}} ^T\bbeta_{0,\mathcal{B}}+\bz_{1,\,\mathcal{I}}^T\bgamma_{0,\mathcal{I}})\varepsilon_1|\mathbb{I}_{\Psi_1^c}].
   \end{align*}
It follows from the Cauchy-Schwarz inequality and \eqref{eq: Ymean-bound1} that
   \begin{align*}
      & E[X_{1k}^2|(\bx_{1,\,\mathcal{B}} ^T\bbeta_{0,\mathcal{B}}+\bz_{1,\,\mathcal{I}}^T\bgamma_{0,\mathcal{I}})\varepsilon_1|\mathbb{I}_{\Psi_i^c}]
  \leq  \{E[X_{1k}^4(\bx_{1,\,\mathcal{B}} ^T\bbeta_{0,\mathcal{B}}+\bz_{1,\,\mathcal{I}}^T\bgamma_{0,\mathcal{I}})^2\varepsilon_1^2]P(\Psi_1^c)\}^{1/2} \\
  \leq &  \{2C_0^2\left[s_2E(X_{1k}^4\|\bx_{1,\,\mathcal{B}}\|^2\varepsilon_1^2)+s_1E(X_{1k}^4\|\bz_{1,\,\mathcal{I}}\|^2\varepsilon_1^2)\right]P(\Psi_1^c)\}^{1/2}.
   \end{align*}
Applying the Cauchy-Schwarz inequality again gives
   \begin{align*}
      E(X_{1k}^4\|\bx_{1,\,\mathcal{B}}\|^2\varepsilon_1^2)
    \leq &   \left[E(X_{1k}^8\|\bx_{1,\,\mathcal{B}}\|^4) E(\varepsilon_1^4)\right]^{1/2}
    \leq \left[s_2\sum_{j\in\mathcal{B}}E(X_{1k}^8X_{1j}^4)\right]^{1/2}\left[E(\varepsilon_1^4)\right]^{1/2} \\
    \leq & \left\{2^{-1}s_2\sum_{j\in\mathcal{B}}[E(X_{1k}^{16})+E(X_{1j}^8)]\right\}^{1/2}\left[E(\varepsilon_1^4)\right]^{1/2}
     \leq \widetilde{C}s_2,
   \end{align*}
where the last inequality follows from Condition \ref{con: XYtail-new} and Lemma \ref{lem: Sub-exp-bound}.  Similarly, we can show that $E(X_{1k}^4\|\bz_{1,\,\mathcal{I}}\|^2\varepsilon_1^2)\leq \widetilde{C}s_1$.  By Condition \ref{con: XYtail-new} and the union bound, we deduce
$P(\Psi_1^c)= P(|X_{ij}|> M_2\,\mbox{for some}\, j\in\mathcal{M}\cup\{k\})\leq (1+2s_1+s_2)c_1{\exp(-M_2^{\alpha_1}/c_1)}$.
This together with the above inequalities entails that
   \begin{align*}
      P(|T_{k5}|\geq \delta)
   \leq \delta^{-1}\{2C_0^2\widetilde{C}(s_1^2+s_2^2)(1+2s_1+s_2)c_1\exp(-M_2^{\alpha_1}/c_1)\}^{1/2}.
   \end{align*}
Recall that $M_2=n^{\eta_2}$. Under Condition \ref{con: sparsity}, taking $\delta=Cn^{-\kappa_1}/16$ yields
   \begin{align}\label{eq: Tk5-prob}
      & P(\max_{1\leq k\leq p}|T_{k5}|\geq Cn^{-\kappa_1}/16)
     \leq  \sum_{k=1}^p P(|T_{k5}|\geq Cn^{-\kappa_1}/16) \nonumber\\
 \leq & 16pC^{-1}n^{\kappa_1}\{2C_0^2\widetilde{C}c_1(n^{2\xi_1}+n^{2\xi_2})(1+2n^{\xi_1}+n^{\xi_2})\}^{1/2}\exp[-n^{\alpha_1\eta_2}/(2c_1)].
   \end{align}

Combining \eqref{eq: Sk1-part2-prob}, \eqref{eq: Tk3-prob}, \eqref{eq: Tk4-prob}, and \eqref{eq: Tk5-prob} yields that for sufficiently large $n$,
   \begin{align}\label{eq: Sk1-part2-bound}
      & P(\max_{1\leq k\leq p}|S_{k1,2}-E(S_{k1, 2})|\geq Cn^{-\kappa_1}/4) \nonumber\\
     \leq  & 2p\exp\left(-\widetilde{C}n^{1-2\kappa_1-6\eta_2-2\eta_3-2\xi_2}\right) + 2p\exp\left(-\widetilde{C}n^{1-2\kappa_1-8\eta_2-2\eta_3-2\xi_1}\right)\nonumber\\
    & + 16pC^{-1}n^{\kappa_1}\{2C_0^2\widetilde{C}c_1(n^{2\xi_1}+n^{2\xi_2})(1+2n^{\xi_1}+n^{\xi_2})\}^{1/2}\exp[-n^{\alpha_1\eta_2}/(2c_1)] \nonumber\\
    & + 16C^{-1}C_0\widetilde{C}n^{3\eta_2+\kappa_1}(n^{\xi_2}+n^{\xi_1+\eta_2})\exp[-n^{\alpha_2\eta_3}/(2c_1)].
   \end{align}
Let $\eta_2=\eta_3=\min\{(1-2\kappa_1-2\xi_2)/(8+\alpha_1), (1-2\kappa_1-2\xi_1)/(10+\alpha_1)\}$. Then \eqref{eq: Sk1-part2-bound} becomes
   \begin{align}\label{eq: Sk1-part2-final}
      & P(\max_{1\leq k\leq p}|S_{k1,2}-E(S_{k1, 2})|\geq Cn^{-\kappa_1}/4) \nonumber\\
     & \leq   p\widetilde{C}_7\exp\left(-\widetilde{C}_8n^{\alpha_1\eta_2}\right)
    + \widetilde{C}_9\exp[-\widetilde{C}_{10}n^{\alpha_2\eta_2}].
   \end{align}
 for all $n$ sufficiently large, where $\widetilde{C}_7$, $\widetilde{C}_8$, $\widetilde{C}_9$, and $\widetilde{C}_{10}$ are some positive constants.

%

{\bf Step 3.} We establish the probability bound for $\max_{1\leq k\leq p}|S_{k1, 3}-E(S_{k1, 3})|$.
Define
  \begin{align*}
      & T_{k6}=n^{-1}\sum_{i=1}^n X_{ik}^2\varepsilon_i ^2\mathbb{I}(|X_{ik}|\leq M_4)\mathbb{I}(|\varepsilon_i|\leq M_5), \\
     & T_{k7}=n^{-1}\sum_{i=1}^n X_{ik}^2\varepsilon_i ^2\mathbb{I}(|X_{ik}|\leq M_4)\mathbb{I}(|\varepsilon_i|> M_5), \\
     & T_{k8}=n^{-1}\sum_{i=1}^n X_{ik}^2\varepsilon_i ^2\mathbb{I}(|X_{ik}|> M_4),
   \end{align*}
where $M_4$ and $M_5$ are two large positive numbers whose values will be specified later.  Then $S_{k1, 3}=T_{k6}+T_{k7}+T_{k8}$.  Similarly,  $E(S_{k1, 3})$ can be written as
$E(S_{k1, 3})=E(T_{k6})+E(T_{k7})+E(T_{k8})$ with $E(T_{k6})=E[X_{1k}^2\varepsilon_1^2\mathbb{I}(|X_{1k}|\leq M_4)\mathbb{I}(|\varepsilon_1|\leq M_5)]$, $E(T_{k7})=E[X_{1k}^2\varepsilon_1^2\mathbb{I}(|X_{1k}|\leq M_4)\mathbb{I}(|\varepsilon_1|> M_5)]$, and $E(T_{k8})=E[X_{1k}^2\varepsilon_1^2\mathbb{I}(|X_{1k}|> M_4)]$.  Thus
$S_{k1, 3}-E(S_{k1, 3})$ can be expressed as
  \begin{equation}\label{eq: Sk1-part3-decomposition}
       S_{k1, 3}-E(S_{k1, 3})
     =[T_{k6}-E(T_{k6})] + T_{k7} + T_{k8} - [ E(T_{k7})+ E(T_{k8})].
   \end{equation}

First consider the last two terms $E(T_{k7})$ and $E(T_{k8})$.  It follows from $0\leq X_{1k}^2\varepsilon_1^2\mathbb{I}(|X_{1k}|\leq M_4)\mathbb{I}(|\varepsilon_1|> M_5)\leq M_4^2\varepsilon_1^2\mathbb{I}(|\varepsilon_1|>M_5)$ that
  \begin{equation}\label{eq: Tk7-expectation}
      0\leq E(T_{k7})
       \leq  M_4^2E[\varepsilon_1^2\mathbb{I}(|\varepsilon_1|>M_5)].
   \end{equation}
An application of the Cauchy-Schwarz inequality leads to  $E[\varepsilon_1^2\mathbb{I}(|\varepsilon_1|>M_5)]\leq [E(\varepsilon_1^4)P(|\varepsilon_1|> M_5)]^{1/2}$. By Condition \ref{con: XYtail-new} and Lemma \ref{lem: Sub-exp-bound}, we have
  \begin{equation}\label{eq: error-tail-M5}
      E[\varepsilon_1^2\mathbb{I}(|\varepsilon_1|>M_5)]
     \leq \{E(\varepsilon_1^4)c_1\}^{1/2}\exp(-c_1^{-1}M_5^{\alpha_2}/2)
     \leq\widetilde{C}\exp[-M_5^{\alpha_2}/(2c_1)]
   \end{equation}
Combining \eqref{eq: Tk7-expectation} with \eqref{eq: error-tail-M5} yields
  \begin{equation}\label{eq: Tk7-expectation-bound}
       |E(T_{k7})|
   \leq \widetilde{C}M_4^2\exp[-M_5^{\alpha_2}/(2c_1)].
   \end{equation}
Similarly, by the Cauchy-Schwarz inequality  and Lemma \ref{lem: Sub-exp-bound} we obtain
  \begin{align}\label{eq: Tk8-expectation-bound}
       |E(T_{k8})|
     & = E[X_{1k}^2\varepsilon_1^2\mathbb{I}(|X_{1k}|> M_4)]
      \leq \{E(X_{1k}^4\varepsilon_1^4)P(|X_{1k}|> M_4)]\}^{1/2}\nonumber\\
      &\leq \big\{\frac{c_1}{2 }[E(X_{1k}^8)+E(\varepsilon_1^8)]\big\}^{1/2}\exp[-M_4^{\alpha_1}/(2c_1)]
       \leq \widetilde{C}\exp[-M_4^{\alpha_1}/(2c_1)].
   \end{align}
Combining \eqref{eq: Tk7-expectation-bound} and
\eqref{eq: Tk8-expectation-bound} results in
  \begin{equation*}
      |E(T_{k7})+ E(T_{k8})|
    \leq \widetilde{C}M_4^2\exp[-M_5^{\alpha_2}/(2c_1)]+\widetilde{C}\exp[-M_4^{\alpha_1}/(2c_1)].
   \end{equation*}
If we choose $M_4=n^{\eta_4}$ and $M_5=n^{\eta_5}$ with $\eta_4>0$ and $\eta_5>0$, then
for any positive constant $C$,  when $n$ is sufficiently large,
  \begin{equation*}
    |E(T_{k7})+ E(T_{k8})|
\leq \widetilde{C}n^{2\eta_4}\exp[-n^{\alpha_2\eta_5}/(2c_1)] + \widetilde{C}\exp[-n^{\alpha_1\eta_4}/(2c_1)]
<Cn^{-\kappa_1}/16
   \end{equation*}
holds uniformly for all $1\leq k\leq p$.  The above inequality together with \eqref{eq: Sk1-part3-decomposition} ensures that
  \begin{align}\label{eq: Sk1-part3-prob}
       &P(\max_{1\leq k\leq p}|S_{k1, 3}-E(S_{k1, 3})| \geq Cn^{-\kappa_1}/4) \nonumber\\
      & \leq  P(\max_{1\leq k\leq p}|T_{k6}-E(T_{k6})| \geq Cn^{-\kappa_1}/16)
         + P(\max_{1\leq k\leq p}|T_{k7}| \geq Cn^{-\kappa_1}/16)   \nonumber\\
       &  \quad +P(\max_{1\leq k\leq p}|T_{k8}| \geq Cn^{-\kappa_1}/16)
   \end{align}
for all $n$ sufficiently large.

In what follows, we will provide details on establishing the probability bound for each term on the right hand side of \eqref{eq: Sk1-part3-prob}.   First consider $\max_{1\leq k\leq p}|T_{k6}-E(T_{k6})|$.  Since $0\leq X_{ik}^2\varepsilon_i^2\mathbb{I}(|X_{ik}|\leq M_4)\mathbb{I}(|\varepsilon_i|\leq M_5)\leq M_4^2M_5^2$, by Hoeffding's inequality \citep{hoeffding1963probability} we have for any $\delta>0$ that
  \begin{eqnarray*}
       P(|T_{k6}-E(T_{k6})|\geq \delta)
     \leq 2\exp\left(-\frac{2n\delta^2}{M_4^4M_5^4}\right)
     = 2\exp\left(-2n^{1-4\eta_4-4\eta_5}\delta^2\right),
   \end{eqnarray*}
by 
noting that $M_4=n^{\eta_4}$ and $M_5=n^{\eta_5}$.  Thus, taking $\delta=Cn^{-\kappa_1}/16$ gives
  \begin{align}\label{eq: max-Tk6}
    & P(\max_{1\leq k\leq p}|T_{k6}-E(T_{k6})| \geq Cn^{-\kappa_1}/16)
    \leq \sum_{k=1}^p P(|T_{k6}-E(T_{k6})| \geq Cn^{-\kappa_1}/16) \nonumber \\
    & \leq  2p\exp\left(-\widetilde{C}n^{1-2\kappa_1-4\eta_4-4\eta_5}\right).
   \end{align}

Next we handle $\max_{1\leq k\leq p}|T_{k7}|$.  Since
$\max_{1\leq k\leq p}|T_{k7}|\leq n^{-1}M_4^2\sum_{i=1}^n\varepsilon_i^2\mathbb{I}(|\varepsilon_i|>M_5)$,
it follows from Markov's inequality and \eqref{eq: error-tail-M5} that for any $\delta>0$,
  \begin{align*}
        \nonumber  P(\max_{1\leq k\leq p}|T_{k7}|\geq \delta)
       \leq &  P\{n^{-1}M_4^2\sum_{i=1}^n\varepsilon_i^2\mathbb{I}(|\varepsilon_i|>M_5)\geq \delta\}
     \leq \delta^{-1}E[n^{-1}M_4^2\sum_{i=1}^n\varepsilon_i^2\mathbb{I}(|\varepsilon_i|>M_5)]  \nonumber\\
        =& \delta^{-1}M_4^2E[\varepsilon_1^2\mathbb{I}(|\varepsilon_1|>M_5)]
       \leq  \widetilde{C}\delta^{-1}M_4^2\exp[-M_5^{\alpha_2}/(2c_1)].
   \end{align*}
Recall that $M_4=n^{\eta_4}$ and $M_5=n^{\eta_5}$.  Setting $\delta=Cn^{-\kappa_1}/16$ in the above inequality entails
  \begin{equation}\label{eq: max-Tk7}
        P(\max_{1\leq j\leq p}|T_{k7}|\geq Cn^{-\kappa_1}/16)
        \leq  16C^{-1}\widetilde{C}n^{2\eta_4+\kappa_1}\exp[-n^{\alpha_2\eta_5}/(2c_1)].
   \end{equation}

We then consider $\max_{1\leq k\leq p}|T_{k8}|$.  By Markov's inequality and \eqref{eq: Tk8-expectation-bound}, for any $\delta>0$,
  \begin{align}\label{eq: Tk8-bound}
       \nonumber  P(|T_{k8}|\geq \delta)
    &\leq \delta^{-1}E[n^{-1}\sum_{i=1}^nX_{ik}^2\varepsilon_i^2\mathbb{I}(|X_{ik}|>M_4)]
       =\delta^{-1}E[X_{1k}^2\varepsilon_1^2\mathbb{I}(|X_{1k}|>M_4)]  \\
     &\leq  \delta^{-1}\widetilde{C}\exp[-M_4^{\alpha_1}/(2c_1)].
   \end{align}
Recall that $M_4=n^{\eta_1}$. In view of
\eqref{eq: Tk8-bound}, taking $\delta=Cn^{-\kappa_1}/16$ leads to
  \begin{align}\label{eq: max-Tk8}
        P(\max_{1\leq k\leq p}|T_{k8}|\geq Cn^{-\kappa_1}/16)
        \leq & \sum_{k=1}^p P(|T_{k8}|\geq Cn^{-\kappa_1}/16) \nonumber\\
        \leq & 16pC^{-1}\widetilde{C} n^{\kappa_1} \exp[-n^{\alpha_1\eta_4}/(2c_1)].
   \end{align}

Combining
\eqref{eq: Sk1-part3-prob}, \eqref{eq: max-Tk6}, \eqref{eq: max-Tk7} with \eqref{eq: max-Tk8}
yields that for sufficiently large $n$,
  \begin{align}\label{eq: Sk1-part3-bound}
    & P(\max_{1\leq k\leq p}|S_{k1, 3}-E(S_{k1, 3})| \geq Cn^{-\kappa_1}/4)
   \leq 2p\exp\left(-\widetilde{C}n^{1-2\kappa_1-4\eta_4-4\eta_5}\right) \nonumber\\
     &\quad\quad\quad+ 16pC^{-1}\widetilde{C} n^{\kappa_1} \exp[-n^{\alpha_1\eta_4}/(2c_1)]
       + 16C^{-1}\widetilde{C}n^{2\eta_4+\kappa_1}\exp[-n^{\alpha_2\eta_5}/(2c_1)].
   \end{align}
Let $\eta_4=\eta_5=(1-2\kappa_1)/(8+\alpha_1)$. Then \eqref{eq: Sk1-part3-bound} becomes
  \begin{align}\label{eq: Sk1-part3-final}
     & P(\max_{1\leq k\leq p}|S_{k1, 3}-E(S_{k1, 3})| \geq Cn^{-\kappa_1}/4)  \nonumber\\
  &\leq  p\widetilde{C}_{11}\exp[-\widetilde{C}_{12}n^{\alpha_1\eta_4}]
   +\widetilde{C}_{13}\exp[-\widetilde{C}_{14}n^{\alpha_2\eta_4}]
   \end{align}
 for all $n$ sufficiently large, where $\widetilde{C}_{11}$, $\widetilde{C}_{12}$, $\widetilde{C}_{13}$, and $\widetilde{C}_{14}$ are some positive constants.


Since $0< \eta_1< \eta_2=\eta_3$ and $\eta_1\leq \eta_4$, it follows from \eqref{eq: Sk1-bound}, \eqref{eq: Sk1-part1-final}, \eqref{eq: Sk1-part2-final}, and \eqref{eq: Sk1-part3-final} that there exist some positive constants $\widetilde{C}_{1}, \cdots, \widetilde{C}_{4}$ such that
 \begin{align*}
       P(\max_{1\leq k\leq p}|S_{k1}-E(S_{k1})| \geq Cn^{-\kappa_1})
   \leq    p\widetilde{C}_1\exp\left(-\widetilde{C}_2n^{\alpha_1\eta_1}\right)
       +\widetilde{C}_3\exp\left(-\widetilde{C}_4n^{\alpha_2\eta_1}\right)
  \end{align*}
for all $n$ sufficiently large.  This concludes the proof of part a) of Theorem \ref{Th: Sure Screening-new}.

\subsection*{D.3. Proof of part b) of Theorem \ref{Th: Sure Screening-new}}

We recall that $\omega_j^{\ast}=E(X_jY)$ and $\widehat{\omega}_j^{\ast}=n^{-1}\sum\limits_{i=1}^n X_{ij}Y_i$.  Note that $Y_i=\beta_0+\bx_i^T\bbeta_0+\bz_i^T\bgamma_0+\varepsilon_i=\beta_0+\bx_{i,\,\mathcal{B}} ^T\bbeta_{0,\mathcal{B}}+\bz_{i,\,\mathcal{I}}^T\bgamma_{0,\mathcal{I}}+\varepsilon_i$, where $\bx_{i}=(X_{i1}, \cdots, X_{ip})^T$, $\bz_{i}=(X_{i1}X_{i2}, \cdots, X_{i, p-1}X_{i, p})^T$, $\bx_{i,\,\mathcal{B}}=(X_{i\ell}, \ell\in \mathcal{B})^T$, $\bz_{i,\,\mathcal{I}}=(X_{ik}X_{i\ell}, (k, \ell)\in \mathcal{I})^T$, $\bbeta_{0,\mathcal{B}}=(\beta_{\ell}^0, \ell\in \mathcal{B})^T$, and $\bgamma_{0,\mathcal{I}}=(\gamma_{k\ell}, (k,\ell)\in \mathcal{I})^T$. To simplify the proof,  we assume that the intercept $\beta_0$ is zero without loss of generality. Thus
   \begin{align*}
      \widehat{\omega}_j^{\ast}
   =n^{-1}\sum\limits_{i=1}^n X_{ij}Y_i
   =n^{-1}\sum\limits_{i=1}^n X_{ij}(\bx_{i,\,\mathcal{B}} ^T\bbeta_{0,\mathcal{B}}+\bz_{i,\,\mathcal{I}}^T\bgamma_{0,\mathcal{I}})
+ n^{-1}\sum\limits_{i=1}^n X_{ij}\varepsilon_i
  \triangleq S_{j1} + S_{j2}.
   \end{align*}
Similarly, $\omega_j^{\ast}$ can be written as $\omega_j^{\ast}=E(X_jY)=E(S_{j1})+E(S_{j2})$. So $\widehat{\omega}_j^{\ast}-\omega_j^{\ast}$ can be expressed as $\widehat{\omega}_j^{\ast}-\omega_j^{\ast}
     =[S_{j1}-E(S_{j1})] + [S_{j2}-E(S_{j2})]$.
By the triangle inequality and the union bound, it holds that
  \begin{align}\label{eq: omega-star-max}
      & P(\max_{1\leq j\leq p}|\widehat{\omega}_j^{\ast}-\omega_j^{\ast}|\geq Cn^{-\kappa_2}) \nonumber\\
   \leq & P(\max_{1\leq j\leq p}|S_{j1}-E(S_{j1})|\geq Cn^{-\kappa_2}/2)
        +P(\max_{1\leq j\leq p}|S_{j2}-E(S_{j2})|\geq Cn^{-\kappa_2}/2).
   \end{align}
In what follows, we will provide details on deriving an exponential tail probability bound for each term on the right hand side above.  To enhance readability, we split the proof into two steps.

{\bf Step 1.} We start with the first term $\max_{1\leq k\leq p}|S_{j1}-E(S_{j1})|$.
Define the event $\Phi_i=\{|X_{i\ell}|\leq M_6\,\mbox{for all}\, \ell\in\mathcal{M}\cup\{j\}\}$ with $\mathcal{M}=\mathcal{A}\cup \mathcal{B}$ and $M_6$ a large positive number that will be specified later.
Let $T_{j1}=n^{-1}\sum\limits_{i=1}^n X_{ij}(\bx_{i,\,\mathcal{B}} ^T\bbeta_{0,\mathcal{B}}+\bz_{i,\,\mathcal{I}}^T\bgamma_{0,\mathcal{I}})\mathbb{I}_{\Phi_i}$ and $T_{j2}=n^{-1}\sum\limits_{i=1}^n X_{ij}(\bx_{i,\,\mathcal{B}} ^T\bbeta_{0,\mathcal{B}}+\bz_{i,\,\mathcal{I}}^T\bgamma_{0,\mathcal{I}})\mathbb{I}_{\Phi_i^c}$,
where $\mathbb{I}(\cdot)$ is the indicator function and $\Phi_i^c$ is the complement of
the set $\Phi_i$.  Then an application of the triangle inequality yields
  \begin{align}\label{eq: Sj1-decomp}
     |S_{j1}-E(S_{j1})|
   =& |[T_{j1} -E(T_{j1})] + T_{j2} -E(T_{j2})|
    \leq |T_{j1} -E(T_{j1})| + |T_{j2}|+|E(T_{j2})| \nonumber\\
    \leq &  |T_{j1} -E(T_{j1})| + |T_{j2}|+E(|T_{j2}|).
   \end{align}
Note that $|T_{j2}|\leq n^{-1}\sum\limits_{i=1}^n |X_{ij}(\bx_{i,\,\mathcal{B}} ^T\bbeta_{0,\mathcal{B}}+\bz_{i,\,\mathcal{I}}^T\bgamma_{0,\mathcal{I}})|\mathbb{I}_{\Phi_i^c}$ and thus
$E(|T_{j2}|)\leq E[|X_{1j}(\bx_{1,\,\mathcal{B}} ^T\bbeta_{0,\mathcal{B}}+\bz_{1,\,\mathcal{I}}^T\bgamma_{0,\mathcal{I}})|\mathbb{I}_{\Phi_1^c}]$.  By the triangle inequality
and 
Condition \ref{con: sparsity}, we have
  \begin{eqnarray}\label{eq: Ymean-bound2}
     |X_{1j}(\bx_{1,\,\mathcal{B}} ^T\bbeta_{0,\mathcal{B}}+\bz_{1,\,\mathcal{I}}^T\bgamma_{0,\mathcal{I}})|
   \leq C_0(|X_{1j}|\|\bx_{1,\,\mathcal{B}}\|_1+|X_{1j}|\|\bz_{1,\,\mathcal{I}}\|_1),
   \end{eqnarray}
which ensures that 
$E(|T_{j2}|)$ is bounded by $C_0 [E(|X_{1j}|\|\bx_{1,\,\mathcal{B}}\|_1\mathbb{I}_{\Omega_1^c})+E(|X_{1j}|\|\bz_{1,\,\mathcal{I}}\|_1\mathbb{I}_{\Omega_1^c})]$.  Here $\|\cdot\|_1$ is the $L_1$ norm.
By the Cauchy-Schwarz inequality and the triangular inequality, we deduce
  \begin{align*}
     E(|X_{1j}|\|\bx_{1,\,\mathcal{B}}\|_1\mathbb{I}_{\Phi_1^c})
    \leq & \left[E(X_{1j}^2\|\bx_{1,\,\mathcal{B}}\|_1^2)P(\Phi_1^c)\right]^{1/2}
    \leq \left\{\left[s_2\sum_{\ell\in\mathcal{B}}E(X_{1j}^2X_{1\ell}^2)\right]P(\Phi_1^c)\right\}^{1/2} \\
    \leq & \left\{2^{-1}s_2\sum_{\ell\in\mathcal{B}}[E(X_{1j}^4)+E(X_{1\ell}^4)]\right\}^{1/2}\left[\sum_{\ell\in  \mathcal{M}\cup\{j\}}P(|X_{i\ell}|>M_6)\right]^{1/2} \\
     \leq &\widetilde{C}s_2(1+s_2+2s_1)^{1/2}\exp[-M_6^{\alpha_1}/(2c_1)]
   \end{align*}
for some positive constant $\widetilde{C}$, where the last inequality follows from Condition \ref{con: XYtail-new} and Lemma  \ref{lem: Sub-exp-bound}. Similarly, we have
$E(|X_{1j}|\|\bz_{1,\,\mathcal{I}}\|_1\mathbb{I}_{\Phi_1^c})\leq\widetilde{C}s_1(1+s_2+2s_1)^{1/2}\exp[-M_6^{\alpha_1}/(2c_1)]$.
This together with the above inequalities entails that
  \begin{align*}
      E(|T_{j2}|)
  \leq C_0\widetilde{C}(s_1+s_2)(1+s_2+2s_1)^{1/2}\exp[-M_6^{\alpha_1}/(2c_1)].
   \end{align*}
If we choose $M_6=n^{\eta_6}$ with $\eta_6>0$, then by Condition \ref{con: sparsity}, for any positive constant $C$, when $n$ is sufficiently large,
   \begin{eqnarray}\label{eq: Tj2-expectation}
     E(|T_{j2}|)
  \leq C_0\widetilde{C}(n^{\xi_1}+n^{\xi_2})(1+n^{\xi_2}+2n^{\xi_1})^{1/2}\exp[-n^{\alpha_1\eta_6}/(2c_1)]<Cn^{-\kappa_2}/6
   \end{eqnarray}
holds uniformly for all $1\leq j\leq p$.  The above inequality together with \eqref{eq: Sj1-decomp} ensures that
  \begin{align}\label{eq: Sj1-prob}
          & P(\max_{1\leq j\leq p}|S_{j1}-E(S_{j1})|\geq Cn^{-\kappa_2}/2) \nonumber\\
   \leq & P(\max_{1\leq j\leq p}|T_{j1}-E(T_{j1})|\geq Cn^{-\kappa_2}/6)
          + P(\max_{1\leq j\leq p}|T_{j2}|\geq Cn^{-\kappa_2}/6)
   \end{align}
for all $n$ is sufficiently large. Thus we only need to establish the probability bound for each term on the right hand side of \eqref{eq: Sj1-prob}.

First consider $\max_{1\leq j\leq p}|T_{j1}-E(T_{j1})|$.
Using similar arguments as for proving \eqref{eq: Ymean-bound2}, we have
  \begin{align*}
      |X_{ij}(\bx_{i,\,\mathcal{B}} ^T\bbeta_{0,\mathcal{B}}+\bz_{i,\,\mathcal{I}}^T\bgamma_{0,\mathcal{I}})\mathbb{I}_{\Phi_i}|
  \leq C_0(|X_{ij}|\|\bx_{i,\,\mathcal{B}}\|_1+|X_{ij}|\|\bz_{i,\,\mathcal{I}}\|_1)\mathbb{I}_{\Phi_i}
\leq C_0(s_2M_6^2+s_1M_6^3).
   \end{align*}
For any $\delta>0$, an application of Hoeffding's inequality \citep{hoeffding1963probability} gives
  \begin{align*}
       P(|T_{j1}-E(T_{j1})|\geq \delta)
     \leq & 2\exp\left[-\frac{n\delta^2}{2C_0^2M_6^4(s_2+s_1M_6)^2}\right]
     \leq 2\exp\left[-\frac{n\delta^2}{4C_0^2M_6^4(s_2^2+s_1^2M_6^2)}\right] \\
     \leq & 2\exp\left(-\frac{n\delta^2}{8C_0^2M_6^4s_2^2}\right) + 2\exp\left(-\frac{n\delta^2}{8C_0^2M_6^{6}s_1^2}\right),
   \end{align*}
where we have used the fact that $(a+b)^2\leq 2(a^2+b^2)$ for any real numbers $a$ and $b$, and $\exp[-c/(a+b)]\leq \exp[-c/(2a)]+\exp[-c/(2b)]$ for any $a, b, c>0$.  Recall that $M_6=n^{\eta_6}$. Under Condition \ref{con: sparsity}, taking $\delta=Cn^{-\kappa_2}/6$ results in
  \begin{align}\label{eq: Tj1-prob}
       & P(\max_{1\leq j\leq p}|T_{j1}-E(T_{j1})|\geq  Cn^{-\kappa_2}/6)
     \leq  \sum_{j=1}^pP(|T_{j1}-E(T_{j1})|\geq Cn^{-\kappa_2}/6) \nonumber\\
     \leq & 2p\exp\left(-\widetilde{C}n^{1-2\kappa_2-4\eta_6-2\xi_2}\right) + 2p\exp\left(-\widetilde{C}n^{1-2\kappa_2-6\eta_6-2\xi_1}\right).
   \end{align}

Next, consider $\max_{1\leq j\leq p}|T_{j2}|$. By Markov's inequality, for any $\delta>0$, we have $ P(|T_{j2}|\geq \delta)\leq \delta^{-1}E(|T_{j2}|)$. In view of the first inequality in \eqref{eq: Tj2-expectation}, taking $\delta=Cn^{-\kappa_2}/6$ gives that
   \begin{align*}
      P(|T_{j2}|\geq Cn^{-\kappa_2}/6)
 \leq & 6C^{-1}C_0\widetilde{C}n^{\kappa_2}(n^{\xi_1}+n^{\xi_2})(1+n^{\xi_2}+2n^{\xi_1})^{1/2}\exp[-n^{\alpha_1\eta_6}/(2c_1)]
   \end{align*}
for all $1\leq j\leq p$. Therefore,
   \begin{align}\label{eq: Tj2-prob}
      & P(\max_{1\leq j\leq p}|T_{j2}|\geq Cn^{-\kappa_2}/6)
  \leq \sum_{j=1}^pP(|T_{j2}|\geq Cn^{-\kappa_2}/6) \nonumber\\
 \leq & 6pC^{-1}C_0\widetilde{C}n^{\kappa_2}(n^{\xi_1}+n^{\xi_2})(1+n^{\xi_2}+2n^{\xi_1})^{1/2}\exp[-n^{\alpha_1\eta_6}/(2c_1)].
   \end{align}

Combining \eqref{eq: Sj1-prob}, \eqref{eq: Tj1-prob}, and \eqref{eq: Tj2-prob} yields that for sufficiently large $n$,
   \begin{align}\label{eq: Sj1-bound}
      & P(\max_{1\leq j\leq p}|S_{j1}-E(S_{j1})|\geq Cn^{-\kappa_2}/2) \nonumber\\
     &\leq   2p\exp\left(-\widetilde{C}n^{1-2\kappa_2-4\eta_6-2\xi_2}\right) + 2p\exp\left(-\widetilde{C}n^{1-2\kappa_2-6\eta_6-2\xi_1}\right) \nonumber\\
     &\quad \quad+ 6pC^{-1}C_0\widetilde{C}n^{\kappa_2}(n^{\xi_1}+n^{\xi_2})(1+n^{\xi_2}+2n^{\xi_1})^{1/2}\exp[-n^{\alpha_1\eta_6}/(2c_1)].
   \end{align}
To balance the three terms on the right hand side of \eqref{eq: Sj1-bound}, we choose $\eta_6=\min\{(1-2\kappa_2-2\xi_2)/(4+\alpha_1),\,(1-2\kappa_2-2\xi_1)/(6+\alpha_1)\}>0$ and the probability bound
\eqref{eq: Sj1-bound} then becomes
   \begin{align}\label{eq: Sj1-final}
      P(\max_{1\leq j\leq p}|S_{j1}-E(S_{j1})|\geq Cn^{-\kappa_2}/2)
     \leq p\widetilde{C}_1\exp\left(-\widetilde{C}_2n^{\alpha_1\eta_6}\right)
   \end{align}
 for all $n$ sufficiently large, where $\widetilde{C}_1$ and $\widetilde{C}_2$ are two positive constants.

{\bf Step 2.} We establish the probability bound for $\max_{1\leq j\leq p}|S_{j2}-E(S_{j2})|$.
Define
  \begin{align*}
      & T_{j3}=n^{-1}\sum_{i=1}^n X_{ij}\varepsilon_i\mathbb{I}(|X_{ij}|\leq M_7)\mathbb{I}(|\varepsilon_i|\leq M_8), \\
     & T_{j4}=n^{-1}\sum_{i=1}^n X_{ij}\varepsilon_i\mathbb{I}(|X_{ij}|\leq M_7)\mathbb{I}(|\varepsilon_i|> M_8), \\
     & T_{j5}=n^{-1}\sum_{i=1}^n X_{ij}\varepsilon_i\mathbb{I}(|X_{ij}|> M_7),
   \end{align*}
where $M_7$ and $M_8$ are two large positive numbers whose values will be specified later.  Then $S_{j2}=T_{j3}+T_{j4}+T_{j5}$.  
Similarly,  $E(S_{j2})$ can be written as
$E(S_{j2})=E(T_{j3})+E(T_{j4})+E(T_{j5})$.
Since $\varepsilon_1$ has mean zero and is independent of 
$X_{1, 1}, \cdots, X_{1, p}$, we have
$E(T_{j5})=E[X_{1j}\varepsilon_1\mathbb{I}(|X_{1j}|> M_7)]=E[X_{1j}\mathbb{I}(|X_{1j}|> M_7)]E(\varepsilon_1)=0$. Thus
$S_{j2}-E(S_{j2})$ can be expressed as
$S_{j2}-E(S_{j2})= [T_{j3}-E(T_{j3})] + T_{j4} + T_{j5} - E(T_{j4})$.
An application of the triangle inequality yields
  \begin{align}\label{eq: Sj2-decomposition}
      |S_{j2}-E(S_{j2})|
     \leq & |T_{j3}-E(T_{j3})| + |T_{j4}| + |T_{j5}|+ |E(T_{j4})| \nonumber\\
     \leq & |T_{j3}-E(T_{j3})| + |T_{j4}| + |T_{j5}|+ E(|T_{j4}|).
   \end{align}

First consider the last term $E(|T_{j4}|)$. Note that $|T_{j4}|\leq n^{-1}\sum_{i=1}^n |X_{ij}\varepsilon_i|\mathbb{I}(|X_{ij}|\leq M_7)\mathbb{I}(|\varepsilon_i|> M_8)$ and thus
  \begin{equation}\label{eq: Tj4-expectation}
     E(|T_{j4}|)
   \leq E[|X_{1j}\varepsilon_1|\mathbb{I}(|X_{1j}|\leq M_7)\mathbb{I}(|\varepsilon_1|> M_8)]
   \leq M_7E[|\varepsilon_1|\mathbb{I}(|\varepsilon_1|>M_8)].
   \end{equation}
An application of the Cauchy-Schwarz inequality gives $E[|\varepsilon_1|\mathbb{I}(|\varepsilon_1|>M_8)]\leq [E(\varepsilon_1^2)P(|\varepsilon_1|> M_8)]^{1/2}$. By Condition \ref{con: XYtail-new} and Lemma \ref{lem: Sub-exp-bound}, we have
  \begin{equation}\label{eq: error-tail-M8}
      E[|\varepsilon_1|\mathbb{I}(|\varepsilon_1|>M_8)]
     \leq \{E(\varepsilon_1^2)c_1\}^{1/2}\exp(-c_1^{-1}M_8^{\alpha_2}/2)
     \leq\widetilde{C}\exp[-M_8^{\alpha_2}/(2c_1)]
   \end{equation}
Combining \eqref{eq: Tj4-expectation} with \eqref{eq: error-tail-M8} yields
  \begin{equation}\label{eq: Tj4-expectation-bound}
       E(|T_{j4}|)
   \leq \widetilde{C}M_7\exp[-M_8^{\alpha_2}/(2c_1)].
   \end{equation}
If we choose $M_7=n^{\eta_7}$ and $M_8=n^{\eta_8}$ with $\eta_7>0$ and $\eta_8>0$, then
for any positive constant $C$,  when $n$ is sufficiently large,
  \begin{equation*}
    E(|T_{j4}|)
\leq \widetilde{C}n^{\eta_7}\exp[-n^{\alpha_2\eta_8}/(2c_1)]
<Cn^{-\kappa_2}/8
   \end{equation*}
holds uniformly for all $1\leq j\leq p$.  The above inequality together with \eqref{eq: Sj2-decomposition} ensures that
  \begin{align}\label{eq: Sj2-prob}
       &P(\max_{1\leq j\leq p}|S_{j2}-E(S_{j2})| \geq Cn^{-\kappa_2}/2) \nonumber\\
      \leq & P(\max_{1\leq j\leq p}|T_{j3}-E(T_{j3})| \geq Cn^{-\kappa_2}/8)
         + P(\max_{1\leq j\leq p}|T_{j4}| \geq Cn^{-\kappa_2}/8)   \nonumber\\
       &  \quad +P(\max_{1\leq j\leq p}|T_{j5}| \geq Cn^{-\kappa_2}/8)
   \end{align}
for all $n$ sufficiently large.

In what follows, we will provide details on establishing the probability bound for each term on the right hand side of \eqref{eq: Sj2-prob}.   First consider $\max_{1\leq j\leq p}|T_{j3}-E(T_{j3})|$.  Since $|X_{ij}\varepsilon_i\mathbb{I}(|X_{ij}|\leq M_7)\mathbb{I}(|\varepsilon_i|\leq M_8)|\leq M_7M_8$, for any $\delta>0$, by Hoeffding's inequality \citep{hoeffding1963probability} we obtain
  \begin{eqnarray*}
       P(|T_{j3}-E(T_{j3})|\geq \delta)
     \leq 2\exp\left(-\frac{n\delta^2}{2M_7^2M_8^2}\right)
     = 2\exp\left(-2^{-1}n^{1-2\eta_7-2\eta_8}\delta^2\right),
   \end{eqnarray*}
by 
noting that $M_7=n^{\eta_7}$ and $M_8=n^{\eta_8}$.  Thus, taking $\delta=Cn^{-\kappa_2}/8$ gives
  \begin{align}\label{eq: max-Tj3}
    & P(\max_{1\leq j\leq p}|T_{j3}-E(T_{j3})| \geq Cn^{-\kappa_2}/8)
    \leq \sum_{j=1}^p P(|T_{j3}-E(T_{j3})| \geq Cn^{-\kappa_2}/8) \nonumber \\
    \leq & 2p\exp\left(-\widetilde{C}n^{1-2\kappa_2-2\eta_7-2\eta_8}\right).
   \end{align}

Next we handle $\max_{1\leq j\leq p}|T_{j4}|$.  Since
$\max_{1\leq j\leq p}|T_{j4}|\leq n^{-1}M_7\sum_{i=1}^n|\varepsilon_i|\mathbb{I}(|\varepsilon_i|>M_8)$,
it follows from Markov's inequality and \eqref{eq: error-tail-M8} that for any $\delta>0$,
  \begin{align*}
        \nonumber  P(\max_{1\leq j\leq p}|T_{j4}|\geq \delta)
       \leq &  P\{n^{-1}M_7\sum_{i=1}^n|\varepsilon_i|\mathbb{I}(|\varepsilon_i|>M_8)\geq \delta\}
     \leq \delta^{-1}E[n^{-1}M_7\sum_{i=1}^n|\varepsilon_i|\mathbb{I}(|\varepsilon_i|>M_8)]  \nonumber\\
        =& \delta^{-1}M_7E[|\varepsilon_1|\mathbb{I}(|\varepsilon_1|>M_8)]
       \leq  \widetilde{C}\delta^{-1}M_7\exp[-M_8^{\alpha_2}/(2c_1)].
   \end{align*}
Recall that $M_7=n^{\eta_7}$ and $M_8=n^{\eta_8}$.  Setting $\delta=Cn^{-\kappa_2}/8$ in the above inequality entails
  \begin{equation}\label{eq: max-Tj4}
        P(\max_{1\leq j\leq p}|T_{j4}|\geq Cn^{-\kappa_2}/8)
        \leq  16C^{-1}\widetilde{C}n^{\eta_7+\kappa_2}\exp[-n^{\alpha_2\eta_8}/(2c_1)].
   \end{equation}


We now consider $\max_{1\leq j\leq p}|T_{j5}|$. By the Cauchy-Schwarz inequality and Lemma \ref{lem: Sub-exp-bound} we deduce that
  \begin{align*}
     & E|T_{j5}|
      = E|X_{1j}\varepsilon_1\mathbb{I}(|X_{1j}|> M_7)|
      \leq \{E(X_{1j}^2\varepsilon_1^2)P(|X_{1j}|> M_7)]\}^{1/2}\nonumber\\
      \leq & \big\{\frac{c_1}{2 }[E(X_{1k}^4)+E(\varepsilon_1^4)]\big\}^{1/2}\exp[-M_7^{\alpha_1}/(2c_1)]
       \leq \widetilde{C}\exp[-M_7^{\alpha_1}/(2c_1)].
   \end{align*}
An application of Markov's inequality yields
  \begin{align}\label{eq: Tj5-bound}
     P(|T_{j5}|\geq \delta)
    \leq \delta^{-1}E|T_{j5}|
     \leq  \delta^{-1}\widetilde{C}\exp[-M_7^{\alpha_1}/(2c_1)]
   \end{align}
for any $\delta>0$.  Recall that $M_7=n^{\eta_7}$. In view of
\eqref{eq: Tj5-bound}, taking $\delta=Cn^{-\kappa_2}/8$ gives that
  \begin{align}\label{eq: max-Tj5}
        P(\max_{1\leq j\leq p}|T_{j5}|\geq Cn^{-\kappa_2}/8)
        \leq & \sum_{j=1}^p P(|T_{j5}|\geq Cn^{-\kappa_2}/8) \nonumber\\
        \leq & 8pC^{-1}\widetilde{C} n^{\kappa_2} \exp[-n^{\alpha_1\eta_7}/(2c_1)].
   \end{align}

Combining
\eqref{eq: Sj2-prob}, \eqref{eq: max-Tj3}, \eqref{eq: max-Tj4}, and \eqref{eq: max-Tj5}
yields that for sufficiently large $n$,
  \begin{align}\label{eq: Sj2-bound}
    & P(\max_{1\leq j\leq p}|S_{j2}-E(S_{j2})| \geq Cn^{-\kappa_2}/2)
   \leq 2p\exp\left(-\widetilde{C}n^{1-2\kappa_2-2\eta_7-2\eta_8}\right) \nonumber\\
     &\quad\quad\quad+ 8pC^{-1}\widetilde{C} n^{\kappa_2} \exp[-n^{\alpha_1\eta_7}/(2c_1)]
       + 16C^{-1}\widetilde{C}n^{\eta_7+\kappa_2}\exp[-n^{\alpha_2\eta_8}/(2c_1)].
   \end{align}
Let $\eta_7=\eta_8=(1-2\kappa_2)/(4+\alpha_1)$. Then \eqref{eq: Sj2-bound} becomes
  \begin{align}\label{eq: Sj2-final}
     & P(\max_{1\leq j\leq p}|S_{j2}-E(S_{j2})| \geq Cn^{-\kappa_1}/2)  \nonumber\\
  \leq & p\widetilde{C}_{3}\exp[-\widetilde{C}_{4}n^{\alpha_1\eta_7}]
   +\widetilde{C}_{5}\exp[-\widetilde{C}_{6}n^{\alpha_2\eta_7}]
   \end{align}
 for all $n$ sufficiently large, where $\widetilde{C}_{3}$, $\widetilde{C}_{4}$, $\widetilde{C}_{5}$, and $\widetilde{C}_{6}$ are some positive constants.

Since $0< \eta_6< \eta_7$, it follows from \eqref{eq: omega-star-max}, \eqref{eq: Sj1-final}, and \eqref{eq: Sj2-final} that
 \begin{align*}
       P(\max_{1\leq j\leq p}|\widehat{\omega}_j^{\ast}-\omega_j^{\ast}|\geq Cn^{-\kappa_2})
   \leq  &  p\widetilde{C}_1\exp\left(-\widetilde{C}_2n^{\alpha_1\eta_6}\right)
        +  p\widetilde{C}_{3}\exp[-\widetilde{C}_{4}n^{\alpha_1\eta_7}]
   +\widetilde{C}_{5}\exp[-\widetilde{C}_{6}n^{\alpha_2\eta_7}]\\
   \leq & p\widetilde{C}_7\exp\left(-\widetilde{C}_8n^{\alpha_1\eta_6}\right)
   +\widetilde{C}_{5}\exp[-\widetilde{C}_{6}n^{\alpha_2\eta_6}]
  \end{align*}
with $\widetilde{C}_7=\widetilde{C}_1+\widetilde{C}_3$ and $\widetilde{C}_8=\min\{\widetilde{C}_2, \widetilde{C}_4\}$ for all $n$ sufficiently large.
If $\log p=o(n^{\alpha_1\eta'})$ with $\eta'=\min\{(1-2\kappa_2-2\xi_2)/(4+\alpha_1), (1-2\kappa_2-2\xi_1)/(6+\alpha_1)\}>0$, then for any positive constant $C$, there exists some arbitrarily large positive constant $C_2$ such that
  \begin{equation*}
      P(\max_{1\leq j\leq p}|\widehat{\omega}_j^{\ast}-\omega_j^{\ast}|\geq Cn^{-\kappa_2})
   \leq  o(n^{-C_2})
   \end{equation*}
for all $n$ sufficiently large, which completes the proof of part b) of Theorem \ref{Th: Sure Screening-new}.

\subsection*{D.4. Proof of part c) of Theorem \ref{Th: Sure Screening-new}}

%
%
The main idea of the proof is to find probability bounds for the two events $\{\mathcal{I}\subset \widehat{\mathcal{I}}\}$ and $\{\mathcal{M}\subset \widehat{\mathcal{M}}\}$, respectively. First note that conditional on the event $\{\mathcal{A}\subset \widehat{\mathcal{A}}\}$, we have $\{\mathcal{I}\subset \widehat{\mathcal{I}}\}$. Thus it holds that
   \begin{equation}\label{eq: prob-inter-1}
     P(\mathcal{I}\subset \widehat{\mathcal{I}})\geq P(\mathcal{A}\subset \widehat{\mathcal{A}}).
   \end{equation}
Define the event $\mathcal{E}_1=\{\max_{k\in \mathcal{A}}|\hat{\omega}_k-\omega_k|<2^{-1}c_2n^{-\kappa_1}\}$. Then, with $\tau = c_2 n^{-\kappa_1}$, the event
$\mathcal{E}_1$ ensures that $\mathcal{A}\subset \widehat{\mathcal{A}}$. Thus,
   \begin{equation*}
     P(\mathcal{A}\subset \widehat{\mathcal{A}})
   \geq P(\mathcal{E}_1)
  = 1-P(\mathcal{E}_1^c)
  =1-P(\max_{k\in \mathcal{A}}|\hat{\omega}_k-\omega_k|\geq 2^{-1}c_2n^{-\kappa_1}).
   \end{equation*}
Following similar arguments as for proving \eqref{eq:omega-max}, it can be shown that there exist some constants $\widetilde{C}_1>0$ and $\widetilde{C}_{2}>0$ such that for all $n$ sufficiently large,
  \begin{align}\label{eq: prob-omega-A}
      P(\max_{k\in \mathcal{A}}|\hat{\omega}_k-\omega_k|\geq 2^{-1}c_2n^{-\kappa_1})
  & \leq 2s_1\widetilde{C}_1\exp[-\widetilde{C}_2n^{\min\{\alpha_1, \alpha_2\}r_1}].
  \end{align}
Note that the right hand side of \eqref{eq: prob-omega-A} can be bounded by $o(n^{-C_1})$ for some arbitrarily large positive constant $C_1$.  This gives
  \begin{equation}\label{eq: prob-A}
     P(\mathcal{A}\subset \widehat{\mathcal{A}})\geq 1-o(n^{-C_1}).
  \end{equation}
Thus combining \eqref{eq: prob-inter-1} and \eqref{eq: prob-A} yields
  \begin{equation}\label{eq: prob-I}
     P(\mathcal{I}\subset \widehat{\mathcal{I}})
     \geq 1-o(n^{-C_1}).
  \end{equation}

Using similar arguments as for proving part b) of Theorem \ref{Th: Sure Screening-new} and \eqref{eq: prob-A}, we can show that there exist some positive constants $\widetilde{C}_1$, $\widetilde{C}_{2}$, and $C_2$ such that for all $n$ sufficiently large,
\begin{align}\label{eq: prob-B}
               P(\mathcal{B}\subset \widehat{\mathcal{B}})
    &\geq  P(\max_{j\in \mathcal{B}}|\hat{\omega}_j^*-\omega_j^*|< 2^{-1}c_2n^{-\kappa_2}) 
    \geq 1-s_2\widetilde{C}_1\exp(-\widetilde{C}_2n^{\alpha_1r_2})\nonumber     \\
     & \geq  1-o(n^{-C_2}),
\end{align}
Combining \eqref{eq: prob-A} and \eqref{eq: prob-B} leads to
  \begin{align}\label{eq: prob-SetM}
    P(\mathcal{M}\subset \widehat{\mathcal{M}})
    \geq & P(\mathcal{A}\subset\widehat{\mathcal{A}} \ \text{ and } \ \mathcal{B}\subset\widehat{\mathcal{B}})
    \geq   P(\mathcal{A}\subset\widehat{\mathcal{A}})+P(\mathcal{B}\subset\widehat{\mathcal{B}})-1 \nonumber\\
     \geq & 1-o(n^{-\min\{C_1, C_2\}}).
  \end{align}
In view of \eqref{eq: prob-I} and \eqref{eq: prob-SetM}, we obtain
  \begin{equation*}
      P(\mathcal{I}\subset\widehat{\mathcal{I}} \ \text{ and } \ \mathcal{M}\subset\widehat{\mathcal{M}})
    \geq   P(\mathcal{I}\subset\widehat{\mathcal{I}})+P(\mathcal{M}\subset\widehat{\mathcal{M}})-1
     \geq  1-o(n^{-\min\{C_1, C_2\}})
  \end{equation*}
for all $n$ sufficiently large.  
This completes the proof for the first part of  Theorem \ref{Th: Sure Screening-new} c).

\bigskip

We proceed to prove the second part of part c) of Theorem \ref{Th: Sure Screening-new}.
The main idea is to establish the probability bounds for two events $\{|\widehat{\mathcal{A}}|= O[n^{2\kappa_1}\lambda_{\max}(\bSig^{\ast})]\}$ and
$\{|\widehat{\mathcal{B}}|= O[n^{2\kappa_2}\lambda_{\max}(\bSig)]\}$, respectively.  If we can show that 
  \begin{align}
     P\left\{|\widehat{\mathcal{A}}|= O[n^{2\kappa_1}\lambda_{\max}(\bSig^{\ast})]\right\}
     \geq &  1-o(n^{-C_1}),  \label{eq: prob-sizeA} \\
      P\left\{|\widehat{\mathcal{B}}|= O[n^{2\kappa_2}\lambda_{\max}(\bSig)]\right\}
     \geq & 1-o(n^{-C_2})   \label{eq: prob-sizeB}
  \end{align}
with $C_1$ and $C_2$ defined in \eqref{eq: bound-omega-new} and \eqref{eq: bound-omega-star-new}, respectively,
then it holds that
  \begin{align*}
   P\left\{|\widehat{\mathcal{I}}|= O\left[n^{4\kappa_1} \lambda_{\max}^2({\bSig}^{\ast})\right] \right\}
   \geq  P\left\{|\widehat{\mathcal{A}}|= O\left[n^{2\kappa_1} \lambda_{\max}({\bSig}^{\ast})\right] \right\}\geq 1-o(n^{-C_1})
  \end{align*}
and
  \begin{align*}
     & P\left\{|\widehat{\mathcal{M}}|= O\left[n^{2\kappa_1} \lambda_{\max}({\bSig}^{\ast})+n^{2\kappa_2}\lambda_{\max}(\bSig)\right]\right\} \\
   \geq &  P\left\{|\widehat{\mathcal{A}}|= O[n^{2\kappa_1}\lambda_{\max}(\bSig^{\ast})]\ \text{and} \ |\widehat{\mathcal{B}}|= O[n^{2\kappa_2}\lambda_{\max}(\bSig)]\right\}
     \geq   1-o(n^{-\min\{C_1, C_2\}}).
  \end{align*}
Combining these two results yields
\begin{align*}
      & P\left(|\widehat{\mathcal{I}}|= O\{n^{4\kappa_1} \lambda_{\max}^2(\bSig^{\ast})\} \text{ and }
                 |\widehat{\mathcal{M}}|
 =O\{n^{2\kappa_1}\lambda_{\max}(\bSig^{\ast})+n^{2\kappa_2} \lambda_{\max}(\bSig)\} \right)  \nonumber\\
  =& 1-o\left(n^{-\min\{C_1, C_2\}}\right) \label{eq: model-size-1}.
\end{align*}

It thus remains to prove \eqref{eq: prob-sizeA} and \eqref{eq: prob-sizeB}.  We begin with showing \eqref{eq: prob-sizeB}.
The key step is to show that
  \begin{equation} \label{eq: sum-Omega-square-bound}
     \sum_{j=1}^p {(\omega_j^\ast)}^2
   =\|E(\bx Y)\|^2_2
  \leq \widetilde{C}_3\lambda_{\max}(\bSig)
  \end{equation}
for some constant $\widetilde{C}_3>0$. If so, conditional on the event $ \mathcal{E}_2=\left\{\max\limits_{1\leq j\leq p}|\widehat{\omega}_j^{\ast}-\omega_j^{\ast}|\leq 2^{-1}c_2n^{-\kappa_2}\right\}$,
the number of variables in $\widehat{\mathcal{B}} = \{j: |\widehat{\omega}_j^{\ast}|> c_2n^{-\kappa_2}\}$ cannot exceed the number of variables in $\{j: |\omega_j^{\ast}|> 2^{-1}c_2n^{-\kappa_2}\}$,
which is bounded by 
 $4\widetilde{C}_3c_2^{-2}n^{2\kappa_2}\lambda_{\max}(\bSig)$. Thus it follows from \eqref{eq: bound-omega-star-new} that for all $n$ sufficiently large,
    \begin{equation}\label{eq: main-size}
    P\left\{|\widehat{\mathcal{B}}|\leq    4\widetilde{C}_3c_2^{-2}n^{2\kappa_2} \lambda_{\max}(\bSig)  \right\}
     \geq  P( \mathcal{E}_2)
    =1-  P( \mathcal{E}_2^c)
\geq 1-o(n^{-C_2}).
   \end{equation}

Now we further prove \eqref{eq: sum-Omega-square-bound}.  Let $\boldsymbol u_0=\argmin_{\boldsymbol u}E\left(Y-\mathbf{x}^T\boldsymbol u\right)^2$.  Then the first order equation $E[\mathbf{x}(Y-\mathbf{x}^T\boldsymbol u_0)]=0$
gives $E(\bx Y)=[E(\mathbf{x}\mathbf{x}^T)]\boldsymbol u_0=\bSig \boldsymbol u_0$.
Thus
  \begin{equation}\label{eq: norm-bound-ExY}
     \|E(\bx Y)\|^2_2
 =\boldsymbol u_0^T\bSig^2\boldsymbol u_0
 \leq \lambda_{\max}(\bSig){\boldsymbol u_0}^T{\bSig}\boldsymbol u_0
  =\lambda_{\max}({\bSig})\mbox{var}\left(\mathbf{x}^T{}\boldsymbol u_0\right).
  \end{equation}
It follows from the orthogonal decomposition that
  \begin{equation*}
     \mbox{var}\left(Y\right)
   =\mbox{var}\left(\mathbf{x}^T{\boldsymbol u_0}\right)
         +\mbox{var}\left(Y-\mathbf{x}^T{\boldsymbol u_0}\right)\geq \mbox{var}\left(\mathbf{x}^T{\boldsymbol u_0}\right) .
  \end{equation*}
 Since $E^2(Y^2)\leq E(Y^4)=O(1)$, we have $\var(Y)\leq E(Y^2)=O(1)$. Then the above inequality ensures that $\mbox{var}\left(\mathbf{x}^T{\boldsymbol u_0}\right)\leq \widetilde{C}_3$ for some constant $\widetilde{C}_3>0$. This together with \eqref{eq: norm-bound-ExY} completes the proof of \eqref{eq: sum-Omega-square-bound}.

We next prove \eqref{eq: prob-sizeA}. Recall that $Y^{\ast}=Y^2$ and $X_k^{\ast}={[X_k^2-E(X_k^2)]}/\sqrt{\var(X_k^2)}$. Then from  the definition of $\omega_k$ in Section \ref{sec: IP}, we have
$\omega_k=E(X_k^{\ast}Y^{\ast})$. Following similar arguments as for proving \eqref{eq: sum-Omega-square-bound}, it can be shown that
  \begin{equation} \label{eq: sum-Omega-bound}
     \sum_{k=1}^p {\omega}_k^2
   =\sum_{k=1}^p{E^2(X_k^{\ast}Y^{\ast})}
   =\|{E(\bx^{\ast} Y^{\ast})}\|^2_2
 \leq \widetilde{C}_4\lambda_{\max}({\bSig}^{\ast}),
  \end{equation}
 where $\widetilde{C}_4$ is some positive constant, $\bx^{\ast}=(X_1^{\ast}, \cdots, X_p^{\ast})^T$, and ${\bSig}^{\ast}=\cov(\bx^{\ast})$.
Then,  on the event $ \mathcal{E}_3=\left\{\max_{1\leq k\leq p}|\widehat{\omega}_k-\omega_k|\leq 2^{-1}c_2n^{-\kappa_1}\right\}$,
the cardinality of $\{k: |\widehat{\omega}_k|>  c_2n^{-\kappa_1}\}$ cannot exceed that of $\{k: |\omega_k|> 2^{-1}c_2n^{-\kappa_1}\}$,
which is bounded by 
 $4\widetilde{C}_4c_2^{-2} n^{2\kappa_1}\lambda_{\max}({\bSig}^{\ast})$. Thus, we have
    \begin{equation*}
         P\left\{|\widehat{\mathcal{A}}|\leq  4\widetilde{C}_4c_2^{-2}n^{2\kappa_1} \lambda_{\max}({\bSig}^{\ast}) \right\}
     \geq  P( \mathcal{E}_3)
    =1-  P( \mathcal{E}_3^c)
 \geq 1-o(n^{-C_1}),
   \end{equation*}
where the last equality follows from \eqref{eq: bound-omega-new}.  
This concludes the proof of part c) of Theorem \ref{Th: Sure Screening-new} and thus Theorem \ref{Th: Sure Screening-new} is proved.

\subsection*{D.5. Proof of Theorem \ref{Th:global}}

Recall that $\widetilde{\bX}=(\widetilde{\bx}_1, \cdots, \widetilde{\bx}_{\widetilde{p}})$ is the corresponding $n \times \widetilde{p}$ augmented design matrix incorporating the covariate vectors for $X_j$'s and their interactions in columns, where $\widetilde{\bx}_j = (X_{1j},\cdots, X_{nj})\t$ for $1\leq j\leq p$ is the $j$th covariate vector and $\widetilde{\bx}_j$ for $p+1\leq j\leq \widetilde{p} = p(p+1)/2$ is $\tbx_k \circ \tbx_{\ell}$ with some $1 \leq k< \ell \leq p$ and $\circ$ denoting the Hadamard (componentwise) product. We rescale the design matrix $\widetilde{\bX}$ such that each column has $L_2$-norm $n^{1/2}$, and denote by $\widetilde{\bZ}=\widetilde{\bX}\bD^{-1}$ the resulting matrix, where $\bD =\diag\{\bD_{11}, \cdots, \bD_{\widetilde{p}\widetilde{p}}\}$ with $\bD_{mm} = n^{-1/2} \|\widetilde{\bx}_m\|_2$ is a diagonal scale matrix.

Define the event $\mathcal{E}_4 = \{L_1\leq \min_{1\leq j\leq\widetilde p}|\bD_{jj}| \leq \max_{1\leq j\leq\widetilde p}|\bD_{jj}| \leq L_2\}$, where $L_1$ and $L_2$ are two positive constants defined in Condition \ref{con:RE-new}. Then by the assumption in Condition \ref{con:RE-new}, event $\mathcal{E}_4$ holds with probability at least $1-a_n$. In what follows, we will condition on the event $\mathcal{E}_4$.

Note that conditional on $\mathcal{E}_4$, we have
  \begin{equation}
\|\widetilde{\bX}\bdelta\|_2 \sim \|\widetilde{\bZ}\bdelta\|_2,
  \end{equation}
where the notation $f_n \sim g_n$ means that the ratio $f_n/g_n$ is bounded between two positive constants. Thus, conditional on $\mathcal{E}_4$, Condition \ref{con:RE-new} holds with matrix $\widetilde \bX$ replaced with $\widetilde{\bZ}$. More specifically, with probability at least $1-a_n$, it holds that
  \begin{equation*}
\min_{\|\bdelta\|_2=1,\, \|\bdelta\|_0<2s} n^{-1/2}\|\widetilde{\bZ}\bdelta\|_2 \geq \widetilde\kappa_0, \quad \min\limits_{\bdelta\neq0,\, \|\bdelta_2\|_1\leq 7\|\bdelta_1\|_1}
        \left\{n^{-1/2}\|\widetilde{\bZ}\bdelta\|_2/(\|\bdelta_1\|_2\vee \|\widetilde{\bdelta}_2\|_2)\right\} \geq \widetilde\kappa,
  \end{equation*}
where $\widetilde\kappa_0$ and $\widetilde\kappa$ are two positive constants depending only on $\kappa$, $\kappa_0$, $L_1$, and $L_2$. In addition, conditional on $\mathcal{E}_4$, the desired results in Theorem \ref{Th:global} are equivalent to those with $\widetilde \bX$ and $\btheta$ replaced by $\widetilde \bZ$ and $\btheta^*= \bD \btheta$, respectively. Thus, we only need to work with the design matrix $\widetilde \bZ$ and reparameterized parameter vector $\btheta^*$.

By examining the proof of Theorem 1 in \citet{fan2014asymptotic}, in order to prove Theorem \ref{Th:global} in our paper, it suffices to show that 
 the following inequality
     \begin{equation}\label{eq: bZ-err-bound}
       \|n^{-1}\widetilde{\bZ}^T\bveps\|_{\infty}>\lambda_0/2
   \end{equation}
holds with probability at most $a_n+o(p^{-c_4})$, where 
 $\lambda_0=\widetilde{c}_0\{(\log p)/n^{\alpha_1\alpha_2/(\alpha_1+2\alpha_2)}\}^{1/2}$
for some constant $\widetilde{c}_0>0$ and 
 $c_4$ is some arbitrarily large positive constant depending on $\widetilde{c}_0$. Then with \eqref{eq: bZ-err-bound}, following the proof of Theorem 1 in \citet{fan2014asymptotic}, we can obtain that all results in Theorem \ref{Th:global} hold with probability at least 
 $1-a_n-o(p^{-c_4})$.

It remains to prove \eqref{eq: bZ-err-bound}.  We first show that $ \|n^{-1}\widetilde{\bX}^T\bveps\|_{\infty}>L_1\lambda_0/2$ holds with an overwhelming probability. To this end, note that an application
of the Bonferroni inequality gives
     \begin{align}\label{eq: bX-err-bound-1}
       P(\|n^{-1}\widetilde{\bX}^T\bveps\|_{\infty}>L_1\lambda_0/2)
    \leq \sum_{j=1}^{\widetilde{p}}P(\|n^{-1}\widetilde{\bx}_j^T\bveps\|_{\infty}>L_1\lambda_0/2)
   \end{align}
for any $\lambda_0>0$. The key idea is to construct an upper bound for $P(\|n^{-1}\widetilde{\bx}_j^T\bveps\|_{\infty}>L_1\lambda_0/2)$.  We claim that such an upper bound is $\widetilde{C}_1\exp\{-\widetilde{C}_2n^{\alpha_1\alpha_2/(\alpha_1+2\alpha_2)}\lambda_0^2\}$ for any 
 $0<L_1\lambda_0<2$, where $\widetilde{C}_1$ and $\widetilde{C}_2$ are some positive constants.
To prove this, we consider the following two cases.

{\bf Case 1:} $1\leq j\leq p$.  In this case, $\widetilde{\bx}_j=(X_{1j}, \cdots, X_{nj})^T$.  Thus  $n^{-1}\widetilde{\bx}_j^T\bveps=n^{-1}\sum_{i=1}^n X_{ij} \varepsilon_i$.  By Lemma \ref{lem: sub-exp}, we have
$P(|X_{ij}\varepsilon_i|>t)\leq 2c_1\exp\{-c_1^{-1}t^{\alpha_1\alpha_2/(\alpha_1+\alpha_2)}\}$
for all $1\leq i\leq n$ and $1\leq j\leq p$.  Note that $E(X_{ij}\varepsilon_i)=0$. Thus it follows from Lemma \ref{lemma-Hao-Zhang-extend} that there exist some positive constants $\widetilde{C}_3$ and $\widetilde{C}_4$ such that
    \begin{align*}
     P(|n^{-1}\widetilde{\bx}_j^T\bveps|>L_1\lambda_0/2)
 \leq \widetilde{C}_3\exp\{-\widetilde{C}_4n^{\min\{\alpha_1\alpha_2/(\alpha_1+\alpha_2), 1\}}\lambda_0^2\} 
   \end{align*}
for all $0<L_1\lambda_0<2$.

{\bf Case 2:} $p+1\leq j\leq \widetilde{p}$. In this case, $\widetilde{\bx}_j=(X_{1k}X_{1\ell}, \cdots, X_{nk}X_{n\ell})^T$.  Thus $n^{-1}\widetilde{\bx}_j^T\bveps=n^{-1}\sum_{i=1}^n X_{ik}X_{i\ell}\varepsilon_i$ with some $1\leq k<\ell\leq p$ if $p+1\leq j\leq \widetilde{p}$.
By Lemma \ref{lem: sub-exp}, we have
$P(|X_{ik}X_{i\ell}\varepsilon_i|>t)\leq 4c_1\exp\{-c_1^{-1}t^{\alpha_1\alpha_2/(\alpha_1+2\alpha_2)}\}$
for all $1\leq i\leq n$ and $1\leq k<j\leq p$.  Note that $E(X_{ik}X_{i\ell}\varepsilon_i)=0$. Thus it follows from Lemma \ref{lemma-Hao-Zhang-extend} and $\alpha_1\alpha_2/(\alpha_1+2\alpha_2) \leq 1$ that there exist some positive constants $\widetilde{C}_5$ and $\widetilde{C}_6$ such that
    \begin{align*}
        P(|n^{-1}\widetilde{\bx}_j^T\bveps|>L_1\lambda_0/2)\leq \widetilde{C}_5\exp\{-\widetilde{C}_6n^{\alpha_1\alpha_2/(\alpha_1+2\alpha_2)}\lambda_0^2\}
   \end{align*}
 for all $0<L_1\lambda_0<2$.

Under the assumption that $\alpha_1\alpha_2/(\alpha_1+2\alpha_2)\leq 1$, we have  $\alpha_1\alpha_2/(\alpha_1+2\alpha_2)\leq \min\{\alpha_1\alpha_2/(\alpha_1+\alpha_2), 1\}$. Thus combining Cases 1 and 2 above along with \eqref{eq: bX-err-bound-1} leads to
    \begin{align*}
        P(\|n^{-1}\widetilde{\bX}^T\bveps\|_{\infty}>L_1\lambda_0/2)
  \leq \sum_{j=1}^{\widetilde{p}}P(|n^{-1}\widetilde{\bx}_j^T\bveps|>L_1\lambda_0/2)
  \leq \widetilde{C}_1p^2\exp\{-\widetilde{C}_2n^{\alpha_1\alpha_2/(\alpha_1+2\alpha_2)}\lambda_0^2\}
   \end{align*}
for all $0<L_1\lambda_0<2$, where $\widetilde{C}_1=\max\{\widetilde{C}_3, \widetilde{C}_5\}$ and $\widetilde{C}_2=\min\{\widetilde{C}_4, \widetilde{C}_6\}$.  Here we have used the fact that $\widetilde{p}=p(p+1)/2\leq p^2$.  Set $\lambda_0=\widetilde{c}_0\{\log p/ n^{\alpha_1\alpha_2/(\alpha_1+2\alpha_2)}\}^{1/2}$ with $\widetilde{c}_0$ some positive constant. Then $0<L_1\lambda_0<2$ for all $n$ sufficiently large.  Thus, with the above choice of $\lambda_0$, it holds that
    \begin{align*}
        P(\|n^{-1}\widetilde{\bX}^T\bveps\|_{\infty}>L_1\lambda_0/2)
  \leq o(p^{-c_4}),
   \end{align*}
where $c_4$ is some positive constant. Note that $P(A) \leq P(A|B) + P(B^c) $ and $P(A|B) \leq P(A)/P(B)$ for any events $A$ and $B$ with $P(B) > 0$. Thus,
\begin{align*}
 P(\|n^{-1}\widetilde{\bZ}^T\bveps\|_{\infty}>\lambda_0/2) & \leq P(\|n^{-1}\widetilde{\bZ}^T\bveps\|_{\infty}>\lambda_0/2|\mathcal{E}_4 ) + P(\mathcal{E}_4^c)\\
&\leq  P(\|n^{-1}\widetilde{\bX}^T\bveps\|_{\infty}>L_1\lambda_0/2| \mathcal{E}_4 ) + P(\mathcal{E}_4^c) \\
&\leq P(\|n^{-1}\widetilde{\bX}^T\bveps\|_{\infty}>L_1\lambda_0/2)/P(\mathcal{E}_4 ) + P(\mathcal{E}_4^c)\\
&\leq o(p^{-c_4}) + a_n,
\end{align*}
which completes the proof of Theorem \ref{Th:global}.

\subsection*{D.6. Proof of Theorem \ref{Th:RE condition}}

We first prove that the diagonal entries $\bD_{mm}$'s of the scale matrix $\bD$ are bounded between two positive constants $L_1\leq L_2$ with significant probability.  
Since $P(|X_{ij}|> t) \leq c_1\exp(-c_1^{-1}t^{\alpha_1})$ for any $t>0$ and  all $1\leq i\leq n$ and $1\leq j\leq p$,
by Lemma \ref{lemma-X-concentration} and noting that 
 $EX_{ij}^2=1$,
there exist some positive constants $\widetilde C_1$ and $\widetilde C_2$ such that
     \begin{align}\label{eq: norm-sub-Gaussian}
    P&(1/2 \leq n^{-1/2}\|\widetilde\bx_j\|_2 \leq \sqrt{7}/2)
    = P\{-3/4 \leq n^{-1}\sum_{i=1}^n[ EX_{ij}^2-X_{ij}^2] \leq 3/4 \} \nonumber\\
    & =  1-P\{|n^{-1}\sum_{i=1}^n[X_{ij}^2- EX_{ij}^2]|> 3/4\}
  \geq 1-\widetilde C_1\exp(-\widetilde C_2n^{\min\{\alpha_1/2, 1\}})
   \end{align}
for all $1\leq j\leq p$.

Since $\var(X_{ik}X_{i\ell})$ is a diagonal entry of the population covariance matrix $\widetilde \bSig$, it follows from Condition \ref{con: eigen} that $\var(X_{ik}X_{i\ell}) \geq K>0$ for all $1\leq k <\ell \leq p$. Thus,  there exists a constant $0<K_0\leq 1$ such that $E(X_{ik}^2X_{i\ell}^2) \geq \var(X_{ik}X_{i\ell}) \geq K>K_0$ for all $1\leq k <\ell \leq p$. Meanwhile, it follows from  $X_{ik}^2X_{i\ell}^2 \leq (X_{ik}^4 + X_{i\ell}^4)/2$ and Lemma \ref{lem: Sub-exp-bound} that $E(X_{ik}^2X_{i\ell}^2) \leq \widetilde C_3$, where  $\widetilde C_3 \geq K_0$ is some positive constant.     
Note that $P(|X_{ij}|> t) \leq c_1\exp(-c_1^{-1}t^{\alpha_1})$ for any $t>0$ and all $1\leq i\leq n$ and $1\leq j\leq p$.  Thus it follows from Lemma \ref{lemma-X-concentration} that there exist some positive constants $\widetilde C_4$ and $\widetilde C_5$ such that for all $1\leq k<\ell\leq p$,
     \begin{align}\label{eq: norm-exp}
   \nonumber & P\Big( \sqrt{K_0}/2  \leq n^{-1/2}\|\widetilde\bx_k\circ\widetilde\bx_{\ell}\|_2 \leq \sqrt{7\widetilde C_3}/2\Big) \\
   \nonumber &\geq   P\Big( \sqrt{K_0}/2  \leq  n^{-1/2}\|\widetilde\bx_k\circ\widetilde\bx_{\ell}\|_2 \leq   \sqrt{3K_0/4 + \widetilde C_3}\Big)\\
    \nonumber & \geq    1-P\Big\{\big|n^{-1}\sum_{i=1}^n[X_{ik}^2X_{i\ell}^2-E(X_{ik}^2X_{i\ell}^2)]\big|> 3K_0/4\Big\}\\
   &\geq   1- \widetilde C_4\exp(-\widetilde C_5n^{\min\{\alpha_1/4, 1\}}).
   \end{align}

Let $L_1=2^{-1}\min\{1,K_0^{1/2}\}=\sqrt{K_0}/2$ and $L_2=\sqrt{7}/2\max\{1, \widetilde{C}_3^{1/2}\}$.  Then combining \eqref{eq: norm-sub-Gaussian} with \eqref{eq: norm-exp} yields that with probability at least $ 1-\widetilde{C}_1p\exp(-\widetilde{C}_2n^{\min\{\alpha_1/2, 1\}})- \widetilde{C}_4p^2\exp(-\widetilde{C}_5n^{\min\{\alpha_1/4, 1\}})$, it holds that
     \begin{align}\label{eq: D-diag}
   \L_1\leq \min_{1\leq j\leq\widetilde p}|\bD_{jj}| \leq \max_{1\leq j\leq\widetilde p}|\bD_{jj}| \leq L_2,
     \end{align}
which shows that $\bD_{mm}$'s are bounded away from zero and infinity with large probability.

We proceed to show that the first two parts of Theorem \ref{Th:RE condition} hold with significant probability.
For any $0<\epsilon<1$, define an event 
$\mathcal{E}_5=\{\|n^{-1}\widetilde{\bX}^T\widetilde{\bX}-\widetilde{\bSig}\|_\infty \leq \epsilon\}$,
where $\|\cdot\|_\infty$ stands for the entrywise matrix infinity norm and $\widetilde{\bX}$ and $\widetilde{\bSig}$ are defined in Section \ref{sec: RE condition}. Recall that $\widetilde{p}=p(p+1)/2$. 
Since $P(|X_{ij}|> t) \leq c_1\exp(-c_1^{-1}t^{\alpha_1})$ for any $t>0$ and  all $1\leq i\leq n$ and $1\leq j\leq p$, it follows from Lemma \ref{lemma-X-concentration} that there exist some positive constants $\widetilde{C}_6$ and $\widetilde{C}_7$ such that
  \begin{align}\label{eq: prob-Omega}
     P(\mathcal{E}_5)
  &= 1-P(|(n^{-1}\widetilde{\bX}^T\widetilde{\bX}-\widetilde{\bSig})_{jk}|> \epsilon\,\,\mbox{for some}\,\, (j, k)\,\,\mbox{with}\,\, 1\leq j, k\leq \widetilde{p}) \nonumber\\
  &\geq 1-\sum_{j=1}^{\widetilde{p}}\sum_{k=1}^{\widetilde{p}}P(|(n^{-1}\widetilde{\bX}^T\widetilde{\bX}-\widetilde{\bSig})_{jk}|> \epsilon) \nonumber\\
 & \geq 1- \widetilde{C}_6\widetilde{p}^2\exp(-\widetilde{C}_7n^{\min\{\alpha_1/4, 1\}}\epsilon^2)
  \end{align}
for any $0< \epsilon< 1$, where $\bA_{jk}$ denotes the $(j,k)$-entry of a matrix $\bA$.

Next, we show that conditional on the event 
$\mathcal{E}_5$, the desired inequalities in Theorem \ref{Th:RE condition} hold.   From now on, we condition on the event 
$\mathcal{E}_5$.  Note that $ (n^{-1/2}\|\widetilde{\bX}\bdelta\|_2)^2=\bdelta^T(n^{-1}\widetilde{\bX}^T\widetilde{\bX}-\widetilde{\bSig})\bdelta + \bdelta^T\widetilde{\bSig}\bdelta$.
Let $\bdelta_J$ be the subvector of $\bdelta$ formed by putting all nonzero components of $\bdelta$ together.
For any $\bdelta$ satisfying $\|\bdelta\|_2=1$ and $\|\bdelta\|_0<2s$, by the Cauchy-Schwarz inequality we have
  \begin{equation}\label{eq:errbound1}
     |\bdelta^T(n^{-1}\widetilde{\bX}^T\widetilde{\bX}-\widetilde{\bSig})\bdelta |
    \leq \epsilon \|\bdelta\|_1^2 =\epsilon \|\bdelta_J\|_1^2\leq \epsilon \|\bdelta_J\|_0\|\bdelta_J\|_2^2
    =\epsilon \|\bdelta\|_0\|\bdelta\|_2^2 < 2s\epsilon.
  \end{equation}
It follows that $ (n^{-1/2}\|\widetilde{\bX}\bdelta\|_2)^2>\bdelta^T\widetilde{\bSig}\bdelta-2s\epsilon$ for any $\bdelta$ satisfying $\|\bdelta\|_2=1$ and $\|\bdelta\|_0<2s$.  Thus we derive
\begin{equation}\label{eq1}
     \min_{\|\bdelta\|_2=1, \|\bdelta\|_0<2s}(n^{-1/2}\|\widetilde{\bX}\bdelta\|_2)^2
    \geq  \min_{\|\bdelta\|_2=1, \|\bdelta\|_0<2s}(\bdelta^T\widetilde{\bSig}\bdelta)-2s\epsilon
    \geq  K-2s\epsilon,
\end{equation}
where the last inequality follows from Condition \ref{con: eigen}.

Meanwhile, for any $\bdelta\neq \bzero$ we have
  \begin{align*}
     \left(\frac{n^{-1/2}\|\widetilde{\bX}\bdelta\|_2}{\|\bdelta_1\|_2\vee \|\widetilde{\bdelta}_2\|_2}\right)^2
     &= \frac{\bdelta^T(n^{-1}\widetilde{\bX}^T\widetilde{\bX}-\widetilde{\bSig})\bdelta}{\|\bdelta_1\|_2^2\vee \|\widetilde{\bdelta}_2\|_2^2}
        +\frac{\bdelta^T\widetilde{\bSig}\bdelta}{\|\bdelta_1\|_2^2\vee \|\widetilde{\bdelta}_2\|_2^2} \nonumber\\
     &\geq \frac{\bdelta^T(n^{-1}\widetilde{\bX}^T\widetilde{\bX}-\widetilde{\bSig})\bdelta}{\|\bdelta_1\|_2^2\vee \|\widetilde{\bdelta}_2\|_2^2}
        +\frac{\bdelta^T\widetilde{\bSig}\bdelta}{\|\bdelta\|_2^2}.
  \end{align*}
Under the additional condition $\|\bdelta_2\|_1\leq 7\|\bdelta_1\|_1$, by the first inequality of \eqref{eq:errbound1} it holds that
  \begin{eqnarray*}
      \left|\frac{\bdelta^T(n^{-1}\widetilde{\bX}^T\widetilde{\bX}-\widetilde{\bSig})\bdelta}{\|\bdelta_1\|_2^2\vee \|\widetilde{\bdelta}_2\|_2^2}\right|
  \leq \frac{\epsilon \|\bdelta\|_1^2}{\|\bdelta_1\|_2^2}
  = \frac{\epsilon (\|\bdelta_1\|_1+\|\bdelta_2\|_1)^2}{\|\bdelta_1\|_2^2}
  \leq  \frac{64\epsilon \|\bdelta_1\|_1^2}{\|\bdelta_1\|_2^2}
  \leq 64s\epsilon,
  \end{eqnarray*}
where the last inequality follows from the Cauchy-Schwarz inequality.
This entails that for any $\bdelta\neq 0$ with $\|\bdelta_2\|_1\leq 7\|\bdelta_1\|_1$,
  \begin{eqnarray*}
     \left(\frac{n^{-1/2}\|\widetilde{\bX}\bdelta\|_2}{\|\bdelta_1\|_2\vee \|\widetilde{\bdelta}_2\|_2}\right)^2
    = \frac{n^{-1}\bdelta^T\widetilde{\bX}^T\widetilde{\bX}\bdelta}{\|\bdelta_1\|_2^2\vee \|\widetilde{\bdelta}_2\|_2^2} \geq \frac{\bdelta^T\widetilde{\bSig}\bdelta}{\|\bdelta\|_2^2} - 64s\epsilon.
  \end{eqnarray*}
 Thus, by Condition \ref{con: eigen} we have
  \begin{align}\label{eq2}
  \nonumber
      \min_{\bdelta\neq 0,\, \|\bdelta_2\|_1\leq 7\|\bdelta_1\|_1} \left(\frac{n^{-1/2}\|\widetilde{\bX}\bdelta\|_2}{\|\bdelta_1\|_2\vee \|\widetilde{\bdelta}_2\|_2}\right)^2
      &\geq \min_{\bdelta\neq 0,\, \|\bdelta_2\|_1\leq 7\|\bdelta_1\|_1} \frac{\bdelta^T\widetilde{\bSig}\bdelta}{\|\bdelta\|_2^2}
      -64s\epsilon\\
      &\geq K-64s\epsilon.
  \end{align}


Recall that $s=O(n^{\xi_0})$ with $0\leq \xi_0<\min\{\alpha_1/8, 1/2\}$ by assumption and thus $s\leq \widetilde{C}_{8}n^{\xi_0}$ for some positive constant $\widetilde{C}_{8}$. Take $\epsilon=Kn^{-\xi_0}/\widetilde{C}_{9}$ with $\widetilde{C}_{9}$ some sufficiently large positive constant such that $\epsilon\in (0, 1)$ and $K-64s\epsilon>0$.   In view of \eqref{eq: D-diag}, \eqref{eq: prob-Omega}, \eqref{eq1}, and \eqref{eq2}, since $\log p = o(n^{\min\{\alpha_1/4, 1\}-2\xi_0})$ by assumption, we obtain that
  \begin{align*}
    a_n &=\widetilde{C}_1p\exp(-\widetilde{C}_2n^{\min\{\alpha_1/2, 1\}}) + \widetilde{C}_4p^2\exp(-\widetilde{C}_5n^{\min\{\alpha_1/4, 1\}}) \\
      & \quad+\widetilde{C}_6\widetilde{p}^2\exp(-\widetilde{C}_7K^2\widetilde{C}_{9}^{-2}n^{\min\{\alpha_1/4, 1\}-2\xi_0})=o(1)
  \end{align*}
with the above choice of $\epsilon$, and that with probability at least $1-a_n$, the desired results in the theorem hold with $\kappa_0 = K(1-2\widetilde{C}_{8}/\widetilde{C}_{9})$ and $\kappa = K(1-64\widetilde{C}_{8}/\widetilde{C}_{9})$. This concludes the proof of Theorem \ref{Th:RE condition}.

\section*{Appendix E: Some technical lemmas and their proofs} \label{AppE}

\renewcommand{\theequation}{E.\arabic{equation}}
\setcounter{equation}{0}

\begin{lemma}\label{lem: sub-exp}
Let $W_1$ and $W_2$ be two random variables such that
$ P(|W_1|>t)\leq \widetilde{C}_1\exp(-\widetilde{C}_2t^{\alpha_1})$ and $P(|W_2|>t)\leq \widetilde{C}_3\exp(-\widetilde{C}_4t^{\alpha_2})$
for all $t>0$, where $\alpha_1$, $\alpha_2$, and $\widetilde{C}_i$'s are some positive constants. Then
$ P(|W_1W_2|>t)\leq \widetilde{C}_5\exp(-\widetilde{C}_6t^{\alpha_1\alpha_2/(\alpha_1+\alpha_2)})$ for all $t>0$,
with $\widetilde{C}_5=\widetilde{C}_1+\widetilde{C}_3$ and $\widetilde{C}_6=\min\{\widetilde{C}_2, \widetilde{C}_4\}$.
\end{lemma}

{\bf Proof.} For any $t>0$, we have
\begin{align*}
    & P(|W_1W_2|>t)
   \leq  P(|W_1|>t^{\alpha_2/(\alpha_1+\alpha_2)}) + P(|W_2|>t^{\alpha_1/(\alpha_1+\alpha_2)})  \\
   & \leq  \widetilde{C}_1\exp(-\widetilde{C}_2t^{\alpha_1\alpha_2/(\alpha_1+\alpha_2)}) + \widetilde{C}_3\exp(-\widetilde{C}_4t^{\alpha_1\alpha_2/(\alpha_1+\alpha_2)}) \\
   & \leq  \widetilde{C}_5\exp(-\widetilde{C}_6t^{\alpha_1\alpha_2/(\alpha_1+\alpha_2)})
\end{align*}
by setting $\widetilde{C}_5=\widetilde{C}_1+\widetilde{C}_3$ and $\widetilde{C}_6=\min\{\widetilde{C}_2, \widetilde{C}_4\}$.

\medskip

\begin{lemma}\label{lem: Sub-exp-bound}
Let $W$ be a nonnegative random variable such that
$P(W>t)\leq \widetilde{C}_1\exp(-\widetilde{C}_2t^{\alpha})$ for all $t>0$, where $\alpha$ and $\widetilde{C}_i$'s are some positive constants. Then it holds that $E(e^{\widetilde{C}_3W^{\alpha}}) \leq \widetilde{C}_4$,
$E(W^{\alpha m}) \leq \widetilde{C}_3^{-m}\widetilde{C}_4m!$ for any integer $m \geq 0$ with $\widetilde{C}_3=\widetilde{C}_2/2$ and $\widetilde{C}_4=1+\widetilde{C}_1$, and $E(W^k)\leq \widetilde{C}_5$ for any integer $k\geq 1$, where constant $\widetilde{C}_5$ depends on $k$  and $\alpha$.
\end{lemma}

{\bf Proof.} Let $F(t)$ be the cumulative distribution function of $W$.  Then for all $t>0$, $1-F(W) = P(W>t) \leq \widetilde{C}_1\exp(-\widetilde{C}_2t^{\alpha})$.  Recall that $W$ is a nonnegative random variable.  Thus, for any $0<T<\widetilde{C}_2$, by integration by parts we have
  \begin{align*}
       E(e^{TW^{\alpha}})
    &= -\int_0^{\infty} e^{Tt^{\alpha}} d[1-F(t)]
    =  1+ \int_0^{\infty} T\alpha  t^{\alpha-1} e^{Tt^{\alpha}}[1-F(t)]\,dt \\
 &  \leq  1+ \int_0^{\infty} T\alpha  t^{\alpha-1}\cdot \widetilde{C}_1e^{-(\widetilde{C}_2-T)t^{\alpha}}\,dt
    = 1+ \frac{{T}\widetilde{C}_1}{\widetilde{C}_2-T}.
  \end{align*}
Then, taking $\widetilde{C}_3=T=\widetilde{C}_2/2$ and $\widetilde{C}_4=1+\widetilde{C}_1$ proves the first desired result.

Note that $\widetilde{C}_3^mE(W^{\alpha m})/m!\leq \sum_{k=0}^{\infty}\widetilde{C}_3^kE(W^{\alpha k})/k!= E(e^{\widetilde{C}_3W^{\alpha}})$ for any nonnegative integer $m$.  Thus $E(W^{\alpha m})\leq \widetilde{C}_3^{-m}\widetilde{C}_4m!$, which proves the second desired result.

For any integer $k\geq 1$, there exists an integer $m\geq 1$ such that $k< \alpha m$. Then applying H\"{o}lder's inequality gives
  \begin{align*}
     E(W^k) 
 \leq & \left\{E[(W^k)^{\alpha m/k}]\right\}^{k/(\alpha m)} \left\{E[1^{\alpha m/(\alpha m-k)}]\right\}^{(\alpha m - k)/(\alpha m)} \\
  = & \left\{E(W^{\alpha m})\right\}^{k/(\alpha m)}
  \leq \left(\widetilde{C}_3^{-m}\widetilde{C}_4m!\right)^{k/(\alpha m)}.
  \end{align*}
Thus the $k$th moment of $W$ is bounded by a constant $\widetilde{C}_5$, which depends on $k$ and $\alpha$.  This proves the third desired result.

\medskip
\begin{lemma}\label{lem: Sub-exp-large}
Let $W$ be a nonnegative random variable with tail probability $P(W>t)\leq \widetilde{C}_1\exp(-\widetilde{C}_2t^{\alpha})$
for all $t>0$, where $\alpha$ and $\widetilde{C}_i$'s are some positive constants.  If constant $\alpha \geq 1$, then
$E(e^{\widetilde{C}_3W}) \leq \widetilde{C}_4$ and
$E(W^{ m}) \leq \widetilde{C}_3^{-m}\widetilde{C}_4m!$ for any integer $m \geq 0$ with $\widetilde{C}_3=\widetilde{C}_2/2$ and $\widetilde{C}_4=e^{\widetilde{C}_2/2}+\widetilde{C}_1e^{-\widetilde{C}_2/2}$.
\end{lemma}

{\bf Proof.} Let $F(t)$ be the cumulative distribution function of nonnegative random variable $W$. Then $1-F(t)=P(W>t)\leq \widetilde{C}_1\exp(-\widetilde{C}_2t^{\alpha})$ for all $t\geq 1$.   If $\alpha\geq 1$, then $t\leq t^{\alpha}$ for all $t\geq 1$ and thus $1-F(t)\leq \widetilde{C}_1\exp(-\widetilde{C}_2t)$ for all $t\geq 1$.  Define $\widetilde{C}_3=\widetilde{C}_2/2$ and $\widetilde{C}_4=e^{\widetilde{C}_2/2} + \widetilde{C}_1e^{-\widetilde{C}_2/2}$. By integration by parts, we deduce
  \begin{align*}
     E(e^{\widetilde{C}_3 W})
   =& -\int_0^{\infty} e^{\widetilde{C}_3 t} d [1-F(t)]
   = 1+ \int_0^{\infty} \widetilde{C}_3e^{\widetilde{C}_3 t}[1-F(t)]dt \\
   = & 1+ \int_0^{1} \widetilde{C}_3e^{\widetilde{C}_3 t}[1-F(t)]dt  +  \int_1^{\infty} \widetilde{C}_3e^{\widetilde{C}_3 t}[1-F(t)]dt\\
   \leq & 1+ \int_0^{1} \widetilde{C}_3e^{\widetilde{C}_3 t}dt  +  \int_1^{\infty} \widetilde{C}_1\widetilde{C}_3e^{(\widetilde{C}_3-\widetilde{C}_2) t}dt
    =e^{\widetilde{C}_2/2} + \widetilde{C}_1e^{-\widetilde{C}_2/2}
    =\widetilde{C}_4,
  \end{align*}
which proves the first desired result.

Note that $\widetilde{C}_3^mE(W^{m})/m!\leq \sum_{k=0}^{\infty}\widetilde{C}_3^kE(W^{k})/k!= E(e^{\widetilde{C}_3W})$ for any nonnegative integer $m$.  Thus $E(W^{m})\leq \widetilde{C}_3^{-m}\widetilde{C}_4m!$, which proves the second desired result.

\medskip
\begin{lemma}\label{lem: inequality}
For any real numbers $b_1, b_2\geq 0$ and $\alpha>0$, it holds that  $(b_1+b_2)^{\alpha}\leq C_{\alpha}(b_1^{\alpha}+b_2^{\alpha})$ with $C_{\alpha}=1$ if $0<\alpha\leq 1$ and $2^{\alpha-1}$ if $\alpha>1$.
\end{lemma}

{\bf Proof.} We first consider the case of $0<\alpha\leq 1$.  It is trivial if $b_1=0$ or $b_2=0$.  Assume that both $b_1$ and $b_2$ are positive.  Since $0< b_1/(b_1+b_2)< 1$, we have $[b_1/(b_1+b_2)]^{\alpha}\geq b_1/(b_1+b_2)$. Similarly, it holds that $[b_2/(b_1+b_2)]^{\alpha}\geq b_2/(b_1+b_2)$. Combining these two results yields
  \begin{eqnarray*}
      \left(\frac{b_1}{b_1+b_2}\right)^{\alpha} + \left(\frac{b_2}{b_1+b_2}\right)^{\alpha}
    \geq \frac{b_1}{b_1+b_2} + \frac{b_2}{b_1+b_2}=1,
  \end{eqnarray*}
which implies that $ (b_1+b_2)^{\alpha} \leq b_1^{\alpha}+b_2^{\alpha}$.

Next, we deal with the case of $\alpha>1$.  Since $x^{\alpha}$ is a convex function on $[0, \infty)$ for a given $\alpha>1$, we have $[(b_1+b_2)/2]^{\alpha}\leq (b_1^{\alpha}+b_2^{\alpha})/2 $, which ensures that $(b_1+b_2)^{\alpha}\leq 2^{\alpha-1}(b_1^{\alpha}+b_2^{\alpha})$. Combining the two cases above leads to the desired result.  

\medskip
\begin{lemma}[Lemma B.4 in \citet{hao2014interaction}]\label{lemma-Hao-Zhang}
Let $W_1, \cdots, W_n$ be independent random variables with $E W_i = 0$ and $E e^{T|W_i|^{\alpha}} \leq A$ for some constants $T, A>0$ and $0<\alpha\leq 1$. Then for $0<\epsilon\leq 1$,
$P(|n^{-1}\sum_{i=1}^nW_i|> \epsilon) \leq \widetilde{C}_1\exp(-\widetilde{C}_2n^{\alpha}\epsilon^2)$ with $\widetilde{C}_1, \widetilde{C}_2 > 0$ some constants.
\end{lemma}

\medskip
\begin{lemma}\label{lemma-Hao-Zhang-extend}
Let $W_1, \cdots, W_n$ be independent random variables with tail probability $P(|W_i|>t)\leq \widetilde{C}_1\exp(-\widetilde{C}_2t^{\alpha})$ for all $t>0$, where $\alpha$ and $\widetilde{C}_i$'s are some positive constants.  Then there exist some positive constants $\widetilde{C}_3$ and $\widetilde{C}_4$ such that
  \begin{align}
     P\{|n^{-1}\sum_{i=1}^n(W_i-EW_i)|> \epsilon\}
\leq \widetilde{C}_3\exp(-\widetilde{C}_4n^{\min\{\alpha, 1\}}\epsilon^2)
  \end{align}
for $0<\epsilon\leq 1$.
\end{lemma}

{\bf Proof.} 
Define $\widetilde{W}_i=W_i-EW_i$. Then by the triangle inequality and the property of expectation, we have
  \begin{align}\label{eq: Wi-triangle}
    |\widetilde{W}_i|
   =|W_i-EW_i|
    \leq |W_i|+ |EW_i|
    \leq |W_i| + E|W_i|.
  \end{align}
Next, we consider two cases.

Case 1: $0<\alpha\leq 1$.  It follows from Lemma \ref{lem: Sub-exp-bound} that $E(e^{T|W_{i}|^{\alpha}})\leq 1+\widetilde{C}_1$ and 
$E|W_{i}|\leq C_0$ for all $1\leq i\leq n$, where $T=\widetilde{C}_2/2$ and $C_0$ is some positive constant.
In view of \eqref{eq: Wi-triangle} and by Lemma \ref{lem: inequality}, we have
$ |\widetilde{W}_i|^{\alpha}\leq (|W_i|+ E|W_i|) ^{\alpha}\leq |W_i|^{\alpha}+ (E|W_i|)^{\alpha}$.
This ensures
  \begin{eqnarray*}
   E(e^{T|\widetilde{W}_i|^{\alpha}})
  \leq e^{T(E|W_i|)^{\alpha}}E(e^{T|W_i|^{\alpha}})
  \leq  e^{TC_0^{\alpha}}(1+\widetilde{C}_1).
  \end{eqnarray*}
Thus, by Lemma \ref{lemma-Hao-Zhang}, there exist some positive constants  $\widetilde{C}_5$ and $\widetilde{C}_6$ such that
  \begin{align}\label{eq: W-Case1}
      P(|n^{-1}\sum_{i=1}^n [W_i-EW_i]|>\epsilon)
   = P(|n^{-1}\sum_{i=1}^n \widetilde{W}_i|>\epsilon)
   \leq \widetilde{C}_5\exp\left(-\widetilde{C}_6n^{\alpha}\epsilon^2\right)
  \end{align}
for any $0<\epsilon\leq 1$.

Case 2: $\alpha>1$.  In view of \eqref{eq: Wi-triangle}, it follows from Lemma \ref{lem: inequality} and Jensen's inequality that for each integer $m\geq 2$,
  \begin{align}\label{eq: W4}
      & E(|\widetilde{W}_i|^m)
   \leq  E[(|W_i| + E|W_i|)^m]
  \leq   2^{m-1}E\left[|W_i|^m+(E|W_i|)^m\right] \nonumber\\
  = & 2^{m-1} [E(|W_i|^m)+(E|W_i|)^m]
  \leq  2^{m-1} [E(|W_i|^m)+E(|W_i|^m)]
      = 2^{m}E(|W_i|^m).
  \end{align}
Recall that $P(|W_i|>t)\leq \widetilde{C}_1\exp(-\widetilde{C}_2 t^{\alpha})$ for all $t>0$ and $\alpha>1$. By Lemma \ref{lem: Sub-exp-large}, there exist some positive constants  $\widetilde{C}_7$ and $\widetilde{C}_8$ such that 
$E(|W_i|^m)\leq m!\widetilde{C}_7^m \widetilde{C}_8$.
This together with \eqref{eq: W4} gives
  \begin{align*}
      E(|\widetilde{W}_i|^m)\leq m! (2\widetilde{C}_7)^{m-2} (8\widetilde{C}_7^2\widetilde{C}_8)/2
  \end{align*}
for all $m\geq 2$. Thus an application of Bernstein's inequality (Lemma 2.2.11 in \citet{van1996weak}) yields
  \begin{align}\label{eq: W-Case2}
      & P\{|n^{-1}\sum_{i=1}^n (W_i-EW_i)|>\epsilon\}
   = P(|n^{-1}\sum_{i=1}^n \widetilde{W}_i|>\epsilon) \nonumber \\
   & \leq  2\exp\left(-\frac{n\epsilon^2}{16\widetilde{C}_7^2\widetilde{C}_8 + 4\widetilde{C}_7\epsilon}\right)
   \leq 2\exp\left(-\frac{n\epsilon^2}{16\widetilde{C}_7^2\widetilde{C}_8 + 4\widetilde{C}_7}\right)
  \end{align}
for any $0<\epsilon<1$. Let $\widetilde{C}_3=\max\{\widetilde{C}_5, 2\}$ and $\widetilde{C}_4=\min\{\widetilde{C}_6, (16\widetilde{C}_7^2\widetilde{C}_8 + 4\widetilde{C}_7)^{-1}\}$. Combining \eqref{eq: W-Case1} and \eqref{eq: W-Case2} completes the proof of Lemma \ref{lemma-Hao-Zhang-extend}.

\medskip
\begin{lemma}\label{lemma-X-concentration}
Assume that for each $1 \leq j \leq p$, $X_{1j}, \cdots, X_{nj}$ are $n$ i.i.d. random variables satisfying
$P(|X_{1j}|>t)\leq \widetilde{C}_1\exp(-\widetilde{C}_2t^{\alpha_1})$ for any $t>0$, where
$\widetilde{C}_1, \widetilde{C}_2$ and $\alpha_1$ are some positive constants. Then for any $0<\epsilon<1$, we have
\begin{align}
& P\left\{\left|n^{-1}\sum_{i=1}^n \left[X_{ij}X_{ik}-E(X_{ij}X_{ik})\right]\right|> \epsilon\right\} \leq \widetilde{C}_3\exp(-\widetilde{C}_4n^{\min\{\alpha_1/2, 1\}}\epsilon^2), \label{eq: inter2}\\
& P\left\{\left|n^{-1}\sum_{i=1}^n \left[X_{ij}X_{ik}X_{i\ell}-E(X_{ij}X_{ik}X_{i\ell})\right]\right|> \epsilon\right\} \leq\widetilde{C}_5\exp(-\widetilde{C}_6n^{\min\{\alpha_1/3, 1\}}\epsilon^2), \label{eq: inter3}\\
& P\left\{\left|n^{-1}\sum_{i=1}^n \left[X_{ik}X_{i\ell}X_{ik'}X_{i\ell'}-E(X_{ik}X_{i\ell}X_{ik'}X_{i\ell'})\right]\right|> \epsilon\right\} \nonumber \\
& \quad\quad\leq \widetilde{C}_7\exp(-\widetilde{C}_8n^{\min\{\alpha_1/4, 1\}}\epsilon^2), \label{eq: inter4}
  \end{align}
where $1\leq j, k, \ell, k', \ell' \leq p$ and $\widetilde{C}_i$'s are some positive constants.
\end{lemma}

{\bf Proof.} The proofs for inequalities \eqref{eq: inter2}--\eqref{eq: inter4} are similar. To save space, we only show the inequality \eqref{eq: inter4} here.
Since $P(|X_{ij}|>t)\leq \widetilde{C}_1\exp(-\widetilde{C}_2 t^{\alpha_1})$ for all $t>0$ and all $i$ and $j$, it follows from Lemma \ref{lem: sub-exp} that $X_{ik}X_{i\ell}X_{ik'}X_{i\ell'}$ admits tail probability $P(|X_{ik}X_{i\ell}X_{ik'}X_{i\ell'}|>t)\leq 4\widetilde{C}_1\exp(-\widetilde{C}_2t^{\alpha_1/4})$.  By Lemma \ref{lemma-Hao-Zhang-extend}, there exist some positive constants $\widetilde{C}_3$ and $\widetilde{C}_4$ such that
 \begin{align*}
     P(|n^{-1}\sum_{i=1}^n [X_{ik}X_{i\ell}X_{ik'}X_{i\ell'}-E(X_{ik}X_{i\ell}X_{ik'}X_{i\ell'})]|>\epsilon)
   \leq \widetilde{C}_3\exp\left(-\widetilde{C}_4n^{\min\{\alpha_1/4, 1\}}\epsilon^2\right)
  \end{align*}
for any $0<\epsilon<1$, which concludes the proof of \eqref{eq: inter4}.

\medskip
\begin{lemma}\label{lem: Ajsq}
Let $A_j$'s with $j \in \mathcal{D} \subset \{1, \cdots, p\}$ satisfy $\max_{j\in\mathcal{D}}|A_j|\leq L_3$ for some constant $L_3 > 0$, and $\widehat{A}_j$ be an estimate of $A_j$ based on a sample of size $n$ for each $j\in \mathcal{D}$. Assume that for any constant $C > 0$, there exist constants $\widetilde{C}_1, \widetilde{C}_2 > 0$ such that
  \begin{eqnarray*}
       P\left(\max_{j\in\mathcal{D}} |\widehat{A}_j-A_j|\geq Cn^{-\kappa_1}\right)\leq |\mathcal{D}|\widetilde{C}_1\exp\left\{-\widetilde{C}_2n^{f(\kappa_1)}\right\}
  \end{eqnarray*}
with $f(\kappa_1)$ some function of $\kappa_1$.
Then for any constant $C > 0$, there exist constants $\widetilde{C}_3, \widetilde{C}_4 > 0$ such that
  \begin{eqnarray*}
      P\left(\max_{j\in\mathcal{D}} |\widehat{A}_j^2-A_j^2|\geq Cn^{-\kappa_1}\right)
         \leq |\mathcal{D}| \widetilde{C}_3\exp\left\{-\widetilde{C}_4n^{f(\kappa_1)}\right\}.
  \end{eqnarray*}
\end{lemma}

{\bf Proof.}
Note that $\max_{j\in\mathcal{D}} |\widehat{A}_j^2-A_j^2|\leq \max_{j\in\mathcal{D}} |\widehat{A}_j(\widehat{A}_j-A_j)| + \max_{j\in\mathcal{D}} |(\widehat{A}_j-A_j)A_j| $. Therefore, for any positive constant $C$,
  \begin{align} \label{eq: Ajsq}
      P(\max_{j\in\mathcal{D}} |\widehat{A}_j^2-A_j^2|\geq Cn^{-\kappa_1})
     &\leq P(\max_{j\in\mathcal{D}} |\widehat{A}_j(\widehat{A}_j-A_j)|\geq Cn^{-\kappa_1}/2)   \nonumber\\
     &\quad + P(\max_{j\in\mathcal{D}} |(\widehat{A}_j-A_j)A_j|\geq Cn^{-\kappa_1}/2).
  \end{align}
We first deal with the second term on the right hand side of \eqref{eq: Ajsq}.  Since $\max_{j\in\mathcal{D}}|A_j|\leq L_3$, we have
  \begin{align} \label{eq: Ajsq-part2}
     & P(\max_{j\in\mathcal{D}} |(\widehat{A}_j-A_j)A_j|\geq Cn^{-\kappa_1}/2)
    \leq  P(\max_{j\in\mathcal{D}} |\widehat{A}_j-A_j|L_3\geq Cn^{-\kappa_1}/2)  \nonumber\\
    =& P\{\max_{j\in\mathcal{D}} |\widehat{A}_j-A_j|\geq (2L_3)^{-1}Cn^{-\kappa_1}\}
    \leq  |\mathcal{D}|\widetilde{C}_1\exp\left\{-\widetilde{C}_2n^{f(\kappa_1)}\right\},
  \end{align}
where $\widetilde{C}_1$ and $\widetilde{C}_2$ are two positive constants.

Next, we consider the first term on the right hand side of \eqref{eq: Ajsq}. Note that
  \begin{align} \label{eq: Ajsq-part1}
  & P(\max_{j\in\mathcal{D}} |\widehat{A}_j(\widehat{A}_j-A_j)|\geq Cn^{-\kappa_1}/2) \nonumber\\
   \leq&   P(\max_{j\in\mathcal{D}} |\widehat{A}_j(\widehat{A}_j-A_j)|\geq Cn^{-\kappa_1}/2, \max_{j\in\mathcal{D}}|\widehat{A}_j|\geq L_3+Cn^{-\kappa_1}/2)   \nonumber\\
   & \quad+ P(\max_{j\in\mathcal{D}} |\widehat{A}_j(\widehat{A}_j-A_j)|\geq Cn^{-\kappa_1}/2, \max_{j\in\mathcal{D}}|\widehat{A}_j|< L_3+Cn^{-\kappa_1}/2)  \nonumber\\
   \leq&   P(\max_{j\in\mathcal{D}}|\widehat{A}_j|\geq L_3+Cn^{-\kappa_1}/2)
   + P(\max_{j\in\mathcal{D}} |\widehat{A}_j(\widehat{A}_j-A_j)|\geq Cn^{-\kappa_1}/2, \max_{j\in\mathcal{D}}|\widehat{A}_j|< L_3+C)  \nonumber\\
   \leq&   P(\max_{j\in\mathcal{D}}|\widehat{A}_j|\geq L_3+Cn^{-\kappa_1}/2)
   + P(\max_{j\in\mathcal{D}} |(L_3+C)(\widehat{A}_j-A_j)|\geq Cn^{-\kappa_1}/2).
  \end{align}
Let us bound the two terms on the right hand side of \eqref{eq: Ajsq-part1} one by one.  Since $\max_{j\in\mathcal{D}}|A_j|\leq L_3$, we have
  \begin{align}\label{eq: Ajsq-part1-1}
      & P(\max_{j\in\mathcal{D}}|\widehat{A}_j|\geq L_3+Cn^{-\kappa_1}/2)
    \leq  P(\max_{j\in\mathcal{D}}|\widehat{A}_j-A_j| +\max_{j\in\mathcal{D}}|A_j| \geq L_3+Cn^{-\kappa_1}/2)  \nonumber\\
    \leq&  P(\max_{j\in\mathcal{D}}|\widehat{A}_j-A_j| \geq 2^{-1}Cn^{-\kappa_1})
    \leq  |\mathcal{D}|\widetilde{C}_5\exp\left\{-\widetilde{C}_6n^{f(\kappa_1)}\right\},
  \end{align}
where $\widetilde{C}_5$ and $\widetilde{C}_6$ are two positive constants.  It also holds that
  \begin{align*}
      & P(\max_{j\in\mathcal{D}} |(L_3+C)(\widehat{A}_j-A_j)|\geq Cn^{-\kappa_1}/2)
     =  P\{\max_{j\in\mathcal{D}} |\widehat{A}_j-A_j|\geq (2L_3+2C)^{-1}Cn^{-\kappa_1}\} \nonumber\\
     \leq&  |\mathcal{D}|\widetilde{C}_7\exp\left\{-\widetilde{C}_8n^{f(\kappa_1)}\right\},
  \end{align*}
where $\widetilde{C}_7$ and $\widetilde{C}_8$ are two positive constants.  This, together with \eqref{eq: Ajsq}--\eqref{eq: Ajsq-part1-1},  entails
  \begin{align*}
      & P(\max_{j\in\mathcal{D}} |\widehat{A}_j^2-A_j^2|\geq Cn^{-\kappa_1})
     \leq |\mathcal{D}|\widetilde{C}_1\exp\left\{-\widetilde{C}_2n^{f(\kappa_1)}\right\}
        + |\mathcal{D}|\widetilde{C}_5\exp\left\{-\widetilde{C}_6n^{f(\kappa_1)}\right\}  \nonumber\\
       &\quad   + |\mathcal{D}|\widetilde{C}_7\exp\left\{-\widetilde{C}_8n^{f(\kappa_1)}\right\}
    \leq  |\mathcal{D}|\widetilde{C}_3\exp\left\{-\widetilde{C}_4n^{f(\kappa_1)}\right\},
  \end{align*}
where $\widetilde{C}_3=\widetilde{C}_1+\widetilde{C}_5+\widetilde{C}_7>0$ and $\widetilde{C}_4=\min\{\widetilde{C}_2, \widetilde{C}_6, \widetilde{C}_8\}>0$.

\medskip
\begin{lemma}\label{lem: AjBj-diff}
Let $\widehat{A}_j$ and $\widehat{B}_j$ be estimates of $A_j$ and $B_j$, respectively, based on a sample of size $n$ for each $j \in \mathcal{D} \subset \{1, \cdots, p\}$. Assume that for any constant $C > 0$, there exist constants $\widetilde{C}_1, \cdots, \widetilde{C}_8> 0$ except $\widetilde{C}_3,  \widetilde{C}_7\geq 0$ such that
  \begin{align*}
       & P\left(\max_{j\in\mathcal{D}} |\widehat{A}_j-A_j|\geq Cn^{-\kappa_1}\right)
            \leq |\mathcal{D}|\widetilde{C}_1\exp\left\{-\widetilde{C}_2n^{f(\kappa_1)}\right\}
   +\widetilde{C}_3\exp\left\{-\widetilde{C}_4n^{g(\kappa_1)}\right\},\\
       & P\left(\max_{j\in\mathcal{D}} |\widehat{B}_j-B_j|\geq Cn^{-\kappa_1}\right)
            \leq |\mathcal{D}|\widetilde{C}_5\exp\left\{-\widetilde{C}_6n^{f(\kappa_1)}\right\}
    + \widetilde{C}_7\exp\left\{-\widetilde{C}_8n^{g(\kappa_1)}\right\}
  \end{align*}
with $f(\kappa_1)$ and $g(\kappa_1)$ some functions of $\kappa_1$.
Then for any constant $C > 0$, there exist constants $\widetilde{C}_9, \cdots, \widetilde{C}_{12} > 0$ except $\widetilde{C}_{11}\geq 0$ such that
   \begin{align*}
      P&\left\{\max_{j\in\mathcal{D}} |(\widehat{A}_j-\widehat{B}_j)-(A_j-B_j)|\geq Cn^{-\kappa_1}\right\} \leq |\mathcal{D}|\widetilde{C}_9\exp\left\{-\widetilde{C}_{10}n^{f(\kappa_1)}\right\} \nonumber \\
         & \quad
   +\widetilde{C}_{11}\exp\left\{-\widetilde{C}_{12}n^{g(\kappa_1)}\right\}.
  \end{align*}
\end{lemma}

{\bf Proof.}  Note that $\max_{j\in\mathcal{D}} |(\widehat{A}_j-\widehat{B}_j)-(A_j-B_j)|\leq \max_{j\in\mathcal{D}} |\widehat{A}_j-A_j|+ \max_{j\in\mathcal{D}} |\widehat{B}_j-B_j|$. Thus, for any positive constant $C$,
    \begin{align*}
      &  P(\max_{j\in\mathcal{D}} |(\widehat{A}_j-\widehat{B}_j)-(A_j-B_j)|\geq Cn^{-\kappa_1})  \\
      \leq& P(\max_{j\in\mathcal{D}} |\widehat{A}_j-A_j|\geq Cn^{-\kappa_1}/2) + P(\max_{j\in\mathcal{D}} |\widehat{B}_j-B_j|\geq Cn^{-\kappa_1}/2)  \\
       \leq&    |\mathcal{D}|\widetilde{C}_1\exp\left\{-\widetilde{C}_2n^{f(\kappa_1)}\right\}
   +\widetilde{C}_3\exp\left\{-\widetilde{C}_4n^{g(\kappa_1)}\right\}
                  + |\mathcal{D}|\widetilde{C}_5\exp\left\{-\widetilde{C}_6n^{f(\kappa_1)}\right\}
   \\
   & +\widetilde{C}_7\exp\left\{-\widetilde{C}_8n^{g(\kappa_1)}\right\}\\
    \leq &|\mathcal{D}|\widetilde{C}_{9}\exp\left\{-\widetilde{C}_{10}n^{f(\kappa_1)}\right\}
   +\widetilde{C}_{11}\exp\left\{-\widetilde{C}_{12}n^{g(\kappa_1)}\right\},
  \end{align*}
where $\widetilde{C}_{9}=\widetilde{C}_1+\widetilde{C}_5>0$, $\widetilde{C}_{10}=\min\{\widetilde{C}_2, \widetilde{C}_6\}>0$,  $\widetilde{C}_{11}=\widetilde{C}_3+\widetilde{C}_7\geq 0$, and $\widetilde{C}_{12}=\min\{\widetilde{C}_4, \widetilde{C}_8\}>0$.

\medskip
\begin{lemma}\label{lem: Bj-root}
Let $B_j$'s with $j \in \mathcal{D} \subset \{1, \cdots, p\}$ satisfy $\min_{j\in\mathcal{D}}B_j\geq L_4$ for some constant $L_4 > 0$, and $\widehat{B}_j$ be an estimate of $B_j$ based on a sample of size $n$ for each $j\in \mathcal{D}$. Assume that for any constant $C > 0$, there exist constants $\widetilde{C}_1, \widetilde{C}_2 > 0$ such that
  \begin{eqnarray*}
      P\left(\max_{j\in\mathcal{D}} |\widehat{B}_j-B_j|\geq Cn^{-\kappa_1}\right)\leq |\mathcal{D}|\widetilde{C}_1\exp\left\{-\widetilde{C}_2n^{f(\kappa_1)}\right\}.
  \end{eqnarray*}
Then for any constant $C > 0$, there exist constants $\widetilde{C}_3, \widetilde{C}_4 > 0$ such that
  \begin{eqnarray*}
      P\left(\max_{j\in\mathcal{D}} \left|\sqrt{\widehat{B}_j}-\sqrt{B_j}\right|\geq Cn^{-\kappa_1}\right)
         \leq |\mathcal{D}| \widetilde{C}_3\exp\left\{-\widetilde{C}_4n^{f(\kappa_1)}\right\}.
  \end{eqnarray*}
\end{lemma}

{\bf Proof.}  Since $\min_{j\in\mathcal{D}}B_j\geq L_4>0$, there exists some constant $L_0$ such that $0<L_0<L_4$. Note that, for any positive constant $C$,
  \begin{align} \label{eq: Bj-root}
      &P(\max_{j\in\mathcal{D}} |\sqrt{\widehat{B}_j}-\sqrt{B_j}|\geq Cn^{-\kappa_1})  \nonumber \\
    \leq&  P(\max_{j\in\mathcal{D}} |\sqrt{\widehat{B}_j}-\sqrt{B_j}|\geq Cn^{-\kappa_1}, \min_{j\in\mathcal{D}} |\widehat{B}_j|\leq L_4-L_0n^{-\kappa_1}) \nonumber\\
      &\quad        +  P(\max_{j\in\mathcal{D}} |\sqrt{\widehat{B}_j}-\sqrt{B_j}|\geq Cn^{-\kappa_1}, \min_{j\in\mathcal{D}} |\widehat{B}_j|> L_4-L_0n^{-\kappa_1})  \nonumber \\
    \leq & P(\min_{j\in\mathcal{D}} |\widehat{B}_j|\leq L_4-L_0n^{-\kappa_1})
        \nonumber \\
        & \quad+ P(\max_{j\in\mathcal{D}} \frac{|\widehat{B}_j-B_j|}{|\sqrt{\widehat{B}_j}+\sqrt{B_j}|}\geq Cn^{-\kappa_1}, \min_{j\in\mathcal{D}} |\widehat{B}_j|> L_4-L_0).
  \end{align}

Consider the first term on the right hand side of \eqref{eq: Bj-root}.  For any positive constant $C$,  we have
  \begin{align} \label{eq: Bj-bound}
       & P(\min_{j\in\mathcal{D}} |\widehat{B}_j|\leq L_4-L_0n^{-\kappa_1})
      \leq   P(\min_{j\in\mathcal{D}} |B_j| - \max_{j\in\mathcal{D}} |\widehat{B}_j-B_j |\leq L_4-L_0n^{-\kappa_1}) \nonumber\\
      \leq &  P( \max_{j\in\mathcal{D}} |\widehat{B}_j-B_j |\geq L_0n^{-\kappa_1})
    \leq |\mathcal{D}|\widetilde{C}_1\exp\left\{-\widetilde{C}_2n^{f(\kappa_1)}\right\},
  \end{align}
by noticing that $\min_{j\in\mathcal{D}}B_j\geq L_4$, where $\widetilde{C}_1$ and $\widetilde{C}_2$ are some positive constants.

Next consider the second term on the right hand side of \eqref{eq: Bj-root}.  Recall that $\min_{j\in\mathcal{D}}B_j\geq L_4$. Then, for any positive constant $C$,
  \begin{align} \label{eq: Bj-root-part2}
      & P(\max_{j\in\mathcal{D}} \frac{|\widehat{B}_j-B_j|}{|\sqrt{\widehat{B}_j}+\sqrt{B_j}|}\geq Cn^{-\kappa_1}, \min_{j\in\mathcal{D}} |\widehat{B}_j|> L_4-L_0)\nonumber\\
    \leq&    P\{\max_{j\in\mathcal{D}}|\widehat{B}_j-B_j| \geq C(\sqrt{L_4-L_0}+\sqrt{L_4})n^{-\kappa_1}\}
    \leq   |\mathcal{D}|\widetilde{C}_5\exp\left\{-\widetilde{C}_6n^{f(\kappa_1)}\right\},
  \end{align}
where $\widetilde{C}_5$ and $\widetilde{C}_6$ are some positive constants.
Combining \eqref{eq: Bj-root}, \eqref{eq: Bj-bound}, and \eqref{eq: Bj-root-part2} gives
  \begin{eqnarray}
           P(\max_{j\in\mathcal{D}} |\sqrt{\widehat{B}_j}-\sqrt{B_j}|\geq Cn^{-\kappa_1})
         \leq |\mathcal{D}|\widetilde{C}_3\exp\left\{-\widetilde{C}_4n^{f(\kappa_1)}\right\},
  \end{eqnarray}
where $\widetilde{C}_3=\widetilde{C}_1+\widetilde{C}_5$ and $\widetilde{C}_4=\min\{\widetilde{C}_2, \widetilde{C}_6\}$.

\medskip
\begin{lemma}\label{lem: AjB-prod}
Let $A_j$'s with $j \in \mathcal{D} \subset \{1, \cdots, p\}$ and $B$ satisfy $\max_{j\in\mathcal{D}} |A_j| \leq L_5$ and $|B|\leq L_6$ for some constants $L_5, L_6 > 0$, and $\widehat{A}_j$ and $\widehat{B}$ be estimates of $A_j$ and $B$, respectively, based on a sample of size $n$ for each $j\in \mathcal{D}$. Assume that for any constant $C > 0$, there exist constants $\widetilde{C}_1, \cdots, \widetilde{C}_8 > 0$ except $\widetilde{C}_3\geq 0$ such that
  \begin{align*}
      & P\left(\max_{j\in\mathcal{D}} |\widehat{A}_j-A_j|\geq Cn^{-\kappa_1}\right)
   \leq |D|\widetilde{C}_1\exp\left\{-\widetilde{C}_2n^{f(\kappa_1)}\right\}
         +\widetilde{C}_3\exp\left\{-\widetilde{C}_4n^{g(\kappa_1)}\right\}, \\
      & P\left( |\widehat{B}-B|\geq Cn^{-\kappa_1}\right)
       \leq \widetilde{C}_5\exp\left\{-\widetilde{C}_6n^{f(\kappa_1)}\right\}
         +\widetilde{C}_7\exp\left\{-\widetilde{C}_8n^{g(\kappa_1)}\right\}
  \end{align*}
with $f(\kappa_1)$ and $g(\kappa_1)$ some functions of $\kappa_1$.
Then for any constant $C > 0$, there exist constants $\widetilde{C}_9, \cdots, \widetilde{C}_{12}> 0$ such that
  \begin{eqnarray*}
      P\left(\max_{j\in\mathcal{D}} |\widehat{A}_j\widehat{B}-A_jB|\geq Cn^{-\kappa_1}\right)
         \leq  |D|\widetilde{C}_9\exp\left\{-\widetilde{C}_{10}n^{f(\kappa_1)}\right\}
              +\widetilde{C}_{11}\exp\left\{-\widetilde{C}_{12}n^{g(\kappa_1)}\right\}.
  \end{eqnarray*}
\end{lemma}

{\bf Proof.}
Note that $\max_{j\in\mathcal{D}} |\widehat{A}_j\widehat{B}-A_jB|\leq \max_{j\in\mathcal{D}} |\widehat{A}_j(\widehat{B}-B)| + \max_{j\in\mathcal{D}} |(\widehat{A}_j-A_j)B| $. Therefore, for any positive constant $C$,
  \begin{align} \label{eq: AjB}
       P(\max_{j\in\mathcal{D}} |\widehat{A}_j\widehat{B}-A_jB|\geq Cn^{-\kappa_1})
     &\leq P(\max_{j\in\mathcal{D}} |\widehat{A}_j(\widehat{B}-B)|\geq Cn^{-\kappa_1}/2)   \nonumber\\
    &\quad + P(\max_{j\in\mathcal{D}} |(\widehat{A}_j-A_j)B|\geq Cn^{-\kappa_1}/2).
  \end{align}
We first deal with the second term on the right hand side of \eqref{eq: AjB}.  Since $|B|\leq L_6$, we have
  \begin{align} \label{eq: AjB-part2}
       & P(\max_{j\in\mathcal{D}} |(\widehat{A}_j-A_j)B|\geq Cn^{-\kappa_1}/2)
      \leq    P(\max_{j\in\mathcal{D}} |\widehat{A}_j-A_j|L_6\geq Cn^{-\kappa_1}/2)  \nonumber\\
    &= P\{\max_{j\in\mathcal{D}} |\widehat{A}_j-A_j|\geq (2L_6)^{-1}Cn^{-\kappa_1}\} \nonumber\\
    &\leq |\mathcal{D}|\widetilde{C}_1\exp\left\{-\widetilde{C}_2n^{f(\kappa_1)}\right\}
     + \widetilde{C}_3\exp\left\{-\widetilde{C}_4n^{g(\kappa_1)}\right\}
  \end{align}
with constants $\widetilde{C}_1, \widetilde{C}_2, \widetilde{C}_4>0$ and $\widetilde{C}_3\geq 0$.

Next, we consider the first term on the right hand side of \eqref{eq: AjB}. Note that
  \begin{align} \label{eq: AjB-part1}
  & P(\max_{j\in\mathcal{D}} |\widehat{A}_j(\widehat{B}-B)|\geq Cn^{-\kappa_1}/2) \nonumber\\
   \leq&   P(\max_{j\in\mathcal{D}} |\widehat{A}_j(\widehat{B}-B)|\geq Cn^{-\kappa_1}/2, \max_{j\in\mathcal{D}}|\widehat{A}_j|\geq L_5+Cn^{-\kappa_1}/2)   \nonumber\\
   & \quad+ P(\max_{j\in\mathcal{D}} |\widehat{A}_j(\widehat{B}-B)|\geq Cn^{-\kappa_1}/2, \max_{j\in\mathcal{D}}|\widehat{A}_j|< L_5+Cn^{-\kappa_1}/2)  \nonumber\\
   \leq&   P(\max_{j\in\mathcal{D}}|\widehat{A}_j|\geq L_5+Cn^{-\kappa_1}/2)  \nonumber \\
   & \quad+ P(\max_{j\in\mathcal{D}} |\widehat{A}_j(\widehat{B}-B)|\geq Cn^{-\kappa_1}/2, \max_{j\in\mathcal{D}}|\widehat{A}_j|< L_5+C)  \nonumber\\
   \leq&   P(\max_{j\in\mathcal{D}}|\widehat{A}_j|\geq L_5+Cn^{-\kappa_1}/2)
   + P\{ (L_5+C)|\widehat{B}-B|\geq Cn^{-\kappa_1}/2\}.
  \end{align}
We will bound the two terms on the right hand side of \eqref{eq: AjB-part1} separately.  Since $\max_{j\in\mathcal{D}}|A_j|\leq L_5$, it holds that
  \begin{align}\label{eq: AjB-part1-1}
     P & (\max_{j\in\mathcal{D}}|\widehat{A}_j|\geq L_5+Cn^{-\kappa_1}/2)
     \leq  P(\max_{j\in\mathcal{D}}|\widehat{A}_j-A_j| +\max_{j\in\mathcal{D}}|A_j| \geq L_5+Cn^{-\kappa_1}/2)  \nonumber\\
    & \leq  P\{\max_{j\in\mathcal{D}}|\widehat{A}_j-A_j| \geq 2^{-1}Cn^{-\kappa_1}\} \nonumber\\
    & \leq  |\mathcal{D}|\widetilde{C}_5\exp\left\{-\widetilde{C}_6n^{f(\kappa_1)}\right\}
        +\widetilde{C}_7\exp\left\{-\widetilde{C}_8n^{g(\kappa_1)}\right\},
  \end{align}
where  $\widetilde{C}_5, \widetilde{C}_6, \widetilde{C}_8>0$ and $\widetilde{C}_7\geq 0$ are some constants.
We also have that
  \begin{align*}
       P & ( (L_5+C)|\widehat{B}-B|\geq Cn^{-\kappa_1}/2)
     =  P\{ |\widehat{B}-B|\geq (2L_5+2C)^{-1}Cn^{-\kappa_1}\} \nonumber \\
     & \leq  \widetilde{C}_{13}\exp\left\{-\widetilde{C}_{14}n^{f(\kappa_1)}\right\}
            + \widetilde{C}_{15}\exp\left\{-\widetilde{C}_{16}n^{g(\kappa_1)}\right\},
  \end{align*}
where $\widetilde{C}_{13}, \cdots, \widetilde{C}_{16}$ are some positive constants.  This, together with \eqref{eq: AjB}--\eqref{eq: AjB-part1-1},  entails that
  \begin{align*}
      & P(\max_{j\in\mathcal{D}} |\widehat{A}_j\widehat{B}-A_jB|\geq Cn^{-\kappa_1}) \\
  \leq& |\mathcal{D}|\widetilde{C}_1\exp\left\{-\widetilde{C}_2n^{f(\kappa_1)}\right\}
     + \widetilde{C}_3\exp\left\{-\widetilde{C}_4n^{g(\kappa_1)}\right\}
    + |\mathcal{D}|\widetilde{C}_5\exp\left\{-\widetilde{C}_6n^{f(\kappa_1)}\right\} \nonumber\\
   &   + \widetilde{C}_7\exp\left\{-\widetilde{C}_8n^{g(\kappa_1)}\right\}
     +  \widetilde{C}_{13}\exp\left\{-\widetilde{C}_{14}n^{f(\kappa_1)}\right\}
            + \widetilde{C}_{15}\exp\left\{-\widetilde{C}_{16}n^{g(\kappa_1)}\right\}\\
  \leq&     |D|\widetilde{C}_9\exp\left\{-\widetilde{C}_{10}n^{f(\kappa_1)}\right\}+ \widetilde{C}_{11}\exp\left\{-\widetilde{C}_{12}n^{g(\kappa_1)}\right\},
  \end{align*}
where $\widetilde{C}_9=\widetilde{C}_1+\widetilde{C}_5+\widetilde{C}_{13}>0$,  $\widetilde{C}_{10}=\min\{\widetilde{C}_2, \widetilde{C}_6, \widetilde{C}_{14}\}>0$,
$\widetilde{C}_{11}=\widetilde{C}_3+\widetilde{C}_7+\widetilde{C}_{15}>0$, and $\widetilde{C}_{12}=\min\{\widetilde{C}_4, \widetilde{C}_8, \widetilde{C}_{16}\}>0$.

\medskip
\begin{lemma}\label{lem: AjBj-ratio}
Let $A_j$'s and $B_j$'s with $j \in \mathcal{D} \subset \{1, \cdots, p\}$ satisfy $\max_{j\in\mathcal{D}}|A_j|\leq L_7$ and $\min_{j\in\mathcal{D}}|B_j|\geq L_8$ for some constants $L_7, L_8 > 0$, and $\widehat{A}_j$ and $\widehat{B}_j$ be estimates of $A_j$ and $B_j$, respectively, based on a sample of size $n$ for each $j\in \mathcal{D}$. Assume that for any constant $C > 0$, there exist constants $\widetilde{C}_1, \cdots, \widetilde{C}_6 > 0$ such that
  \begin{align*}
     & P\left(\max_{j\in\mathcal{D}} |\widehat{A}_j-A_j|\geq Cn^{-\kappa_1}\right)
        \leq  |\mathcal{D}|\widetilde{C}_1\exp\left\{-\widetilde{C}_2n^{f(\kappa_1)}\right\}
   +\widetilde{C}_3\exp\left\{-\widetilde{C}_4n^{g(\kappa_1)}\right\}, \\
     & P\left(\max_{j\in\mathcal{D}} |\widehat{B}_j-B_j|\geq Cn^{-\kappa_1}\right)
      \leq |D|\widetilde{C}_5\exp\left\{-\widetilde{C}_6n^{f(\kappa_1)}\right\}
  \end{align*}
with $f(\kappa_1)$ and $g(\kappa_1)$ some functions of $\kappa_1$.
Then for any constant $C > 0$, there exist constants $\widetilde{C}_7, \cdots, \widetilde{C}_{10} > 0$ such that
  \begin{align*}
      P&\left(\max_{j\in\mathcal{D}} \left|\widehat{A}_j/\widehat{B}_j-A_j/B_j\right|\geq Cn^{-\kappa_1}\right)
         \leq    |\mathcal{D}|\widetilde{C}_7\exp\left\{-\widetilde{C}_8n^{f(\kappa_1)}\right\}\\
         & \quad
   +\widetilde{C}_9\exp\left\{-\widetilde{C}_{10}n^{g(\kappa_1)}\right\}.
  \end{align*}
\end{lemma}

{\bf Proof.}  Since $\min_{j\in\mathcal{D}}B_j\geq L_8>0$, there exists some constant $L_0$ such that $0<L_0<L_8$. Note that, for any positive constant $C$,
  \begin{align} \label{eq: AjBj-ratio}
      & P(\max_{j\in\mathcal{D}} |\frac{\widehat{A}_j}{\widehat{B}_j}-\frac{A_j}{B_j}|\geq Cn^{-\kappa_1})  \nonumber \\
    \leq&  P(\max_{j\in\mathcal{D}} |\frac{\widehat{A}_j}{\widehat{B}_j}-\frac{A_j}{B_j}|\geq Cn^{-\kappa_1}, \min_{j\in\mathcal{D}} |\widehat{B}_j|\leq L_8-L_0n^{-\kappa_1}) \nonumber\\
      &\quad        +  P(\max_{j\in\mathcal{D}} |\frac{\widehat{A}_j}{\widehat{B}_j}-\frac{A_j}{B_j}|\geq Cn^{-\kappa_1}, \min_{j\in\mathcal{D}} |\widehat{B}_j|> L_8-L_0n^{-\kappa_1})  \nonumber \\
    \leq & P(\min_{j\in\mathcal{D}} |\widehat{B}_j|\leq L_8-L_0n^{-\kappa_1})
        + P(\max_{j\in\mathcal{D}} |\frac{\widehat{A}_j}{\widehat{B}_j}-\frac{A_j}{B_j}|\geq Cn^{-\kappa_1}, \min_{j\in\mathcal{D}} |\widehat{B}_j|> L_8-L_0).
  \end{align}

Let us consider the first term on the right hand side of \eqref{eq: AjBj-ratio}. Since $\min_{j\in\mathcal{D}}B_j\geq L_8$,  it holds that for any positive constant $C$,
  \begin{align} \label{eq: AjBj-ratio-part1}
       & P(\min_{j\in\mathcal{D}} |\widehat{B}_j|\leq L_8-L_0n^{-k})
     \leq   P(\min_{j\in\mathcal{D}} |B_j| - \max_{j\in\mathcal{D}} |\widehat{B}_j-B_j |\leq L_8-L_0n^{-\kappa_1})  \nonumber \\
      & \leq   P( \max_{j\in\mathcal{D}} |\widehat{B}_j-B_j |\geq L_0n^{-\kappa_1})
     \leq |\mathcal{D}|\widetilde{C}_1\exp\left\{-\widetilde{C}_2n^{f(\kappa_1)}\right\},
  \end{align}
where $\widetilde{C}_1$ and $\widetilde{C}_2$ are some positive constants.

The second term on the right hand side of \eqref{eq: AjBj-ratio} can be bounded as
  \begin{align} \label{eq: AjBj-ratio-part2}
      & P(\max_{j\in\mathcal{D}} |\frac{\widehat{A}_j}{\widehat{B}_j}-\frac{A_j}{B_j}|\geq Cn^{-\kappa_1}, \,\min_{j\in\mathcal{D}} |\widehat{B}_j|> L_8-L_0) \nonumber \\
     \leq & P(\max_{j\in\mathcal{D}} |\frac{\widehat{A}_j}{\widehat{B}_j}-\frac{A_j}{\widehat{B}_j}|\geq Cn^{-\kappa_1}/2, \,\min_{j\in\mathcal{D}} |\widehat{B}_j|> L_8-L_0)\nonumber\\
     & \quad+ P(\max_{j\in\mathcal{D}} |\frac{A_j}{\widehat{B}_j}-\frac{A_j}{B_j}|\geq Cn^{-\kappa_1}/2, \,\min_{j\in\mathcal{D}} |\widehat{B}_j|> L_8-L_0) \nonumber \\
     \leq & P\{\max_{j\in\mathcal{D}} |\widehat{A}_j-A_j|\geq2^{-1}(L_8-L_0)Cn^{-\kappa_1}\} \nonumber\\
     & \quad+ P\{\max_{j\in\mathcal{D}} |\widehat{B}_j-B_j|\geq(2L_7)^{-1} (L_8-L_0)L_8Cn^{-\kappa_1}\}\nonumber \\
    \leq & |\mathcal{D}|\widetilde{C}_3\exp\left\{-\widetilde{C}_4n^{f(\kappa_1)}\right\}
   +\widetilde{C}_5\exp\left\{-\widetilde{C}_6n^{g(\kappa_1)}\right\}
         +  |\mathcal{D}|\widetilde{C}_{11}\exp\left\{-\widetilde{C}_{12}n^{f(\kappa_1)}\right\},
  \end{align}
where $\widetilde{C}_3, \cdots, \widetilde{C}_6$, $\widetilde{C}_{11}$, and $\widetilde{C}_{12}$ are some positive constants.
Combining \eqref{eq: AjBj-ratio}--\eqref{eq: AjBj-ratio-part2} results in
  \begin{eqnarray*}
       P(\max_{j\in\mathcal{D}} |\widehat{A}_j/\widehat{B}_j-A_j/B_j|\geq Cn^{-\kappa_1})
         \leq   |\mathcal{D}|\widetilde{C}_7\exp\left\{-\widetilde{C}_8n^{f(\kappa_1)}\right\}
   +\widetilde{C}_9\exp\left\{-\widetilde{C}_{10}n^{g(\kappa_1)}\right\},
  \end{eqnarray*}
where $\widetilde{C}_7=\widetilde{C}_1+\widetilde{C}_3+\widetilde{C}_{11}>0$, $\widetilde{C}_8=\min\{\widetilde{C}_2, \widetilde{C}_4, \widetilde{C}_{12}\}>0$,  $\widetilde{C}_9=\widetilde{C}_5>0$, and $\widetilde{C}_{10}=\widetilde{C}_6>0$. This completes the proof of Lemma \ref{lem: AjBj-ratio}.

\newpage

\begin{table}
\caption{The overall and individual signal-to-noise ratios (SNRs) for simulation example in Section A.1 of Supplementary Material. Case 1: $\varepsilon_1 \sim N(0, 3^2)$, $\varepsilon_2 \sim N(0, 2.5^2)$, $\varepsilon_3 \sim N(0, 2.5^2)$, $\varepsilon_4 \sim N(0, 2^2)$; Case 2:  $\varepsilon_1 \sim N(0, 3.5^2)$, $\varepsilon_2 \sim N(0, 3^2)$, $\varepsilon_3 \sim N(0, 3^2)$, $\varepsilon_4 \sim N(0, 2.5^2)$; Case 3: $\varepsilon_1 \sim N(0, 4^2)$, $\varepsilon_2 \sim N(0, 3.5^2)$, $\varepsilon_3 \sim N(0, 3.5^2)$, $\varepsilon_4 \sim N(0, 3^2)$.} \vspace{0.1in}
\centering
\scalebox{0.8}{
\begin{tabular}{ l c c c c c c c c}
  \toprule
  && \multicolumn{2}{c}{Case 1} & \multicolumn{2}{c}{Case 2} & \multicolumn{2}{c}{Case 3}\\
  \cmidrule(lr{.5em}){3-4}  \cmidrule(lr{.5em}){5-6}  \cmidrule(lr{.5em}){7-8}
  && Settings 1,\,3 & Settings 2,\,4 & Settings 1,\,3 & Settings 2,\,4  & Settings 1,\,3 & Settings 2,\,4 \\
  \midrule
 M1 & $X_1$      & 0.44 & 0.44   & 0.33 & 0.33  & 0.25  & 0.25 \\
             & $X_5$       & 0.44 & 0.44  & 0.33 & 0.33  & 0.25  & 0.25 \\
             & $X_1X_5$  & 1 & 1.00      & 0.73 & 0.74  & 0.56  & 0.56\\
             & Overall      & 1.89 & 1.95  & 1.39 & 1.43  & 1.06  & 1.10\\
  \addlinespace
  M2 &  $X_1$     & 0.64 & 0.64   & 0.44 & 0.44  & 0.33  & 0.33\\
              & $X_{10}$  & 0.64 & 0.64  & 0.44 & 0.44  & 0.33  & 0.33\\
              & $X_1X_5$  & 1.44 & 1.45  & 1     & 1.00  & 0.73  & 0.74\\
              & Overall      & 2.72 & 2.73  & 1.89 & 1.89  & 1.39  & 1.39\\
  \addlinespace
  M3 &  $X_{10}$  & 0.64 & 0.64  & 0.44 & 0.44  & 0.33  & 0.33\\
              & $X_{15}$  & 0.64 & 0.64   & 0.44 & 0.44  & 0.33  & 0.33\\
              & $X_1X_5$  & 1.44 & 1.45  & 1 & 1.00  & 0.73  & 0.74\\
              & Overall      & 2.72 & 2.77   & 1.89 & 1.92  & 1.39  & 1.41\\
  \addlinespace
   M4 & $X_1X_5$                   & 2.25 & 2.26  & 1.44 & 1.45  & 1  & 1.00\\
               & $X_{10}X_{15}$  & 2.25 & 2.25  & 1.44 & 1.44  & 1  & 1.00\\
               & Overall                & 4.5   & 4.51 & 2.88 & 2.89   & 2  & 2.00\\
  \midrule
\end{tabular}
\label{tab:snr-supp}
}
\end{table}

\begin{table}
\caption{\label{supp:screen} The percentages of retaining each important interaction or main effect, and all important ones (All) by all the screening methods over different models and settings for simulation example in Section A.1 of Supplementary Material.} \vspace{0.1in}
\centering
\scalebox{0.75}{
\begin{tabular}{ l c c c c c c c c ccccccc }
  \toprule
   Method & \multicolumn{4}{c}{M1} & \multicolumn{4}{c}{M2} & \multicolumn{4}{c}{M3}  & \multicolumn{3}{c}{M4} \\
    \cmidrule(lr{.75em}){2-5}  \cmidrule(lr{.75em}){6-9} \cmidrule(lr|{.75em}){10-13}  \cmidrule(lr|{.75em}){14-16}
    & $X_1$ & $X_5$ &  $X_1X_5$  & All & $X_1$ & $X_{10}$ & $X_1X_5$  &All & $X_{10}$ & $X_{15}$ & $X_1X_5$  & All  & $X_1X_5$ & $X_{10}X_{15}$ & All  \\
  \midrule
    \multicolumn{15}{c}{{Case 1:} $\varepsilon_1 \sim N(0, 3^2)$, $\varepsilon_2 \sim N(0, 2.5^2)$, $\varepsilon_3 \sim N(0, 2.5^2)$, $\varepsilon_4 \sim N(0, 2^2)$} \\
SIS2&	1.00	&	1.00	&	1.00	&	1.00	&	1.00	&	1.00	&	0.15	&	0.15	&	1.00	&	1.00	&	0.00	&	0.00	&	0.01	&	0.04	&	0.00 \\
DC-SIS2&	1.00	&	1.00	&	1.00	&	1.00	&	1.00	&	1.00	&	0.76	&	0.76	&	1.00	&	1.00	&	0.02	&	0.02	&	0.08	&	0.07	&	0.01 \\
 SIRI$^*2$ &	1.00	&	1.00	&	1.00	&	1.00	&	1.00	&	0.99	&	0.60	&	0.59	&	0.99	&	0.99	&	0.07	&	0.07	&	0.26	&	0.23	&	0.07 \\
IP&	1.00	&	1.00	&	0.95	&	0.95	&	1.00	&	1.00	&	0.83	&	0.83	&	1.00	&	1.00	&	0.78	&	0.78	&	0.72	&	0.80	&	0.52 \\
\addlinespace
    \multicolumn{15}{c}{{Case 2:} $\varepsilon_1 \sim N(0, 3.5^2)$, $\varepsilon_2 \sim N(0, 3^2)$, $\varepsilon_3 \sim N(0, 3^2)$, $\varepsilon_4 \sim N(0, 2.5^2)$} \\
SIS2&	1.00	&	1.00	&	1.00	&	1.00	&	1.00	&	1.00	&	0.14	&	0.14	&	1.00	&	1.00	&	0.00	&	0.00	&	0.01	&	0.04	&	0.00\\
DC-SIS2&	1.00	&	1.00	&	1.00	&	1.00	&	1.00	&	1.00	&	0.58	&	0.58	&	1.00	&	1.00	&	0.00	&	0.00	&	0.06	&	0.03	&	0.00\\
SIRI$^*2$&	1.00	&	1.00	&	1.00	&	1.00	&	0.99	&	0.99	&	0.40	&	0.39	&	0.99	&	0.98	&	0.03	&	0.03	&	0.15	&	0.17	&	0.02\\
IP&	1.00	&	1.00	&	0.91	&	0.91	&	1.00	&	1.00	&	0.77	&	0.77	&	1.00	&	1.00	&	0.68	&	0.68	&	0.67	&	0.73	&	0.41\\
\addlinespace
      \multicolumn{15}{c}{{Case 3:} $\varepsilon_1 \sim N(0, 4^2)$, $\varepsilon_2 \sim N(0, 3.5^2)$, $\varepsilon_3 \sim N(0, 3.5^2)$, $\varepsilon_4 \sim N(0, 3^2)$} \\
SIS2&	1.00	&	1.00	&	1.00	&	1.00	&	1.00	&	0.99	&	0.13	&	0.13	&	1.00	&	1.00	&	0.00	&	0.00	&	0.00	&	0.04	&	0.00\\
DC-SIS2&	1.00	&	1.00	&	1.00	&	1.00	&	1.00	&	0.99	&	0.44	&	0.43	&	1.00	&	1.00	&	0.00	&	0.00	&	0.04	&	0.02	&	0.00\\
SIRI$^*2$	& 0.99	&	1.00	&	0.99	&	0.99	&	0.98	&	0.97	&	0.37	&	0.36	&	0.98	&	0.96	&	0.01	&	0.01	&	0.11	&	0.09	&	0.00\\
IP&	1.00	&	1.00	&	0.77	&	0.77	&	1.00	&	0.99	&	0.71	&	0.70	&	0.99	&	0.99	&	0.64	&	0.62	&	0.58	&	0.62	&	0.29\\
  \midrule
\end{tabular}
}
\end{table}

\begin{table}
\caption{The means and standard errors (in parentheses) of computation time in minutes of each method based on 100 replications for simulation example in Section A.2 of Supplementary Material.} \vspace{0.1in}
\centering
\scalebox{1}{
\begin{tabular}{lccc }
\midrule
 Method  &   $p=200$  &  $p=300$ & $p=500$  \\
\midrule
hierNet                  &   46.06 (0.69)  & 103.89 (1.20) & 292.77 (2.47) \\
IP-hierNet       &    5.44 (0.14) & 5.69 (0.11) & 6.05 (0.14) \\
Ratio of mean & 8.46 & 18.25 & 48.42 \\
\midrule
\end{tabular}
\label{tab:comptime}
}
\end{table}

\begin{table}[h]
\caption{\label{tab:no-interaction-screen} The percentages of retaining each important main effect, and all important ones (All) by all the screening methods over different settings for
simulation example M5 in Section A.3 of Supplementary Material.} \vspace{0.1in}
\centering
\scalebox{0.95}{
\begin{tabular}{ l ccccc cccccc}
  \toprule
   Method & \multicolumn{5}{c}{$\varepsilon\sim N(0, 2^2)$} & & \multicolumn{5}{c}{$\varepsilon\sim t_{(3)}$}  \\
    \cmidrule(lr{.75em}){2-6}  \cmidrule(lr{.75em}){8-12}
    & $X_1$ & $X_5$ &  $X_{10}$ & $X_{15}$ & All & & $X_1$ & $X_5$ &  $X_{10}$ & $X_{15}$ & All  \\
  \midrule
    \multicolumn{12}{c}{{Setting 1}: $(n, p, \rho)=(200, 2000, 0)$} \\
  SIS2      & 1.00  & 1.00  & 1.00  & 1.00   & 1.00  &   & 0.99  & 1.00  & 0.99  & 1.00   & 0.99\\
  DC-SIS2   & 1.00  & 1.00  & 1.00  & 1.00   & 1.00  &   & 1.00  & 1.00  & 1.00  & 1.00   & 1.00 \\
  SIRI$^*2$ & 0.95  & 0.94  & 0.93  & 0.94   & 0.77  &   & 1.00  & 1.00  & 0.98  & 0.99   & 0.97\\
  IP        & 1.00  & 1.00  & 1.00  & 1.00   & 1.00  &   & 0.99  & 1.00  & 0.99  & 1.00   & 0.99\\
\addlinespace
    \multicolumn{12}{c}{{Setting 2}: $(n, p, \rho)=(200, 2000, 0.5)$} \\
  SIS2      & 1.00  & 1.00  & 1.00  & 1.00   & 1.00  &   & 0.99  & 1.00  & 0.99  & 0.99   & 0.99\\
  DC-SIS2   & 1.00  & 1.00  & 1.00  & 1.00   & 1.00  &   & 1.00  & 1.00  & 1.00  & 1.00   & 1.00 \\
  SIRI$^*2$ & 0.99  & 1.00  & 0.98  & 0.91   & 0.88  &   & 1.00  & 1.00  & 1.00  & 1.00   & 1.00\\
  IP        & 1.00  & 1.00  & 1.00  & 1.00   & 1.00  &   & 0.99  & 1.00  & 0.99  & 0.99   & 0.99\\
\addlinespace
      \multicolumn{12}{c}{{Setting 3}: $(n, p, \rho)=(300, 5000, 0)$} \\
  SIS2      & 1.00  & 1.00  & 1.00  & 1.00   & 1.00  &   & 0.99  & 1.00  & 0.99  & 0.99   & 0.99\\
  DC-SIS2   & 1.00  & 1.00  & 1.00  & 1.00   & 1.00  &   & 1.00  & 1.00  & 1.00  & 1.00   & 1.00 \\
  SIRI$^*2$ & 1.00  & 1.00  & 0.99  & 0.99   & 0.98  &   & 1.00  & 1.00  & 1.00  & 1.00   & 1.00\\
  IP        & 1.00  & 1.00  & 1.00  & 1.00   & 1.00  &   & 0.99  & 1.00  & 0.99  & 0.99   & 0.99\\
\addlinespace
    \multicolumn{12}{c}{{Setting 4}: $(n, p, \rho)=(300, 5000, 0.5)$} \\
  SIS2      & 1.00  & 1.00  & 1.00  & 1.00   & 1.00  &   & 0.99  & 1.00  & 0.99  & 0.99   & 0.99\\
  DC-SIS2   & 1.00  & 1.00  & 1.00  & 1.00   & 1.00  &   & 1.00  & 1.00  & 1.00  & 1.00   & 1.00 \\
  SIRI$^*2$      & 1.00  & 1.00  & 1.00  & 1.00   & 1.00  &   & 1.00  & 1.00  & 1.00  & 1.00   & 1.00\\
  IP        & 1.00  & 1.00  & 1.00  & 1.00   & 1.00  &   & 0.99  & 1.00  & 0.99  & 0.99   & 0.99\\
  \midrule
\end{tabular}
}
\end{table}

\begin{table}[h]
\caption{\label{tab:screen-CS}The percentages of retaining each important interaction or main effect, and all important ones (All) by all the screening methods for interaction models M3$'$ and M4$'$ in Section A.4 of Supplementary Material.} \vspace{0.1in}
\centering
\begin{tabular}{ l c c c c c c cc }
  \toprule
   Method & \multicolumn{4}{c}{M3$'$} & & \multicolumn{3}{c}{M4$'$} \\
    \cmidrule(lr{.75em}){2-5}  \cmidrule(lr{.75em}){7-9}
    & $X_{10}$ & $X_{15}$ &  $X_1X_5$  & All & & $X_1X_5$  & $X_{10}X_{15}$ & All   \\
  \midrule
  SIS2      & 1.00  & 1.00  & 0.00  & 0.00   & & 0.01 & 0.02 & 0.00 \\
  DC-SIS2   & 1.00  & 1.00  & 0.00  & 0.00   & & 0.41 & 0.35 & 0.16 \\
  SIRI*2      & 1.00  & 1.00  & 0.05  & 0.05   & & 0.40 & 0.46 & 0.20 \\
  IP        & 1.00  & 1.00  & 0.69  & 0.69   & & 0.54 & 0.45 & 0.26 \\
  \midrule
\end{tabular}
\end{table}


\begin{thebibliography}{}
\bibitem[\protect\citeauthoryear{Barut, Fan, and Verhasselt}{Barut
  et~al.}{2016}]{barut2016conditional}
Barut, E., Fan, J. and Verhasselt, A. (2016),
``Conditional sure independence screening,''
\textit{Journal of American Statistical Association}, to appear.


\bibitem[\protect\citeauthoryear{Bickel, Ritov, and Tsybakov}{Bickel
  et~al.}{2009}]{bickel2009simultaneous}
Bickel, P.~J., Ritov, Y. and Tsybakov, A.~B. (2009),
``Simultaneous analysis of lasso and dantzig selector,''
\textit{The Annals of Statistics}, 37, 1705--1732.

\bibitem[\protect\citeauthoryear{Bien, Taylor, and Tibshirani}{Bien
  et~al.}{2013}]{bien2013lasso}
Bien, J., Taylor, J. and Tibshirani, R. (2013),
``A lasso for hierarchical interactions,''
\textit{The Annals of Statistics}, 41, 1111--1141.


\bibitem[\protect\citeauthoryear{Breiman}{Breiman}{1995}]{breiman1995better}
Breiman, L. (1995),
``Better subset regression using the nonnegative garrote,''
\textit{Technometrics}, 37, 373--384.


\bibitem[\protect\citeauthoryear{Candes and Tao}{Candes and
  Tao}{2007}]{candes2007dantzig}
Candes, E. and Tao, T. (2007),
``The Dantzig selector: Statistical estimation when $p$ is much larger than $n$,''
\textit{The Annals of Statistics}, 35, 2313--2351.


\bibitem[\protect\citeauthoryear{Chang, T.~Y et al.}{Chang et al.}{2013}]{chang2013marginal}
Chang, J., Tang, C.~Y. and Wu, Y. (2013),
``Marginal empirical likelihood and sure independence feature screening,''
\textit{The Annals of Statistics}, 41, 2123--2148.


\bibitem[\protect\citeauthoryear{Cheng, M.~Y et al.}{Cheng et al.}{2014}]{cheng2014nonparametric}
Cheng, M.~Y, Honda, T., Li, J. and Peng, H. (2014),
``Nonparametric independence screening and structural identification for ultra-high dimensional longitudinal data,''
\textit{The Annals of Statistics}, 42, 1819--1849.

\bibitem[\protect\citeauthoryear{Cheng, W.~S. et al.}{Cheng et al.}{2003}]{cheng2003}
Cheng, W.~S., Giandomenico, V., Pastan, I. and Essand, M. (2003),
``Characterization of the androgen-regulated prostate-specific T cell receptor gamma-chain alternate reading frame protein (TARP) promoter,''
\textit{Endocrinology}, 144, 3433--3440.


\bibitem[\protect\citeauthoryear{Cho and Fryzlewicz}{Cho and
  Fryzlewicz}{2012}]{cho2012high}
Cho, H. and Fryzlewicz, P. (2012),
``High dimensional variable selection via tilting,''
\textit{Journal of the Royal Statistical Society}, Series B, 74, 593--622.




\bibitem[\protect\citeauthoryear{Choi, Li, and Zhu}{Choi
  et~al.}{2010}]{choi2010variable}
Choi, N.~H., Li, W. and Zhu, J. (2010),
``Variable selection with the strong heredity constraint and its oracle
  property,''
\textit{Journal of the American Statistical Association}, 105, 354--364.



\bibitem[\protect\citeauthoryear{Cordell}{Cordell}{2009}]{cordell2009detecting}
Cordell, H.~J. (2009),
``Detecting gene--gene interactions that underlie human diseases,''
\textit{Nature Reviews Genetics}, 10, 392--404. 


\bibitem[\protect\citeauthoryear{Cui, Hengjian and Li, Runze and Zhong, Wei}{Cui
  et~al.}{2015}]{cui2015model}
Cui, H., Li, R. and Zhong, W. (2015),
``Model-free feature screening for ultrahigh dimensional discriminant analysis,''
\textit{Journal of the American Statistical Association}, 110, 630--641.


\bibitem[\protect\citeauthoryear{Culverhouse, R. et al.}{Culverhouse et al.}{2002}]{Culverhouse2002}
Culverhouse, R., Suarez, B.~K., Lin, J. and Reich, T. (2002),
``A perspective on epistasis: limits of models displaying no main effect,''
\textit{The American Journal of Human Genetics}, 70, 461--471.


\bibitem[\protect\citeauthoryear{Fan and Fan}{Fan and Fan}{2008}]{fan2008high}
Fan, J. and Fan, Y. (2008),
``High dimensional classification using features annealed independence rules,''
\textit{The Annals of Statistics}, 36, 2605--2637.

\bibitem[\protect\citeauthoryear{Fan, Feng, and Song}{Fan
  et~al.}{2011}]{fan2011nonparametric}
Fan, J., Feng, Y. and Song, R. (2011),
``Nonparametric independence screening in sparse ultra-high-dimensional
  additive models,''
\textit{Journal of the American Statistical Association}, 106, 544--557.



\bibitem[\protect\citeauthoryear{Fan and Li}{Fan and
  Li}{2001}]{fan2001variable}
Fan, J. and Li, R. (2001),
``Variable selection via nonconcave penalized likelihood and its oracle
  properties,''
\textit{Journal of the American Statistical Association}, 96, 1348--1360.



\bibitem[\protect\citeauthoryear{Fan and Lv}{Fan and Lv}{2008}]{fan2008sure}
Fan, J. and Lv, J. (2008),
``Sure independence screening for ultrahigh dimensional feature space'' (with discussion),
\textit{Journal of the Royal Statistical Society}, Series B, 70, 849--911.



\bibitem[\protect\citeauthoryear{Fan, Samworth, and Wu}{Fan
  et~al.}{2009}]{fan2009ultrahigh}
Fan, J., Samworth, R. and Wu, Y. (2009),
``Ultrahigh dimensional feature selection: beyond the linear model,''
\textit{Journal of Machine Learning Research}, 10, 2013--2038.



\bibitem[\protect\citeauthoryear{Fan and Song}{Fan and Song}{2010}]{fan2010sure}
Fan, J. and Song, R. (2010),
``Sure independence screening in generalized linear models with NP-dimensionality,''
\textit{The Annals of Statistics}, 38, 3567--3604.





\bibitem[\protect\citeauthoryear{Fan and Lv}{Fan and
  Lv}{2014}]{fan2014asymptotic}
Fan, Y. and Lv, J. (2014),
``Asymptotic properties for combined {$L_1$} and
  concave regularization,''
\textit{Biometrika}, 101, 57--70.




\bibitem[\protect\citeauthoryear{Furusato B, Tan SH, Young D, Dobi A, Sun C, Mohamed AA, Thangapazham R, Chen Y, McMaster G, Sreenath T, Petrovics G, McLeod DG, Srivastava S, and Sesterhenn IA.}{Furusato et al.}{2010}]{Furusato2010}
Furusato, B., Tan, S. H., Young, D., Dobi, A., Sun, C., Mohamed, A. A., Thangapazham, R., Chen, Y., McMaster, G., Sreenath, T., Petrovics, G., McLeod, D. G., Srivastava, S. and Sesterhenn, I. A. (2010),
``ERG oncoprotein expression in prostate cancer: clonal progression of ERG-positive tumor cells and potential for ERG-based stratification,''
\textit{Prostate Cancer and Prostatic Diseases}, 13, 228--237.


%

\bibitem[\protect\citeauthoryear{Hall and Xue}{Hall and
  Xue}{2014}]{hall2014selecting}
Hall, P. and Xue, J.-H. (2014),
``On selecting interacting features from high-dimensional data,''
\textit{Computational Statistics and Data Analysis}, 71, 694--708.




\bibitem[\protect\citeauthoryear{Ning Hao, Yang Feng, Hao Helen Zhang}{Hao et al.}{2015}]{hao2015model}
Hao, N., Feng, Y. and Zhang, H. H. (2015),
``Model selection for high dimensional quadratic regression via regularization,''
\textit{Preprint}, arXiv:1501.00049.



\bibitem[\protect\citeauthoryear{Hao and Zhang}{Hao and
  Zhang}{2014}]{hao2014interaction}
Hao, N. and Zhang, H. H. (2014),
``Interaction screening for ultra-high dimensional data,''
\textit{Journal of the American Statistical Association}, 109, 1285--1301.




\bibitem[\protect\citeauthoryear{He, Xuming and Wang, Lan and Hong, Hyokyoung Grace}{He et al.}{2013}]{he2013quantile}
He, X., Wang, L. and Hong, H. G. (2013),
``Quantile-adaptive model-free variable screening for high-dimensional heterogeneous data,''
\textit{The Annals of Statistics}, 41, 342--369.



\bibitem[\protect\citeauthoryear{Hillerdal, V., Nilsson, B., Carlsson, B., Eriksson, F. and Essand, M.}{Hillerdal et al.}{2012}]{Hillerdal2012}
Hillerdal, V., Nilsson, B., Carlsson, B., Eriksson, F. and Essand, M. (2012),
``T cells engineered with a T cell receptor against the prostate antigen TARP specifically kill HLA-A2+ prostate and breast cancer cells,''
\textit{Proceedings of the National Academy of Sciences}, 109, 15877--15881.

\bibitem[\protect\citeauthoryear{Hoeffding}{Hoeffding}{1963}]{hoeffding1963probability}
Hoeffding, W. (1963),
``Probability inequalities for sums of bounded random variables,''
\textit{Journal of the American Statistical Association}, 58, 13--30.

\bibitem[\protect\citeauthoryear{Holt, S.~K., Kwon, E.~M., Lin, D.~W., Ostrander, E.~A., Stanford J.~L.}{Holt et al.}{2010}]{Holt2010}
Holt, S.~K., Kwon, E.~M., Lin, D.~W., Ostrander, E.~A. and Stanford, J.~L. (2010),
``Association of hepsin gene variants with prostate cancer risk and prognosis,''
\textit{Prostate}, 70, 1012--1019.


\bibitem[\protect\citeauthoryear{Hunter}{Hunter}{2005}]{hunter2005gene}
Hunter, D.~J. (2005),
``Gene--environment interactions in human diseases,''
\textit{Nature Reviews Genetics}, 6, 287--298.

\bibitem[\protect\citeauthoryear{Jiang and Liu}{Jiang and
  Liu}{2014}]{jiang2014sliced}
Jiang, B. and Liu, J.~S. (2014),
``Variable selection for general index models via sliced inverse regression,''
\textit{The Annals of Statistics}, 42, 1751--1786.


\bibitem[\protect\citeauthoryear{Klezovitch O, Risk M, Coleman I, Lucas JM, Null M, True LD, Nelson PS, Vasioukhin V}{Klezovitch et al.}{2008}]{Klezovitch2008}
Klezovitch, O., Risk, M., Coleman, I., Lucas, J. M., Null, M., True, L. D., Nelson, P. S. and Vasioukhin, V.(2008),
``A causal role for ERG in neoplastic transformation of prostate epithelium,''
\textit{Proceedings of the National Academy of Sciences}, 105, 2105--2110.




\bibitem[\protect\citeauthoryear{Li, Zhong, and Zhu}{Li
  et~al.}{2012}]{li2012feature}
Li, R., Zhong, W. and Zhu, L. (2012),
``Feature screening via distance correlation learning,''
\textit{Journal of the American Statistical Association}, 107, 1129--1139.




\bibitem[\protect\citeauthoryear{Lv and Fan}{Lv and Fan}{2009}]{lv2009unified}
Lv, J. and Fan, Y. (2009),
``A unified approach to model selection and sparse recovery using
  regularized least squares,''
\textit{The Annals of Statistics}, 37, 3498--3528.





\bibitem[\protect\citeauthoryear{McCarthy, N.}{Musani}{2013}]{McCarthy2013prostate}
McCarthy, N. (2013),
``Prostate cancer: understanding why,''
\textit{Nature Reviews Cancer}, 13, 754.

\bibitem[\protect\citeauthoryear{Musani, S.~K., Shriner, D., Liu, N., Feng, R., Coffey, C.~S., Yi, N., Tiwari, H.~K. and Allison, D.~B.}{Musani et al.}{2007}]{musani2007detection}
Musani, S.~K., Shriner, D., Liu, N., Feng, R., Coffey, C.~S., Yi, N., Tiwari, H.~K. and Allison, D.~B. (2007),
``Detection of gene$\times$gene interactions in genome-wide association studies of human population data,''
\textit{Human Heredity}, 63, 67--84.









\bibitem[\protect\citeauthoryear{Oh, S., Terabe, M., Pendleton, C.~D., Bhattacharyya, A., Bera, T.~K., Epel, M., Reiter, Y., Phillips, J., Linehan, W.~M., Kasten-Sportes, C., Pastan, I. and Berzofsky, J.~A.}{Oh et al.}{2004}]{Oh2004}
Oh, S., Terabe, M., Pendleton, C.~D., Bhattacharyya, A., Bera, T.~K., Epel, M., Reiter, Y., Phillips, J., Linehan, W.~M., Kasten-Sportes, C., Pastan, I. and Berzofsky, J.~A. (2004),
``Human CTLs to wild-type and enhanced epitopes of a novel prostate and breast tumor-associated protein, TARP, lyse human breast cancer cells,''
\textit{Cancer Research}, 64, 2610--2618. 
\bibitem[\protect\citeauthoryear{Ritchie, M.~D. et al.}{Ritchie et al.}{2001}]{Ritchie2001}
Ritchie, M.~D., Hahn, L.~W., Roodi, N., Bailey, R., Dupont, W.~D., Parl, F.~F. and Moore, J.~H. (2001),
``Multifactor-dimensionality reduction reveals high-order interactions among estrogen-metabolism genes in sporadic breast cancer,''
\textit{The American Journal of Human Genetics}, 69, 138--147.


\bibitem[\protect\citeauthoryear{Saleem, M., Kweon, M.~H., Johnson, J.~J., Adhami, V.~M., Elcheva, I., Khan, N., Bin Hafeez, B., Bhat, K.~M., Sarfaraz, S., Reagan-Shaw, S., Spiegelman, V.~S., Setaluri, V. and Mukhtar, H.}{Saleem et al.}{2006}]{Saleem2006}
Saleem, M., Kweon, M.~H., Johnson, J.~J., Adhami, V.~M., Elcheva, I., Khan, N., Bin Hafeez, B., Bhat, K.~M., Sarfaraz, S., Reagan-Shaw, S., Spiegelman, V.~S., Setaluri, V. and Mukhtar, H. (2006),
``S100A4 accelerates tumorigenesis and invasion of human prostate cancer through the transcriptional regulation of matrix metalloproteinase 9,''
\textit{Proceedings of the National Academy of Sciences}, 103, 14825--14830.


\bibitem[\protect\citeauthoryear{Schwender and Ickstadt}{Schwender and Ickstadt}{2008}]{schwender2008identification}
Schwender, H. and Ickstadt, K. (2008), 
``Identification of SNP interactions using logic regression,''
\textit{Biostatistics}, 9, 187--198.


\bibitem[\protect\citeauthoryear{Shao and Zhang}{Shao and Zhang}{2014}]{shao2014martingale}
Shao, X. and Zhang, J. (2014), 
``Martingale difference correlation and its use in high-dimensional variable screening,''
\textit{Journal of the American Statistical Association}, 109, 1302--1318.


\bibitem[\protect\citeauthoryear{Singh, Febbo, Ross, Jackson, Manola, Ladd,
  Tamayo, Renshaw, D'Amico, Richie, Lander, Loda, Kantoff, Golub, and
  Sellers}{Singh et~al.}{2002}]{singh2002gene}
Singh, D., Febbo, P.~G., Ross, K., Jackson, D.~G., Manola, J., Ladd, C., Tamayo, P.,
 Renshaw, A.~A., D'Amico, A.~V., Richie, J.~P., Lander, E.~S., Loda, M.,
 Kantoff, P.~W., Golub, T.~R. and Sellers, W.~R. (2002),
``Gene expression correlates of clinical prostate cancer behavior,''
\textit{Cancer Cell}, 1, 203--209.





\bibitem[\protect\citeauthoryear{Tibshirani}{Tibshirani}{1996}]{tibshiranit1996regression}
Tibshirani, R. (1996), 
``Regression shrinkage and selection via the Lasso,''
\textit{Journal of the Royal Statistical Society}, Series B, 58, 267--288.

\bibitem[\protect\citeauthoryear{van~der Vaart and Wellner}{van~der Vaart and
  Wellner}{1996}]{van1996weak}
van~der Vaart, A. and Wellner, J.~A. (1996),
\textit{Weak Convergence and Empirical Processes: With Applications to
  Statistics}, New York: Springer.

\bibitem[\protect\citeauthoryear{Wolfgang, C.~D., Essand, M., Vincent, J.~J., Lee, B., Pastan, I.}{Wolfgang et~al.}{2000}]{Wolfgang2000}
Wolfgang, C.~D., Essand, M., Vincent, J.~J., Lee, B. and Pastan, I. (2000),
``TARP: a nuclear protein expressed in prostate and breast cancer cells derived from an alternate reading frame of the T cell receptor gamma chain locus,''
\textit{Proceedings of the National Academy of Sciences}, 97, 9437--9442.


\bibitem[\protect\citeauthoryear{Xu et al.}{Xu et al.}{2004}]{xu2004interaction}
Xu, J., Langefeld, C.~D., Zheng, S.~L., Gillanders, E.~M., Chang, B.-L., Isaacs, S.~D. and others (2004),
``Interaction effect of PTEN and CDKN1B chromosomal regions on prostate cancer linkage,''
\textit{Human Genetics}, 115, 255--262.


\bibitem[\protect\citeauthoryear{Yuan, Joseph, and Zou}{Yuan
  et~al.}{2009}]{yuan2009structured}
Yuan, M., Joseph, V.~R. and Zou, H. (2009),
``Structured variable selection and estimation,''
\textit{Annals of Applied Statistics}, 3, 1738--1757.


\bibitem[\protect\citeauthoryear{Zhang}{Zhang}{2010}]{zhang10mcp}
Zhang, C.-H. (2010),
``Nearly unbiased variable selection under minimax concave penalty,''
\textit{The Annals of Statistics}, 38, 894--942.


\bibitem[\protect\citeauthoryear{Zou}{Zou}{2006}]{zou2006adaptive}
Zou, H. (2006),
``The adaptive lasso and its oracle properties,''
\textit{Journal of the American Statistical Association}, 101, 1418--1429.

\bibitem[\protect\citeauthoryear{Zou and Hastie}{Zou and
  Hastie}{2005}]{zou2005regularization}
Zou, H. and Hastie, T. (2005),
``Regularization and variable selection via the elastic net,''
\textit{Journal of the Royal Statistical Society}, Series B, 67, 301--320.



\end{thebibliography}
\end{document}